\documentclass{sig-alternate}
\usepackage{hyperref}

\newif\ifextended\extendedtrue

\graphicspath{{Figures/}}
\DeclareGraphicsExtensions{.pdf,.jpeg,.png}

\usepackage{amsmath,amssymb}
\usepackage[tight,footnotesize]{subfigure}
\usepackage{url}

\usepackage{listings}
\usepackage{xcolor}

\usepackage{xspace}

\usepackage{triotex}
\newcommand{\avg}[1]{\mathbb{E}\left[#1\right]}
\newcommand{\EV}[1]{\avg{#1}}
\newcommand{\lophi}{\mathcal{L}}
\newcommand{\phim}{\phi_{\text{mean}}^c(k)}
\newcommand{\phimed}{\phi_{\text{median}}^c(k)}
\newcommand{\intau}{\widetilde{\phi}(k)}

\newcommand{\ETH}{${}^*$\xspace}
\newcommand{\ABB}{${}^\dagger$}
\newcommand{\UY}{${}^\ddagger$\xspace}
\newcommand{\ITMO}{${}^\S$\xspace}

\begin{document}

\ifextended
\else
\conferenceinfo{SAC'13}{March 18--22, 2013, Coimbra, Portugal.}
\CopyrightYear{2013}
\crdata{978-1-4503-1656-9/13/03}
\fi

\title{The Search for the Laws of Automatic Random Testing}

\numberofauthors{1}
\let\email=\affaddr  
\author{
\alignauthor
Carlo~A.~Furia\ETH $\cdot$
Bertrand~Meyer\ETH $\cdot$
Manuel Oriol\ABB\UY $\cdot$
Andrey Tikhomirov\ITMO $\cdot$
Yi Wei\ETH \\[2mm]
\begin{tabular}{c@{$\qquad$}c}
       \affaddr{\ETH{}Chair of Software Engineering, ETH Zurich} 
       & \affaddr{\ABB{}\xspace{}ABB Corporate Research, Industrial Software Systems} \\
       \affaddr{Zurich, Switzerland} 
       & \affaddr{Baden-D\"attwil, Switzerland} \\
       \email{\{firstname.lastname@inf.ethz.ch\}} 
       & \email{manuel.oriol@ch.abb.com} \\
       \\[-1mm]
       \affaddr{\UY{Dept.\ of Computer Science, University of York}}
       & \affaddr{\ITMO{Software Engineering Laboratory, ITMO}} \\
       \affaddr{York, UK} 
       & \affaddr{St.~Petersburg, Russia} \\
       & \email{and.tikhomirov@gmail.com}
\end{tabular}
}

\ifextended
\toappear{}
\fi

\maketitle
\begin{abstract}
  Can one estimate the number of remaining faults in a software
  system? A credible estimation technique would be immensely useful to
  project managers as well as customers. It would also be of
  theoretical interest, as a general law of software engineering. We
  investigate possible answers in the context of automated random
  testing, a method that is increasingly accepted as an effective way
  to discover faults. Our experimental results, derived from best-fit
  analysis of a variety of mathematical functions, based on a large
  number of automated tests of library code equipped with automated
  oracles in the form of contracts, suggest a poly-logarithmic
  law. Although further confirmation remains necessary on different
  code bases and testing techniques, we argue that understanding the
  laws of testing may bring significant benefits for estimating the
  number of detectable faults and comparing different projects and
  practices.
\end{abstract}


\section{Introduction}

A scientific discipline is characterized by general laws, such as
Maxwell's equations. Where the topics of discourse involve human
phenomena, and belong to engineering rather than science, the laws
cannot be absolute truths comparable to the laws of nature, and often
involve a probabilistic element; but they should still describe
properties and constraints that govern all or almost all instances of
the phenomena under consideration.

Software engineering is particularly in need of such laws. Some useful
ones have already been identified, in areas such as project
management; an example \cite{Boehm-economics, McConnel-estimation} is
the observation, first expressed by Barry Boehm on the basis of his
study of a large database of projects, that every software project has
a \emph{nominal} cost, deducible from the project's overall characteristics,
from which it is not possible to deviate -- whether by giving the
project more time or by including more developers -- by more than
about 25\%.

One area in which such a general law would be of particular interest
is testing. It has been well known since at least Lehmann and Belady's
seminal work on faults (``bugs'') in successive IBM OS 360
releases~\cite{BeladyL76} that if a project tracks faults in a
sufficiently diligent way the evolution of the number of faults in
successive releases follows regular patterns. Is it possible to turn
this general, informal observation into a precise \emph{law of
  software testing}, on which project managers and developers could
rely to estimate, from the number of faults uncovered so far through
testing, the number of \emph{detectable} faults in an individual module
or an entire system?  This is the question that we set out to explore,
and on which we are reporting our first results.

\subsection{Expected benefits}

Were it available, and backed by experimental evidence covering a
broad enough base of projects and environments to make it credible, a
law describing the rate of fault detection would be of considerable
interest to the industry.

An immediate benefit would be to help project managers answer one of
the most common and also toughest questions they face: \emph{can we
  ship yet}?  A release should be shipped early enough to avoid
missing a market opportunity or being scooped by the competition; but
not too early if this means that so many faults remain as to provoke
an adverse reaction from the market. 
Another application, for individual developers, is to estimate the
amount of testing that remains necessary for a module or a
subsystem.
Yet another benefit would be to help assess how a project's or
organization's fault patterns differ, for better or worse, from
average industry practices.  Such an assessment would be particularly
appropriate in organizations applying strict guidelines, for example
as part of CMMI \cite{CMMI}.
Also of interest is the purely intellectual benefit of gaining
insights into the nature of software development, in the form of a
general law of software engineering that describes fault patterns and
might help in the effort to avoid faults in the first place.

\subsection{Automatic testing}

In the present work, we set out to determine through empirical study
if a general law of testing exists, predicting the number of detectable
faults. The context of the study is \emph{automatic} testing, in which
test cases are not manually prepared by humans but generated
automatically through appropriate algorithms. More specifically, we
rely on automated \emph{random} testing, where these algorithms use
partly random techniques. Once dismissed as ineffective \cite{Myers},
random testing has shown itself, in recent years, to be a viable
technique; such tools as AutoTest~\cite{bertrand2009} for Eiffel and
Randoop~\cite{PachecoLEB07} and YETI~\cite{YETI} for Java and .NET,
which apply automatic random testing, succeed by the criterion that
matters most in practice: finding faults in programs, including
released libraries and systems.

With automated testing, the user selects a number of modules -- in
object-oriented programming, classes -- for testing; the testing tool
then creates objects (class instances) and performs method calls on
them with arguments selected according to a strategy that involves a
random component. It then records any abnormal behavior (failure) that
may be signaling a fault in the software.


The work described here relies on automated random testing as just
described. This approach has the advantage of simplicity, and of not
relying on hard-to-control human actions. Another advantage is
reproducibility (up to the variation caused by the choice of seeds in
random number generation). It also has the disadvantage of limiting
the generalizability of our findings
(Section~\ref{sec:threats-validity}); but it is hardly surprising that
a general empirical understanding of testing is likely to require much
more investigation and varied experiments.

The paper's main experiments involve software equipped with
\emph{contracts} (specification elements embedded in the software and
subject to run-time evaluation, such as pre- and postconditions and
class invariants); in such a context, the testing process focuses on
generating failures due to contract violations.  Since contracts are
meant to express the \emph{specification} of the methods and classes
being tested, they provide testing \emph{oracles}~\cite{StaatsWH11} to
reliably identify \emph{faults} (as opposed to mere failures).  The
bulk of our experiments (Section~\ref{sec:exper-with-eiff}) target
code with contracts and studies the evolution of found faults over
time.  Therefore, the goal is to obtain experimentally a general law
$\mathcal{F}(t)$, where $t$ is the testing time, measured in number of
drawn test cases and $\mathcal{F}(t)$ is the number of unique faults
uncovered by the tests up to time $t$.  

If contracts are not available, a fully automated approach must rely
on failures such as arithmetic overflow or running out of memory. In
these cases, a methodological issue arises in practice since automated
testing finds failures rather than faults; obtaining $\mathcal{F}$
requires a sound policy to determine whether two given failures relate
to the same fault. We hint at some ways to address this problem in
Section~\ref{sec:exper-with-java}.  

\textbf{Extended version.} More detailed data and graphs are available in an extended version of the paper at:
\ifextended
\begin{center}
Appendix: Section~\ref{sec:appendix} of this paper.
\end{center}
\else
\begin{center}
\url{http://arxiv.org/abs/XXXX.XXXX}
\end{center}
\fi

\section{Experiments}
The experiments run repeated random testing sessions on Eiffel classes (Section~\ref{sec:exper-with-eiff}) with AutoTest, and on Java classes (Section~\ref{sec:exper-with-java}) with YETI.
Each testing \emph{session} is characterized by:
\begin{itemize}
\item $c$: the tested \emph{class};
\item \textsc{T}: the number of rounds in the simulation, that is the number of \emph{test cases} (valid and invalid) drawn;
\item $\varphi: [0..\textsc{T}] \rightarrow \naturals$: the function counting the cumulative number of unique failures or faults after each round.
\end{itemize}
The dataset $D^c$ for a class $c$ consists of a collection of counting
functions $\varphi_1, \varphi_2, \ldots, \varphi_{\textsc{S}^c}$,
where \textsc{S}$^c$ is the total number of testing sessions on $c$.
In our experiments, all source code was tested ``as is'', without
injecting any artificial error or modifying it in any way.

Given a dataset $D^c$, consider the mean function of the fault count:
\begin{equation*}
\phim \quad = \quad \frac{1}{\textsc{S}^c} \sum_{1 \leq i \leq \textsc{S}^c} \varphi_i(k)
\end{equation*}
as well as the median $\phimed$.
We considered nine model functions, described in Section~\ref{sec:model-functions}, suggested by various sources, and fit each model function to each of $\phim$ and $\phimed$ from the experimental data.
We used Matlab 2011 for all data analysis and fitting computations.
The following sections illustrate the main results.

\subsection{Model functions}\label{sec:model-functions}
We consider a number of ``reasonable'' fitting functions, suggested either by intuition or by the theoretical model discussed in Section~\ref{sec:coupon-collection}.
Some of the functions are special cases of other functions; nonetheless, it is advisable to try to fit both -- special and general case -- because the performance of the fitting algorithms may be sensitive to the form in which a function is presented.
In all the examples, lower- and upper-case names other than the argument $x$ denote parameters to be instantiated by fitting.

Following an original intuition of analogy with biological phenomena -- suggested in related work~\cite{Oriol12} -- we consider the Michaelis-Menten equation:
\begin{equation}
\Phi_1(x) \quad = \quad
\frac
{a x}
{x + B}
\end{equation}
as well as its generalizations into a rational function of third degree:
\begin{equation}
\Phi_2(x) \quad = \quad
\frac
{a x^3 + b x^2 + c x + d}
{A x^3 + B x^2 + C x + D}
\end{equation}
and a rational function of arbitrary degree:
\begin{equation}
\Phi_3(x) \quad = \quad
\frac
{a x^b + c}
{A x^B + C}
\end{equation}
The model analyzed in Section~\ref{sec:coupon-collection} suggests including functional forms similar to polynomials, as well as logarithms and exponentials.
Hence, we consider logarithms to any power:
\begin{equation}
\Phi_4(x) \quad = \quad
a \log^b (x+1) + c
\end{equation}
and poly-logarithms of third degree:
\begin{equation}
\Phi_5(x) \quad = \quad
a \log^3 (x+1) + b \log^2 (x+1) + c \log (x+1) + d
\end{equation}
where the translation coefficient $+ 1$ is needed because the fault function $\phi$ is such that $\phi(0) = 0$, but $\log (0) = -\infty$.
The base of the logarithm is immaterial because the multiplicative parameters can accommodate any.
We also consider exponentials of powers:
\begin{equation}
\Phi_6(x) \quad = \quad
a b^{x^{1/c}} + d
\end{equation}
Polynomials up to degree three are covered by $\Phi_2$, but we still consider some special cases explicitly; third degree:
\begin{equation}
\Phi_7(x) \quad = \quad
{a x^3 + b x^2 + c x + d}
\end{equation}
arbitrary degree:
\begin{equation}
\Phi_8(x) \quad = \quad
{a x^b + c}
\end{equation}
and third degree with negative powers:
\begin{equation}
\Phi_{9}(x) \quad = \quad
{a x^{-3} + b x^{-2} + c x^{-1} + d}
\end{equation}

\subsection{Experiments with Eiffel code} \label{sec:exper-with-eiff}
Experiments with Eiffel used AutoTest~\cite{bertrand2009}; they targeted 42 Eiffel classes, from the widely used open-source data structure libraries EiffelBase~\cite{EiffelBase} and Gobo~\cite{Gobo}.
Table~\ref{table:at-datastats} reports the statistics about the testing sessions with AutoTest on these classes; for each class, the table lists the number of testing sessions (\textsc{S}), the test cases sampled per session (\textsc{T}), the maximum number of unique faults found (\textsc{F}, corresponding to unique contract violations), the mean sampled standard deviation and skewness ($\EV{\sigma}, \EV{\gamma}$) of the faults found across the class's multiple testing sessions, and the mean and standard deviation ($\EV{\Delta}, \sigma[\Delta]$) of the new faults found with every test case sampled (thus, for example, $\EV{\Delta} = 10^{-3}$ means that a new fault is found every thousand test cases on average).
The bottom rows display mean, median, and standard deviation of the values in each column. 
The data was collected according to the suggested guidelines for empirical evaluation of randomized algorithms~\cite{AB11-icse}; in particular, the very large number of drawn test cases makes the averaged data robust with respect to statistical fluctuations.

\begin{table*}[!t]
\renewcommand{\arraystretch}{0.9}
\caption{Eiffel classes statistics.}
\label{table:at-datastats}
\centering
\scriptsize
\begin{tabular}{ l  r  r@{.}l@{E}l  r  r@{.}l  r@{.}l@{E}l  r@{.}l@{E}l  r@{.}l@{E}l }
 {\textsc{Class}} & {\textsc{S}} & \multicolumn{3}{c}{\textsc{T}} & {\textsc{F}} & \multicolumn{2}{c}{$\avg{\sigma}$} & \multicolumn{3}{c}{$\avg{\gamma}$} & \multicolumn{3}{c}{$\avg{\Delta}$} & \multicolumn{3}{c}{$\sigma[\Delta]$} \\ 
\hline
 ACTIVE\_LIST  &   11  &  2&14&+06  &   18  &  1&0553  &  -4&88&+00  &  8&39&-04  &  2&27&-19 \\
 ARRAY  &   11  &  1&58&+06  &   69  &  5&0972  &  -3&36&+00  &  4&24&-03  &  9&02&-05 \\
 ARRAYED\_CIRCULAR  &   15  &  7&77&+05  &   52  &  5&2882  &  -2&09&+00  &  6&12&-03  &  2&75&-04 \\
 ARRAYED\_LIST  &   14  &  1&25&+06  &   24  &  1&4279  &  -4&06&+00  &  1&70&-03  &  9&89&-05 \\
 ARRAYED\_QUEUE  &   12  &  1&99&+06  &    6  &  0&3742  &  -2&00&+01  &  2&56&-04  &  1&45&-05 \\
 ARRAYED\_SET  &   15  &  2&69&+06  &   15  &  1&0823  &  -3&72&+00  &  5&34&-04  &  2&35&-05 \\
 BINARY\_TREE  &   13  &  2&25&+06  &   40  &  5&3739  &  -2&14&+00  &  1&66&-03  &  7&56&-05 \\
 BOUNDED\_QUEUE  &   14  &  2&34&+06  &    5  &  0&3135  &  -1&46&+01  &  2&14&-04  &  0&00&+00 \\
 COMPACT\_CURSOR\_TREE  &   14  &  5&56&+05  &   67  &  5&6622  &  -2&47&+00  &  1&13&-02  &  3&35&-04 \\
 DS\_ARRAYED\_LIST  &   12  &  2&07&+06  &   91  &  7&0082  &  -3&69&+00  &  4&29&-03  &  6&65&-05 \\
 DS\_AVL\_TREE  &   15  &  5&10&+05  &   16  &  2&0347  &  -1&82&+00  &  2&53&-03  &  2&02&-04 \\
 DS\_BILINKED\_LIST  &   14  &  2&70&+06  &   30  &  2&2035  &  -3&77&+00  &  8&90&-04  &  9&43&-05 \\
 DS\_BINARY\_SEARCH\_TREE  &   11  &  8&95&+05  &   15  &  1&8224  &  -1&93&+00  &  1&51&-03  &  9&16&-05 \\
 DS\_BINARY\_SEARCH\_TREE\_SET  &   12  &  5&96&+05  &    5  &  0&4048  &  -6&31&+00  &  5&18&-04  &  1&67&-04 \\
 DS\_HASH\_SET  &   14  &  1&86&+06  &   19  &  1&5352  &  -1&32&+00  &  6&31&-04  &  1&83&-04 \\
 DS\_HASH\_TABLE  &   14  &  1&66&+06  &   25  &  2&0158  &  -1&39&+00  &  1&07&-03  &  1&88&-04 \\
 DS\_LEFT\_LEANING\_RED\_BLACK\_TREE  &   14  &  7&18&+05  &   33  &  2&9464  &  -3&45&+00  &  4&32&-03  &  1&73&-04 \\
 DS\_LINKED\_LIST  &   14  &  2&17&+06  &   94  &  7&7943  &  -2&67&+00  &  3&88&-03  &  1&88&-04 \\
 DS\_LINKED\_QUEUE  &   14  &  2&23&+06  &    9  &  1&3880  &  -4&98&-01  &  3&05&-04  &  5&89&-05 \\
 DS\_LINKED\_STACK  &   14  &  2&22&+06  &   10  &  1&4648  &  -5&39&-01  &  3&57&-04  &  6&23&-05 \\
 DS\_MULTIARRAYED\_HASH\_SET  &   14  &  2&61&+06  &   22  &  3&7948  &  -7&99&-01  &  7&61&-04  &  5&60&-05 \\
 DS\_RED\_BLACK\_TREE  &   15  &  6&62&+05  &   32  &  2&9321  &  -3&55&+00  &  4&62&-03  &  9&67&-05 \\
 FIXED\_TREE  &   13  &  7&54&+05  &   49  &  4&9311  &  -2&47&+00  &  6&25&-03  &  1&67&-04 \\
 HASH\_TABLE  &   14  &  2&95&+06  &   24  &  3&6472  &  -8&98&-01  &  6&59&-04  &  8&49&-05 \\
 LINKED\_AUTOMATON  &   14  &  4&03&+06  &   14  &  0&4384  &  -1&74&+01  &  3&28&-04  &  1&06&-05 \\
 LINKED\_CIRCULAR  &   15  &  1&50&+06  &   40  &  4&2747  &  -1&92&+00  &  2&50&-03  &  1&10&-04 \\
 LINKED\_CURSOR\_TREE  &   14  &  1&34&+06  &   51  &  4&6304  &  -1&75&+00  &  3&56&-03  &  1&42&-04 \\
 LINKED\_DFA  &   13  &  2&72&+06  &  135  &  12&9783  &  -1&97&+00  &  4&67&-03  &  1&69&-04 \\
 LINKED\_LIST  &   14  &  3&41&+06  &   21  &  1&4436  &  -2&84&+00  &  5&30&-04  &  3&89&-05 \\
 LINKED\_PRIORITY\_QUEUE  &   14  &  1&20&+06  &   23  &  2&1477  &  -2&34&+00  &  1&80&-03  &  6&35&-05 \\
 LINKED\_SET  &   14  &  2&59&+06  &   29  &  2&1800  &  -3&45&+00  &  1&03&-03  &  4&48&-05 \\
 LINKED\_TREE  &   15  &  1&68&+06  &  136  &  22&3981  &  -9&56&-01  &  7&20&-03  &  4&06&-04 \\
 MULTI\_ARRAY\_LIST  &   14  &  4&73&+05  &   35  &  3&4960  &  -3&22&+00  &  6&92&-03  &  3&66&-04 \\
 PART\_SORTED\_SET  &   13  &  3&27&+06  &   33  &  3&6535  &  -2&12&+00  &  9&60&-04  &  3&43&-05 \\
 PART\_SORTED\_TWO\_WAY\_LIST  &   13  &  2&61&+06  &   36  &  3&7867  &  -2&01&+00  &  1&28&-03  &  5&73&-05 \\
 SORTED\_TWO\_WAY\_LIST  &   15  &  2&60&+06  &   34  &  3&3816  &  -2&37&+00  &  1&22&-03  &  4&64&-05 \\
 SUBSET\_STRATEGY\_TREE  &   15  &  3&08&+06  &   20  &  1&8972  &  -2&87&+00  &  5&84&-04  &  3&47&-05 \\
 TWO\_WAY\_CIRCULAR  &   13  &  1&53&+06  &   41  &  4&3326  &  -1&88&+00  &  2&46&-03  &  1&37&-04 \\
 TWO\_WAY\_CURSOR\_TREE  &   13  &  1&35&+06  &   50  &  4&5548  &  -1&86&+00  &  3&56&-03  &  1&11&-04 \\
 TWO\_WAY\_LIST  &   15  &  2&01&+06  &   22  &  1&7119  &  -3&45&+00  &  9&61&-04  &  6&42&-05 \\
 TWO\_WAY\_SORTED\_SET  &   14  &  3&09&+06  &   47  &  4&2784  &  -3&43&+00  &  1&47&-03  &  3&29&-05 \\
 TWO\_WAY\_TREE  &   14  &  2&46&+06  &  163  &  28&6749  &  -1&02&+00  &  6&27&-03  &  1&77&-04 \\
\hline
 {\textbf{Mean}} &    14 &   1&93&+06 &    40 &   4&3299 &   -3&55&+00 &   2&54&-03 &   1&15&-04 \\ 
 {\textbf{Median}} &    14 &   2&04&+06 &    31 &   3&1640 &   -2&42&+00 &   1&49&-03 &   9&09&-05 \\ 
 {\textbf{Stdev}} &     1 &   9&04&+05 &    36 &   5&3955 &   4&08&+00 &   2&49&-03 &   9&59&-05 \\ 
\end{tabular} 
\end{table*}

\subsubsection{Fitting results}
Table~\ref{table:at-bestfits-mean} reports the result of fitting the
models $\Phi_1$--$\Phi_9$ on the mean fault count $\phim$ curves; for
each class, column $\textsc{Best Fit Ranking}$ ranks the $\Phi_i$'s
from the best fitting to the worst, according to the coefficient of
determination $R^2$ (the higher, the better), and the root mean
squared error RMSE (the smaller, the better); the rankings according
to the two measures agreed in all experiments.  The other columns
report the $R^2$ and RMSE scores of the best fit, and the absolute
value of the difference between such scores for the best fit and the
same scores for $\Phi_5$.  The data shows that $\Phi_5$ emerges as
consistently better than the other functions; as the bottom of the
table summarizes, $\Phi_5$ is the best fit in 62\% and one of the top
two fits in 90\% of the cases; when it is not the best, it fares
within 1\% of the best in terms of scores.  A Wilcoxon signed-rank
test confirms that the observed differences between $\Phi_5$ and the
other functions are highly statistically significant: for example, a
comparison of the $R^2$ values indicates that the values for $\Phi_5$
are different (smaller) with high probability ($7.14 \cdot 10^{-7} < p
< 1.65\cdot 10^{-8}$) and large effect size (between $0.54$ and
$0.62$, computed using the $Z$-statistics).  The same data with
respect to the median curves is qualitatively the same as with the
mean; hence we do not report it in detail.
\begin{table*}[!hbt]
\renewcommand{\arraystretch}{0.9}
\caption{Testing of Eiffel classes: best fits with mean.}
\label{table:at-bestfits-mean}
\centering
\scriptsize
\begin{tabular}{ l  c  r@{.}l  r@{.}l  r@{.}l  r@{.}l }
 {\textsc{Class}} & {\textsc{Fit Ranking}} & \multicolumn{2}{c}{$R^2$ (best)} & \multicolumn{2}{c}{\textsc{RMSE} (best)} & \multicolumn{2}{c}{$R^2$ ($\Delta$ best)} & \multicolumn{2}{c}{\textsc{RMSE} ($\Delta$ best)} \\ 
\hline
 ACTIVE\_LIST  &    2  5  4  1  9  6  8  3  7  &  0&9825  &  0&1304  &  0&0108  &  -0&0353 \\
 ARRAY  &    5  4  1  9  8  6  3  7  2  &  0&9894  &  0&5151  &  0&0000  &  0&0000 \\
 ARRAYED\_CIRCULAR  &    5  3  4  8  1  2  9  6  7  &  0&9978  &  0&2461  &  0&0000  &  0&0000 \\
 ARRAYED\_LIST  &    3  5  4  1  8  9  6  7  2  &  0&9877  &  0&1485  &  0&0202  &  -0&0932 \\
 ARRAYED\_QUEUE  &    8  5  4  3  1  9  2  6  7  &  0&7299  &  0&1547  &  0&0027  &  -0&0008 \\
 ARRAYED\_SET  &    5  4  1  2  8  9  6  3  7  &  0&9550  &  0&2078  &  0&0000  &  0&0000 \\
 BINARY\_TREE  &    3  5  4  8  1  7  9  6  2  &  0&9987  &  0&1943  &  0&0030  &  -0&1566 \\
 BOUNDED\_QUEUE  &    5  4  1  3  2  9  6  8  7  &  0&8048  &  0&1043  &  0&0000  &  0&0000 \\
 COMPACT\_CURSOR\_TREE  &    5  4  3  8  1  2  9  6  7  &  0&9830  &  0&7254  &  0&0000  &  0&0000 \\
 DS\_ARRAYED\_LIST  &    5  4  1  8  2  9  6  3  7  &  0&9860  &  0&8201  &  0&0000  &  0&0000 \\
 DS\_AVL\_TREE  &    5  4  8  3  1  2  9  6  7  &  0&9861  &  0&2299  &  0&0000  &  0&0000 \\
 DS\_BILINKED\_LIST  &    5  4  3  1  2  9  6  8  7  &  0&9947  &  0&1528  &  0&0000  &  0&0000 \\
 DS\_BINARY\_SEARCH\_TREE  &    5  3  4  8  1  2  9  6  7  &  0&9913  &  0&1618  &  0&0000  &  0&0000 \\
 DS\_BINARY\_SEARCH\_TREE\_SET  &    8  4  5  3  2  1  9  6  7  &  0&8727  &  0&1145  &  0&0912  &  -0&0355 \\
 DS\_HASH\_SET  &    8  5  4  3  1  2  9  6  7  &  0&9782  &  0&2085  &  0&0021  &  -0&0097 \\
 DS\_HASH\_TABLE  &    8  5  4  3  1  9  6  7  2  &  0&9653  &  0&3536  &  0&0053  &  -0&0258 \\
 DS\_LEFT\_LEANING\_RED\_BLACK\_TREE  &    5  4  3  1  8  9  6  7  2  &  0&9888  &  0&3015  &  0&0000  &  0&0000 \\
 DS\_LINKED\_LIST  &    5  4  8  1  2  9  6  3  7  &  0&9905  &  0&7505  &  0&0000  &  0&0000 \\
 DS\_LINKED\_QUEUE  &    8  4  5  7  3  1  9  6  2  &  0&9940  &  0&0986  &  0&0094  &  -0&0595 \\
 DS\_LINKED\_STACK  &    8  3  4  5  1  9  6  2  7  &  0&9914  &  0&1265  &  0&0077  &  -0&0480 \\
 DS\_MULTIARRAYED\_HASH\_SET  &    5  8  4  3  7  1  2  9  6  &  0&9852  &  0&4465  &  0&0000  &  0&0000 \\
 DS\_RED\_BLACK\_TREE  &    5  4  1  9  6  8  3  2  7  &  0&9915  &  0&2625  &  0&0000  &  0&0000 \\
 FIXED\_TREE  &    5  4  1  8  2  9  6  3  7  &  0&9920  &  0&4335  &  0&0000  &  0&0000 \\
 HASH\_TABLE  &    8  5  4  3  7  1  2  9  6  &  0&9979  &  0&1621  &  0&0010  &  -0&0353 \\
 LINKED\_AUTOMATON  &    1  2  5  9  6  4  8  3  7  &  0&9693  &  0&0669  &  0&2517  &  -0&1361 \\
 LINKED\_CIRCULAR  &    5  3  4  1  8  2  9  6  7  &  0&9982  &  0&1774  &  0&0000  &  0&0000 \\
 LINKED\_CURSOR\_TREE  &    5  8  4  3  1  9  6  7  2  &  0&9904  &  0&4413  &  0&0000  &  0&0000 \\
 LINKED\_DFA  &    5  4  8  3  1  2  9  6  7  &  0&9954  &  0&8751  &  0&0000  &  0&0000 \\
 LINKED\_LIST  &    3  5  4  8  1  2  9  6  7  &  0&9678  &  0&2441  &  0&0005  &  -0&0017 \\
 LINKED\_PRIORITY\_QUEUE  &    5  4  3  8  1  2  9  6  7  &  0&9866  &  0&2391  &  0&0000  &  0&0000 \\
 LINKED\_SET  &    5  4  1  8  2  9  6  3  7  &  0&9749  &  0&3313  &  0&0000  &  0&0000 \\
 LINKED\_TREE  &    5  4  8  3  7  1  2  9  6  &  0&9992  &  0&6490  &  0&0000  &  0&0000 \\
 MULTI\_ARRAY\_LIST  &    5  1  4  8  9  6  3  2  7  &  0&9951  &  0&2402  &  0&0000  &  0&0000 \\
 PART\_SORTED\_SET  &    5  4  3  8  1  2  9  6  7  &  0&9976  &  0&1747  &  0&0000  &  0&0000 \\
 PART\_SORTED\_TWO\_WAY\_LIST  &    5  4  8  1  3  2  9  6  7  &  0&9851  &  0&4510  &  0&0000  &  0&0000 \\
 SORTED\_TWO\_WAY\_LIST  &    5  4  3  8  1  2  9  6  7  &  0&9790  &  0&4763  &  0&0000  &  0&0000 \\
 SUBSET\_STRATEGY\_TREE  &    5  3  4  1  8  2  9  6  7  &  0&9871  &  0&2066  &  0&0000  &  0&0000 \\
 TWO\_WAY\_CIRCULAR  &    5  3  4  8  1  2  9  6  7  &  0&9953  &  0&2898  &  0&0000  &  0&0000 \\
 TWO\_WAY\_CURSOR\_TREE  &    5  4  1  2  9  8  6  3  7  &  0&9897  &  0&4469  &  0&0000  &  0&0000 \\
 TWO\_WAY\_LIST  &    5  4  1  8  2  9  6  3  7  &  0&9862  &  0&1899  &  0&0000  &  0&0000 \\
 TWO\_WAY\_SORTED\_SET  &    5  4  1  8  3  2  9  6  7  &  0&9933  &  0&3427  &  0&0000  &  0&0000 \\
 TWO\_WAY\_TREE  &    4  5  8  3  7  1  2  9  6  &  0&9993  &  0&7493  &  0&0000  &  -0&0191 \\
\hline
 {} &  69\% (5 is best) &  \multicolumn{2}{c}{} &  \multicolumn{2}{c}{\textbf{Mean}} &   0&0097 &   -0&0156 \\ 
\end{tabular} 
\end{table*}

In a few cases, models other than $\Phi_5$ achieve a higher fitting score even if visual inspection of the curves shows that it is obviously unsatisfactory.
The reasons for this behavior are arguably due to numerical errors; there are, however, a few cases of fits with high scores but visibly unsatisfactory, whose fitting process converged without Matlab signaling any numerical approximation problem. In any case, visual inspection confirms that $\Phi_5$ is a visibly proper fit to $\varphi$ in all cases.

\subsubsection{Poly-logarithmic fits}
Once acknowledged that the evidence points towards a poly-logarithmic law, it remains the question of which polynomial degree achieves the best fit.
Consider seven poly-logarithmic model functions $\lophi_k$, $1 \leq k \leq 7$, of various degrees.
$\lophi_1$--$\lophi_5$ have degree equal to the number of their nonzero coefficients plus one: 
\begin{equation}
\lophi_k(x) \quad = \quad
c_0 + \sum_{1 \leq j \leq k} c_j \log^{j}(x+1)\,, 1 \leq k \leq 5
\end{equation}
$\lophi_6$ and $\lophi_7$ are, instead, binomials whose degree is a parameter; namely, $\lophi_6$ equals $\Phi_4$ and:
\begin{equation}
\lophi_7(x) \quad = \quad
a\,\log^{1/b}(x+1) + c
\end{equation}

The results show that it is always the case that $\lophi_5 > \lophi_4 > \lophi_3 > \lophi_2 > \lophi_1$, that is the higher the degree the better the fit; in hindsight, this is an unsurprising consequence of the fact that models of higher degree offer more degrees of freedom, hence they are more flexible.
$\lophi_3 = \Phi_5$, however, still fares quite well on average; the few cases where its performance is as much as 10\% worse than $\lophi_5$ correspond to more irregular curves with fewer faults and hence fewer new datapoints, such as for class \mbox{\lstinline|ARRAYED_QUEUE|.}
Finally, $\lophi_6$ and $\lophi_7$ occasionally rank third; whenever this happens, however, the difference with the best (and with $\Phi_5$) is negligible.
In all, $\Phi_5$ seems to be a reasonable compromise between flexibility and economy, but poly-logarithmic functions of higher order could be considered when useful.
Notice that, even if in principle we could obtain enough degrees of freedom with any function of arbitrarily high degree, only poly-logarithmic works for reasonable degrees; for example, polynomials of fifth degree (which generalize $\Phi_7$) never provide good fits.

\lstset{language=Java,columns=flexible,sensitive,mathescape=true}
\subsection{Experiments with Java code} \label{sec:exper-with-java}

With the goal of confirming (or invalidating) the results of the Eiffel experiments, we used YETI to test 9 Java classes from \mbox{\lstinline|java.|$\!\!$\lstinline|util|} and 29 classes from \lstinline|java.lang|. For lack of space, we only present a summary of the results and highlight the differences in comparison with Eiffel and the questions about the comparison that remain open.

Unlike in Eiffel, where the code has contracts, interpreting a Java failure
to know whether it is a ``real'' fault cannot be done in a fully
automated way in normal Java code without contracts.  To address this
point at least in part, YETI collects all the exceptions triggered
during random testing and considers the following exceptions as
\emph{not} provoked by a fault, but merely accidents of the testing
process: (1) declared exceptions, including \lstinline+RuntimeException+, as client code
knows that they may be thrown and may even be part of a code pattern; (2) \lstinline+InvalidArgumentException+, comparable to precondition violations.
This still falls short of a complete identification of real faults, but it helps to reduce the number of spurious failures. 

Given this filtering performed by YETI, the experiments with the 38 Java classes are in two parts.
The first part targets the 9 \lstinline|java.util| classes and 2 classes from \mbox{\lstinline|java.lang|,} and analyzes all unique \emph{failures} filtered by YETI as described above (two failures are the same if they generate the same stack exception trace).
The second part targets the 29 classes from \lstinline|java.lang| and only reports failures manually pruned by discarding all failures that reflect behavior compatible with the informal documentation (for example, division by zero when the informal API documentation requires an argument to be non-zero).
In this case, we get to a fairly reasonable approximation of real \emph{faults}; hence we will refer to the first part of Java experiments as ``failures'' and to the second as ``faults''.

Tables~\ref{table:j1-datastats}--\ref{table:j2-datastats} report the summary statistics about the testing sessions with YETI on all classes, with the same data as for Eiffel (see Section~\ref{sec:exper-with-eiff}).
The data for faults, where manual pruning eliminated the vast majority of failures as spurious, has far fewer significant datapoints.
Skewness is not computable when there are no faults found, which happens for 15 classes; hence the summary statistics about skewness are immaterial (NaN stands for ``Not a Number'').

\begin{table}[!htb]
\renewcommand{\arraystretch}{0.9}
\setlength{\tabcolsep}{2pt}
\caption{Java classes tested for failures: statistics.}
\label{table:j1-datastats}
\centering
\scriptsize
\begin{tabular}{ l  r  r@{.}l@{E}l  r  r@{.}l  r@{.}l@{E}l  r@{.}l@{E}l  r@{.}l@{E}l }
 {\textsc{Summary}} & {\textsc{S}} & \multicolumn{3}{c}{\textsc{T}} & {\textsc{F}} & \multicolumn{2}{c}{$\avg{\sigma}$} & \multicolumn{3}{c}{$\avg{\gamma}$} & \multicolumn{3}{c}{$\avg{\Delta}$} & \multicolumn{3}{c}{$\sigma[\Delta]$} \\ 
\hline
 {\textbf{Mean}} &    30 &   1&23&+05 &    32 &   3&9044 &   -8&02&+00 &   1&03&-02 &   1&86&-04 \\ 
 {\textbf{Median}} &    30 &   1&18&+05 &    15 &   1&7799 &   -6&10&+00 &   7&16&-03 &   6&68&-05 \\ 
 {\textbf{Stdev}} &   &   1&65&+04 &    29 &   3&8803 &   4&70&+00 &   8&86&-03 &   2&63&-04 \\ 
\end{tabular} 
\end{table}

\begin{table}[!htb]
\renewcommand{\arraystretch}{0.9}
\setlength{\tabcolsep}{2pt}
\caption{Java classes tested for faults: statistics.}
\label{table:j2-datastats}
\centering
\scriptsize
\begin{tabular}{ l  r  r@{.}l@{E}l  r  r@{.}l  r@{.}l@{E}l  r@{.}l@{E}l  r@{.}l@{E}l }
 {\textsc{Summary}} & {\textsc{S}} & \multicolumn{3}{c}{\textsc{T}} & {\textsc{F}} & \multicolumn{2}{c}{$\avg{\sigma}$} & \multicolumn{3}{c}{$\avg{\gamma}$} & \multicolumn{3}{c}{$\avg{\Delta}$} & \multicolumn{3}{c}{$\sigma[\Delta]$} \\ 
\hline
 {\textbf{Mean}} &     4 &   4&68&+05 &     2 &   0&2356 &   \multicolumn{3}{c}{NaN} &   1&73&-03 &   2&51&-05 \\ 
 {\textbf{Median}} &     4 &   2&61&+05 &   &   0&0000 &   \multicolumn{3}{c}{NaN} &   0&00&+00 &   0&00&+00 \\ 
 {\textbf{Stdev}} &     1 &   9&45&+05 &     3 &   0&4094 &   \multicolumn{3}{c}{NaN} &   2&72&-03 &   1&26&-04 \\ 
\end{tabular} 
\end{table}

\subsubsection*{Fitting results}

We tried to fit the models $\Phi_1, \Phi_2, \Phi_4, \Phi_5$ on the mean fault count $\phim$ 
curves, for both Java failures and faults.
The data is somewhat harder to fit than in the Eiffel experiments, and in fact we had to exclude the other models because they could produce converging fixes in only a fraction of the experiments.
In comparison with the Eiffel data, there is a detectable overall difference: $\Phi_2$ and $\Phi_1$ seem to emerge as the best models, whereas $\Phi_5$ is best only in a limited number of cases. Looking at the $R^2$ scores, $\Phi_1$ and $\Phi_2$ alternate as best model for the failure curves, and their difference is often small ($p = 0.067$ and effect size $0.398$); for the fault curves, $\Phi_1$ ranks first in the majority of classes, but its difference w.r.t.\ $\Phi_2$ is statistically insignificant ($p = 0.5$ and effect size $0.095$).
A common phenomenon is that the majority of curves $\phim$ with YETI show an horizontal asymptote, which $\Phi_1$ or $\Phi_2$ can accommodate much better than $\Phi_5$ can.

A closer look suggests that the differences w.r.t.\ the Eiffel experiments may still be reconcilable.
First, in the majority of cases of failures, $\Phi_5$ fits to within the 5\% of the best model; the only exceptions are the \lstinline|java.util| classes \lstinline|ArrayList|, \lstinline|Hashtable|, and \lstinline|LinkedList| where visual inspection shows curves with a steeper initial phase that $\Phi_2$ fits best, and where $\Phi_1$ and $\Phi_5$ behave similarly.
For the fault experiments, the picture is more varied; but $\Phi_5$ fits to within the 20\% of the best models in all classes but three; and the quality of fitting a certain model is much more varied because of the small number of faults found in these experiments (but we excluded from these statistics the classes with no faults found).
Finally, the difference between $\Phi_2$ and $\Phi_5$ is often statistically insignificant; the difference between $\Phi_1$ and $\Phi_5$ is significant ($p \simeq 0.2$) but not large for faults (effect size $0.3$); in contrast, the difference between $\Phi_5$ and $\Phi_4$ is highly statistically significant and large ($p < 10^{-3}$ and effect size from $0.43$ to $0.62$).
In all, further experiments are necessary to conclusively determine what the exact magnitude and the ultimate causes of these differences are: possible candidates are the fault density of the software tested, the details of the algorithms implemented by the testing tools (AutoTest and YETI), and the availability of contracts to detect faults.

\section{Discussion and Threats} \label{sec:discussion-threats}

\subsection{Towards a justification for the law}
\label{sec:coupon-collection}
Arcuri et al.~\cite{ArcuriRT} model random testing as a \emph{coupon collector problem}~\cite{Shioda}: 
``Consider a box which contains $N$ types of numerous objects. 
An object in the box is repeatedly sampled on a random basis.
Let $p_i > 0$ denote the probability that a type-$i$ object is sampled.
The successive samplings are statistically independent and the sampling probabilities, $p_1, p_2, \ldots, p_N$, are fixed.
When a type-$i$ object is sampled for the first time, we say that a type-$i$ object is detected.
To find the number of samplings required for detecting a set of object types (say, object types indexed by $i = 1, \ldots, n \leq N$) is traditionally called \emph{coupon collector problem}.''
%
In random testing, the objects are test cases $U$, and each type $U_i \subseteq U$ is a testing \emph{target}: a unique failure or fault in our experiments.

Following \cite{Shioda}, let $\tau_i$ be a random variable denoting the number of samplings required to draw a test case in target $U_i$, and let $\tau(n)$ be $\max\{\tau_1, \ldots, \tau_n\}$, for $n \leq N$, that is the number of samplings to cover all the first $n$ targets (in any order).
It is possible to prove that the expected value of $\tau(n)$ is:
\begin{equation}
\avg{\tau(n)} \quad=\quad \sum_{1 \leq i \leq n}(-1)^{i+1} \sum_{J ; |J| = i} \frac{1}{\sum_{j \in J}p_j}
\label{eq:expExact}
\end{equation}
The inverse $\intau$ of $\tau(n)$ is a function from the number of test cases to the expected number of failures, hence it corresponds to an analytic version of $\phim$.

It is possible to approximate $\tau(n)$ for two special cases of probabilities $p_i$'s: when they are all equal ($p_1 = p_2 = \ldots = p_N = \theta$), and when they are exponentially decreasing ($p_i = \theta/10^{i-1}$).
We can show that, in the first case, $\tau(n)$ is polynomial in $\Theta(\log^h(n))$ for some constant $k$, hence $\intau$ is $\Theta(\exp(g k))$ for some constant $g$; in the second case, $\tau(n)$ is $\Theta(\exp(h n))$, hence $\intau$ is $\Theta(\log^h(k))$.

This might provide a partial explanation for why the faults in AutoTest follow a poly-logarithmic curve: the distribution of faults has exponentially decreasing probability.
It may also justify some of the moderate differences in the Java experiments, if the fault distribution is different in the Java code with respect to the Eiffel code analyzed.

A more general problem related with this explanation has to do with
how the coupon collector model applies to random testing of
\emph{object-oriented} programs the way it is available in AutoTest
and YETI.  Such tools generate test cases dynamically by incrementally
populating a pool of objects with random
calls. 
This behavior entails that the probability of sampling an object with
certain characteristics (and hence of constructing a test case that
belongs to a certain target) is not fixed but dynamically varies, and
successive draws are not statistically independent.  For example, the
first round can only successfully call creation procedures, and the
probability of constructing a test case involving any other routine is
zero.  Hence, test cases are not really sampled uniformly at random,
and the problem is to determine the connection between the real
process and its representation as a coupon collection process.  All we
can say with the currently available data is that object-oriented
random testing with AutoTest seems to be describable as a coupon
collection process with a probability distribution over faults that is
always exponentially decreasing; this is not quite the same as saying that
the distribution of bugs is exponentially decreasing.

\subsection{Threats to validity} \label{sec:threats-validity}
Threats to \emph{construct validity} are present in the Java experiments where we have no reliable way of reconstructing faults from failures; a similar threat occurs with the Eiffel experiments if the contracts incorrectly capture the designer's intended behavior.
However, even if we may be measuring different things, we are arguably still measuring things that significantly correlate; the results partially confirm so.

The very large number of repeated experiments should have reduced the potential threats to \emph{internal validity} -- referring to the possibility that the study does not support the claimed findings -- to the minimum~\cite{AB11-icse}.
As discussed in Section~\ref{sec:exper-with-java}, however, the Java experiments are somehow less clear-cut than the Eiffel experiments; further experiments will hopefully provide conclusive evidence.

\emph{External validity} -- which refers to the findings' generalizability -- is limited by the focus on random testing and by the availability of software that can be tested with this technique.
This limitation is largely a consequence of the need of designing experiments approachable with the currently available technology.
In future work, we will target more software with contracts (e.g., JML and .NET code equipped with CodeContracts) and more testing frameworks (see Section~\ref{sec:related-work}) to improve the generalizability of our findings.

\section{Related work} \label{sec:related-work}
Automated random testing is a technique that is inexpensive to run and
proved to find bugs in different contexts, including Java libraries and
applications~\cite{Pacheco2005,Csallner2004,SharmaGAFM11}; Eiffel libraries~\cite{CPOLM11}; and Haskell programs~\cite{Claessen00quickcheck:a}.
For brevity, we only succinctly mention the most closely related work.

Yeti and AutoTest are two representatives in a series of random testing tools developed during the last decade, including Randoop~\cite{PachecoLEB07}, JCrasher~\cite{Csallner2004}, Eclat~\cite{Pacheco2005}, Jtest~\cite{Jtest}, Jartege~\cite{Oriat2004},  and RUTE-J~\cite{Andrews2006a}. 

Arcuri et al.'s theoretical analysis of random testing~\cite{ArcuriRT} is an important basis to understand also the practical and empirical side.
Section~\ref{sec:coupon-collection} outlined the connection between the work in \cite{ArcuriRT} and the present paper's, as well as the points that still remain open.
Our preliminary results, hint at a plausible consistency between the two results (the theoretical and the empirical).
To our knowledge, \cite{Oriol12} is the only other empirical result about random testing; \cite{Oriol12} considers only one model, which we have included as $\Phi_1$ in our analysis.

\textbf{Acknowledgements.} We gratefully acknowledge the \linebreak funding by the Swiss
National Science foundation (proj.\ LSAT and ASII); by the Hasler
foundation (proj.~Mancom); by the Swiss National Supercomputing
Centre (proj.~s264).


\ifextended
\clearpage
\newpage
\onecolumn
\section{Appendix} \label{sec:appendix}

\subsection{Experiments with Eiffel code}
Table~\ref{table:app-Eiffel-signedrank} displays the statistics of the
comparison between $\Phi_5$ and the other models according to their
$R^2$ scores in the fittings across all Eiffel classes used in the
experiments.  The $p$-values characterize whether differences are
statistically significant.  The effect size (computed with the $Z$
statistics as $Z/\sqrt{2N}$, where $N$ is the number of classes, and
hence $2N$ is the total sample size) characterizes the magnitude of
the observed differences: an effect size of $0.1$ is small; $0.3$
is medium; $0.5$ is large.

\begin{table*}[!htb]
\renewcommand{\arraystretch}{0.9}
\caption{Eiffel classes: differences between $\Phi_5$ and the other $\Phi_i$'s (Wilcoxon signed-rank test).}
\label{table:app-Eiffel-signedrank}
\centering
\scriptsize
\begin{tabular}{l *{8}{r@{.}l@{E}l}}
\textsc{Statistics} &  \multicolumn{3}{c}{$\Phi_1$} &  \multicolumn{3}{c}{$\Phi_2$} &  \multicolumn{3}{c}{$\Phi_3$} &  \multicolumn{3}{c}{$\Phi_4$} &  \multicolumn{3}{c}{$\Phi_6$} &  \multicolumn{3}{c}{$\Phi_7$} &  \multicolumn{3}{c}{$\Phi_8$} &  \multicolumn{3}{c}{$\Phi_9$} \\
\hline
$p$-value &
2&95&-03  &  28&46&-03  &  19&53&-06  &  1&11&-06  
  &  52&55&-09  &  71&79&-09  &  4&12&-06  &  61&45&-09  \\
effect size &
324&32&-03  &  239&05&-03  &  465&92&-03  &  531&39&-03
  &  593&82&-03  &  587&73&-03  &  502&46&-03  &  590&77&-03
\end{tabular}
\end{table*}

\subsection{Experiments with Java code}
Tables~\ref{table:app-j1-datastats} and~\ref{table:app-j2-datastats} show the complete statistics about the testing sessions with YETI on the Java classes described in Section~\ref{sec:exper-with-java}.
They are the extended versions of Tables~\ref{table:j1-datastats} and~\ref{table:j2-datastats}.

\begin{table*}[!htb]
\renewcommand{\arraystretch}{0.9}
\caption{Java classes tested for failures: statistics per class.}
\label{table:app-j1-datastats}
\centering
\scriptsize
\begin{tabular}{l r r@{.}l@{E}l r r@{.}l@{E}l r@{.}l@{E}l r@{.}l@{E}l r@{.}l@{E}l}
\textsc{Class} & \textsc{S} & \multicolumn{3}{c}{\textsc{T}} & \textsc{F} & \multicolumn{3}{c}{$\EV{\sigma}$} & \multicolumn{3}{c}{$\EV{\gamma}$} & \multicolumn{3}{c}{$\EV{\Delta}$} & \multicolumn{3}{c}{$\sigma[\Delta]$} \\
\hline
 \mbox{\lstinline[language=Java]|java.lang.Character|}  &   30  &  1&29&+05  &   32  &  9&91&-01  &  7&17&-01  &  1&24&-02  &  6&07&-02\\
 \mbox{\lstinline[language=Java]|java.lang.String|}  &   30  &  1&31&+05  &   84  &  7&58&-01  &  -1&34&+00  &  2&15&-02  &  1&04&-01\\
 \mbox{\lstinline[language=Java]|java.util.ArrayList|}  &   30  &  1&08&+05  &   10  &  4&84&-01  &  -1&01&+00  &  1&57&-03  &  1&85&-02\\
 \mbox{\lstinline[language=Java]|java.util.Calendar|}  &   30  &  1&35&+05  &   36  &  1&31&+00  &  4&21&-01  &  1&28&-02  &  3&89&-02\\
 \mbox{\lstinline[language=Java]|java.util.Date|}  &   30  &  1&24&+05  &    8  &  6&69&-02  &  -9&06&-01  &  3&31&-03  &  1&95&-02\\
 \mbox{\lstinline[language=Java]|java.util.HashMap|}  &   31  &  1&13&+05  &    7  &  2&84&-01  &  -3&36&+00  &  9&01&-04  &  8&89&-03\\
 \mbox{\lstinline[language=Java]|java.util.Hashtable|}  &   30  &  1&16&+05  &   14  &  4&42&-01  &  -1&45&+00  &  2&27&-03  &  2&48&-02\\
 \mbox{\lstinline[language=Java]|java.util.LinkedList|}  &   30  &  1&08&+05  &   12  &  6&70&-02  &  -1&31&+00  &  5&46&-03  &  4&55&-02\\
 \mbox{\lstinline[language=Java]|java.util.Properties|}  &   30  &  1&18&+05  &   15  &  1&25&+00  &  5&34&-01  &  7&16&-03  &  2&84&-02\\
 \mbox{\lstinline[language=Java]|java.util.SimpleTimeZone|}  &   30  &  1&63&+05  &   80  &  2&05&+00  &  -6&33&-02  &  1&93&-02  &  7&63&-02\\
 \mbox{\lstinline[language=Java]|java.util.TreeMap|}  &   30  &  1&07&+05  &   58  &  8&56&-01  &  -2&03&+00  &  2&63&-02  &  8&18&-02\\
\hline
\textbf{Mean} &    30  &  1&23&+05  &   32  &  7&78&-01  &  -8&91&-01  &  1&03&-02  &  4&61&-02\\
\textbf{Median} &    30  &  1&18&+05  &   15  &  7&58&-01  &  -1&01&+00  &  7&16&-03  &  3&89&-02\\
\textbf{Stdev} &        &  1&65&+04  &   29  &  6&02&-01  &  1&23&+00  &  8&86&-03  &  3&07&-02\\

\end{tabular}
\end{table*}

\begin{table*}[!htb]
\renewcommand{\arraystretch}{0.9}
\caption{Java classes tested for faults: statistics per class.}
\label{table:app-j2-datastats}
\centering
\scriptsize
\begin{tabular}{l r r@{.}l@{E}l r r@{.}l@{E}l r@{.}l@{E}l r@{.}l@{E}l r@{.}l@{E}l}
\textsc{Class} & \textsc{S} & \multicolumn{3}{c}{\textsc{T}} & \textsc{F} & \multicolumn{3}{c}{$\EV{\sigma}$} & \multicolumn{3}{c}{$\EV{\gamma}$} & \multicolumn{3}{c}{$\EV{\Delta}$} & \multicolumn{3}{c}{$\sigma[\Delta]$} \\
\hline
 \mbox{\lstinline[language=Java]|java.lang.Boolean|}  &    4  &  2&49&+05  &       &  0&00&+00  &  \multicolumn{3}{c}{NaN}  &  0&00&+00  &  0&00&+00\\
 \mbox{\lstinline[language=Java]|java.lang.Byte|}  &    4  &  2&79&+05  &    2  &  4&10&-03  &  0&00&+00  &  3&00&-03  &  2&72&-02\\
 \mbox{\lstinline[language=Java]|java.lang.Character|}  &    4  &  3&67&+05  &       &  0&00&+00  &  \multicolumn{3}{c}{NaN}  &  0&00&+00  &  0&00&+00\\
 \mbox{\lstinline[language=Java]|java.lang.Class|}  &    4  &  2&39&+05  &    8  &  4&39&-01  &  1&07&-01  &  8&79&-03  &  5&34&-02\\
 \mbox{\lstinline[language=Java]|java.lang.ClassLoader|}  &    4  &  2&66&+05  &    8  &  2&82&-02  &  -3&85&-01  &  8&50&-03  &  6&36&-02\\
 \mbox{\lstinline[language=Java]|java.lang.Compiler|}  &    4  &  1&15&+05  &       &  0&00&+00  &  \multicolumn{3}{c}{NaN}  &  0&00&+00  &  0&00&+00\\
 \mbox{\lstinline[language=Java]|java.lang.Double|}  &    4  &  2&70&+05  &    4  &  2&23&-02  &  -1&04&-01  &  7&13&-03  &  5&93&-02\\
 \mbox{\lstinline[language=Java]|java.lang.Enum|}  &    4  &  4&29&+06  &       &  0&00&+00  &  \multicolumn{3}{c}{NaN}  &  0&00&+00  &  0&00&+00\\
 \mbox{\lstinline[language=Java]|java.lang.Float|}  &    4  &  2&71&+05  &    3  &  9&52&-03  &  -4&40&-01  &  5&26&-03  &  5&10&-02\\
 \mbox{\lstinline[language=Java]|java.lang.InheritableThreadLocal|}  &    4  &  2&23&+05  &       &  0&00&+00  &  \multicolumn{3}{c}{NaN}  &  0&00&+00  &  0&00&+00\\
 \mbox{\lstinline[language=Java]|java.lang.Integer|}  &    4  &  3&06&+05  &    2  &  5&95&-03  &  1&10&-01  &  3&16&-03  &  3&43&-02\\
 \mbox{\lstinline[language=Java]|java.lang.Long|}  &    4  &  3&08&+05  &    2  &  8&62&-03  &  -2&66&-01  &  2&86&-03  &  2&66&-02\\
 \mbox{\lstinline[language=Java]|java.lang.Math|}  &    4  &  3&28&+05  &       &  0&00&+00  &  \multicolumn{3}{c}{NaN}  &  0&00&+00  &  0&00&+00\\
 \mbox{\lstinline[language=Java]|java.lang.Number|}  &    4  &  0&00&+00  &       &  0&00&+00  &  \multicolumn{3}{c}{NaN}  &  0&00&+00  &  0&00&+00\\
 \mbox{\lstinline[language=Java]|java.lang.Object|}  &    4  &  2&23&+05  &       &  0&00&+00  &  \multicolumn{3}{c}{NaN}  &  0&00&+00  &  0&00&+00\\
 \mbox{\lstinline[language=Java]|java.lang.Package|}  &    4  &  3&36&+06  &    1  &  4&40&-04  &  0&00&+00  &  3&22&-04  &  8&96&-03\\
 \mbox{\lstinline[language=Java]|java.lang.Process|}  &    4  &  0&00&+00  &       &  0&00&+00  &  \multicolumn{3}{c}{NaN}  &  0&00&+00  &  0&00&+00\\
 \mbox{\lstinline[language=Java]|java.lang.ProcessBuilder|}  &    4  &  2&61&+05  &    3  &  1&07&-02  &  -4&48&-01  &  4&39&-03  &  3&80&-02\\
 \mbox{\lstinline[language=Java]|java.lang.Runtime|}  &    4  &  3&35&+05  &   10  &  7&92&-01  &  -3&78&-01  &  4&80&-04  &  1&74&-02\\
 \mbox{\lstinline[language=Java]|java.lang.RuntimePermission|}  &    4  &  2&52&+05  &       &  0&00&+00  &  \multicolumn{3}{c}{NaN}  &  0&00&+00  &  0&00&+00\\
 \mbox{\lstinline[language=Java]|java.lang.Short|}  &    4  &  2&84&+05  &    2  &  7&31&-03  &  -3&11&-01  &  2&89&-03  &  2&99&-02\\
 \mbox{\lstinline[language=Java]|java.lang.StackTraceElement|}  &    1  &  2&12&+05  &       &  0&00&+00  &  \multicolumn{3}{c}{NaN}  &  0&00&+00  &  0&00&+00\\
 \mbox{\lstinline[language=Java]|java.lang.StrictMath|}  &    4  &  3&28&+05  &       &  0&00&+00  &  \multicolumn{3}{c}{NaN}  &  0&00&+00  &  0&00&+00\\
 \mbox{\lstinline[language=Java]|java.lang.String|}  &    4  &  3&45&+05  &    4  &  2&91&-02  &  -3&32&-01  &  3&31&-03  &  3&03&-02\\
 \mbox{\lstinline[language=Java]|java.lang.StringBuffer|}  &    1  &  7&30&+03  &    4  &  0&00&+00  &  \multicolumn{3}{c}{NaN}  &  0&00&+00  &  2&12&-01\\
 \mbox{\lstinline[language=Java]|java.lang.StringBuilder|}  &    1  &  1&69&+03  &    6  &  0&00&+00  &  \multicolumn{3}{c}{NaN}  &  0&00&+00  &  2&97&-01\\
 \mbox{\lstinline[language=Java]|java.lang.ThreadLocal|}  &    3  &  2&21&+05  &       &  0&00&+00  &  \multicolumn{3}{c}{NaN}  &  0&00&+00  &  0&00&+00\\
 \mbox{\lstinline[language=Java]|java.lang.Throwable|}  &    3  &  2&38&+05  &       &  0&00&+00  &  \multicolumn{3}{c}{NaN}  &  0&00&+00  &  0&00&+00\\
 \mbox{\lstinline[language=Java]|java.lang.Void|}  &    3  &  0&00&+00  &       &  0&00&+00  &  \multicolumn{3}{c}{NaN}  &  0&00&+00  &  0&00&+00\\
\hline
\textbf{Mean} &     4  &  4&68&+05  &    2  &  4&68&-02  &  \multicolumn{3}{c}{NaN}  &  1&73&-03  &  3&27&-02\\
\textbf{Median} &     4  &  2&61&+05  &       &  0&00&+00  &  \multicolumn{3}{c}{NaN}  &  0&00&+00  &  0&00&+00\\
\textbf{Stdev} &     1  &  9&45&+05  &    3  &  1&65&-01  &  \multicolumn{3}{c}{NaN}  &  2&72&-03  &  6&58&-02\\

\end{tabular}
\end{table*}

Tables~\ref{table:app-j1-bestfits-mean} and~\ref{table:app-j2-bestfits-mean} show the result of fitting the models $\Phi_1, \Phi_2, \Phi_4, \Phi_5$ on the mean fault count $\phim$ curves, reporting the same data as for the AutoTest experiments in Section~\ref{sec:exper-with-eiff}.

\begin{table*}[!htb]
\renewcommand{\arraystretch}{0.9}
\caption{Testing of Java classes for failures: best fits with mean.}
\label{table:app-j1-bestfits-mean}
\centering
\scriptsize
\begin{tabular}{l l r@{.}l@{E}l r@{.}l@{E}l r@{.}l@{E}l r@{.}l@{E}l r r@{.}l@{E}l r@{.}l@{E}l r@{.}l@{E}l r@{.}l@{E}l}
\textsc{Class} & \textsc{Best Fit Ranking} & \multicolumn{3}{c}{$R^2_{\text{best}}$} & \multicolumn{3}{c}{\textsc{RMSE}$_{\text{best}}$} & \multicolumn{3}{c}{$\Delta(R^2)$} & \multicolumn{3}{c}{$\Delta$(\textsc{RMSE})} \\
\hline
 \mbox{\lstinline[language=Java]|java.lang.Character|}  &    1  5  4  2  &  9&87&-01  &  3&49&-01  &  4&89&-02  &  4&24&-01\\
 \mbox{\lstinline[language=Java]|java.lang.String|}  &    1  5  4  2  &  9&94&-01  &  7&37&-01  &  3&23&-02  &  1&14&+00\\
 \mbox{\lstinline[language=Java]|java.util.ArrayList|}  &    2  1  5  4  &  9&75&-01  &  9&76&-02  &  2&01&-01  &  1&96&-01\\
 \mbox{\lstinline[language=Java]|java.util.Calendar|}  &    1  5  4  2  &  9&91&-01  &  5&29&-01  &  5&85&-03  &  1&56&-01\\
 \mbox{\lstinline[language=Java]|java.util.Date|}  &    2  1  5  4  &  9&78&-01  &  1&08&-01  &  1&14&-01  &  1&61&-01\\
 \mbox{\lstinline[language=Java]|java.util.HashMap|}  &    1  5  4  2  &  9&09&-01  &  1&73&-01  &  5&71&-02  &  4&78&-02\\
 \mbox{\lstinline[language=Java]|java.util.Hashtable|}  &    2  1  5  4  &  9&62&-01  &  1&44&-01  &  2&54&-01  &  2&54&-01\\
 \mbox{\lstinline[language=Java]|java.util.LinkedList|}  &    2  1  5  4  &  7&41&-01  &  4&26&-01  &  1&59&-01  &  1&15&-01\\
 \mbox{\lstinline[language=Java]|java.util.Properties|}  &    5  1  4  2  &  9&80&-01  &  2&34&-01  &  0&00&+00  &  0&00&+00\\
 \mbox{\lstinline[language=Java]|java.util.SimpleTimeZone|}  &    2  5  1  4  &  9&98&-01  &  3&81&-01  &  8&87&-03  &  5&93&-01\\
 \mbox{\lstinline[language=Java]|java.util.TreeMap|}  &    1  5  2  4  &  9&97&-01  &  5&04&-01  &  3&14&-03  &  2&05&-01\\
\hline
\textbf{Mean} & 9\% ($\phi_5$ is best) & \multicolumn{3}{c}{} & \multicolumn{3}{c}{} &  8&04&-02  &  3&00&-01\\ 
 & 64\% ($\phi_5$ is 2$^{\text{nd}}$ best) 
\end{tabular}
\end{table*}

\begin{table*}[!htb]
  \renewcommand{\arraystretch}{0.9}
  \caption{Testing of Java classes for faults: best fits with mean.}
\label{table:app-j2-bestfits-mean}
\centering
\scriptsize
\begin{tabular}{l l r@{.}l@{E}l r@{.}l@{E}l r@{.}l@{E}l r@{.}l@{E}l r r@{.}l@{E}l r@{.}l@{E}l r@{.}l@{E}l r@{.}l@{E}l}
\textsc{Class} & \textsc{Best Fit Ranking} & \multicolumn{3}{c}{$R^2_{\text{best}}$} & \multicolumn{3}{c}{\textsc{RMSE}$_{\text{best}}$} & \multicolumn{3}{c}{$\Delta(R^2)$} & \multicolumn{3}{c}{$\Delta$(\textsc{RMSE})} \\
\hline
 \mbox{\lstinline[language=Java]|java.lang.Boolean|}  &    1  2  4  5  &  \multicolumn{3}{c}{-Inf}  &  5&23&-10  &  \multicolumn{3}{c}{NaN}  &  4&01&-05\\
 \mbox{\lstinline[language=Java]|java.lang.Byte|}  &    1  5  4  2  &  9&01&-01  &  4&33&-02  &  1&82&-01  &  2&99&-02\\
 \mbox{\lstinline[language=Java]|java.lang.Character|}  &    1  2  4  5  &  \multicolumn{3}{c}{-Inf}  &  4&91&-12  &  \multicolumn{3}{c}{NaN}  &  2&41&-05\\
 \mbox{\lstinline[language=Java]|java.lang.Class|}  &    2  1  5  4  &  9&58&-01  &  1&98&-01  &  1&37&-02  &  2&91&-02\\
 \mbox{\lstinline[language=Java]|java.lang.ClassLoader|}  &    1  5  2  4  &  9&42&-01  &  1&98&-01  &  1&46&-01  &  1&73&-01\\
 \mbox{\lstinline[language=Java]|java.lang.Compiler|}  &    1  2  4  5  &  \multicolumn{3}{c}{-Inf}  &  2&45&-11  &  \multicolumn{3}{c}{NaN}  &  4&00&-05\\
 \mbox{\lstinline[language=Java]|java.lang.Double|}  &    1  5  4  2  &  9&28&-01  &  1&01&-01  &  1&61&-01  &  8&12&-02\\
 \mbox{\lstinline[language=Java]|java.lang.Enum|}  &    1  2  4  5  &  \multicolumn{3}{c}{-Inf}  &  2&52&-11  &  \multicolumn{3}{c}{NaN}  &  8&31&-06\\
 \mbox{\lstinline[language=Java]|java.lang.Float|}  &    2  1  5  4  &  9&80&-01  &  3&01&-02  &  1&97&-01  &  6&91&-02\\
 \mbox{\lstinline[language=Java]|java.lang.InheritableThreadLocal|}  &    1  2  4  5  &  \multicolumn{3}{c}{-Inf}  &  3&60&-12  &  \multicolumn{3}{c}{NaN}  &  4&27&-05\\
 \mbox{\lstinline[language=Java]|java.lang.Integer|}  &    2  1  5  4  &  9&85&-01  &  2&31&-02  &  3&02&-01  &  8&20&-02\\
 \mbox{\lstinline[language=Java]|java.lang.Long|}  &    2  1  5  4  &  9&85&-01  &  2&10&-02  &  2&10&-01  &  6&05&-02\\
 \mbox{\lstinline[language=Java]|java.lang.Math|}  &    1  2  4  5  &  \multicolumn{3}{c}{-Inf}  &  6&39&-11  &  \multicolumn{3}{c}{NaN}  &  4&07&-05\\
 \mbox{\lstinline[language=Java]|java.lang.Number|}  &    1  2  4  5  &  \multicolumn{3}{c}{NaN}  &  0&00&+00  &  \multicolumn{3}{c}{NaN}  &  1&18&-04\\
 \mbox{\lstinline[language=Java]|java.lang.Object|}  &    1  2  4  5  &  \multicolumn{3}{c}{-Inf}  &  3&41&-12  &  \multicolumn{3}{c}{NaN}  &  3&60&-05\\
 \mbox{\lstinline[language=Java]|java.lang.Package|}  &    2  1  5  4  &  9&94&-01  &  1&84&-03  &  2&16&-01  &  9&73&-03\\
 \mbox{\lstinline[language=Java]|java.lang.Process|}  &    1  2  4  5  &  \multicolumn{3}{c}{NaN}  &  0&00&+00  &  \multicolumn{3}{c}{NaN}  &  1&30&-04\\
 \mbox{\lstinline[language=Java]|java.lang.ProcessBuilder|}  &    1  5  4  2  &  9&72&-01  &  4&48&-02  &  1&22&-01  &  5&91&-02\\
 \mbox{\lstinline[language=Java]|java.lang.Runtime|}  &    5  1  4  2  &  8&34&-01  &  1&35&-01  &  0&00&+00  &  0&00&+00\\
 \mbox{\lstinline[language=Java]|java.lang.RuntimePermission|}  &    1  2  4  5  &  \multicolumn{3}{c}{-Inf}  &  1&63&-11  &  \multicolumn{3}{c}{NaN}  &  3&18&-05\\
 \mbox{\lstinline[language=Java]|java.lang.Short|}  &    2  1  5  4  &  9&94&-01  &  1&25&-02  &  2&60&-01  &  7&12&-02\\
 \mbox{\lstinline[language=Java]|java.lang.StackTraceElement|}  &    1  2  4  5  &  \multicolumn{3}{c}{-Inf}  &  4&44&-10  &  \multicolumn{3}{c}{NaN}  &  2&05&-05\\
 \mbox{\lstinline[language=Java]|java.lang.StrictMath|}  &    1  2  4  5  &  \multicolumn{3}{c}{-Inf}  &  2&20&-10  &  \multicolumn{3}{c}{NaN}  &  2&82&-05\\
 \mbox{\lstinline[language=Java]|java.lang.String|}  &    1  5  4  2  &  9&72&-01  &  6&20&-02  &  9&72&-02  &  7&00&-02\\
 \mbox{\lstinline[language=Java]|java.lang.StringBuffer|}  &    2  1  5  4  &  1&00&+00  &  3&58&-02  &  1&08&-03  &  3&34&-02\\
 \mbox{\lstinline[language=Java]|java.lang.StringBuilder|}  &    2  5  1  4  &  9&91&-01  &  7&40&-02  &  1&34&-03  &  4&87&-03\\
 \mbox{\lstinline[language=Java]|java.lang.ThreadLocal|}  &    1  2  4  5  &  \multicolumn{3}{c}{-Inf}  &  5&17&-12  &  \multicolumn{3}{c}{NaN}  &  4&28&-05\\
 \mbox{\lstinline[language=Java]|java.lang.Throwable|}  &    1  2  4  5  &  \multicolumn{3}{c}{-Inf}  &  2&88&-12  &  \multicolumn{3}{c}{NaN}  &  1&71&-05\\
 \mbox{\lstinline[language=Java]|java.lang.Void|}  &    1  2  4  5  &  \multicolumn{3}{c}{NaN}  &  0&00&+00  &  \multicolumn{3}{c}{NaN}  &  1&59&-04\\
\hline
\textbf{Mean} & 3\% ($\phi_5$ is best) & \multicolumn{3}{c}{} & \multicolumn{3}{c}{} &  \multicolumn{3}{c}{NaN}  &  2&67&-02\\ 
 & 24\% ($\phi_5$ is 2$^{\text{nd}}$ best) 
\end{tabular}
\end{table*}

Tables~\ref{table:app-Java-failures-signedrank} and~\ref{table:app-Java-faults-signedrank} report the same statistics as Table~\ref{table:app-Eiffel-signedrank} for the Java classes (failure and fault data).

\begin{table}[!htb]
\renewcommand{\arraystretch}{0.9}
\caption{Java failures: differences between $\Phi_5$ and some other $\Phi_i$'s (Wilcoxon signed-rank tests).}
\label{table:app-Java-failures-signedrank}
\centering
\scriptsize
\begin{tabular}{l *{3}{r@{.}l@{E}l}}
\textsc{Statistics} &  \multicolumn{3}{c}{$\Phi_1$} &  \multicolumn{3}{c}{$\Phi_2$} &  \multicolumn{3}{c}{$\Phi_4$} \\
\hline
$p$-value &
 18&55&-03  &  123&05&-03  &  976&56&-06 \\
effect size &
 492&85&-03  &  341&21&-03  &  625&54&-03
\end{tabular}
\end{table}

\begin{table}[!htb]
\renewcommand{\arraystretch}{0.9}
\caption{Java faults: differences between $\Phi_5$ and some other $\Phi_i$'s (Wilcoxon signed-rank tests).}
\label{table:app-Java-faults-signedrank}
\centering
\scriptsize
\begin{tabular}{l *{3}{r@{.}l@{E}l}}
\textsc{Statistics} &  \multicolumn{3}{c}{$\Phi_1$} &  \multicolumn{3}{c}{$\Phi_2$} &  \multicolumn{3}{c}{$\Phi_4$} \\
\hline
$p$-value &
 20&26&-03  &  855&22&-03  &  122&07&-06 \\
effect size &
 300&87&-03  &  28&85&-03  &  432&75&-03
\end{tabular}
\end{table}

\clearpage

\subsection{Eiffel experiments: all graphs}

Figures~\ref{fig:eiffel-first}--\ref{fig:eiffel-last} display, for
each of the 42 Eiffel classes tested, the mean curve (in black) and
the three models that fit best. Horizontal axes are scaled by millions
of test cases drawn; vertical axes by total number of faults found.

\begin{figure*}[!ht]
  \begin{center}$
  \begin{array}{cc}
    \includegraphics[width={.48\textwidth}]{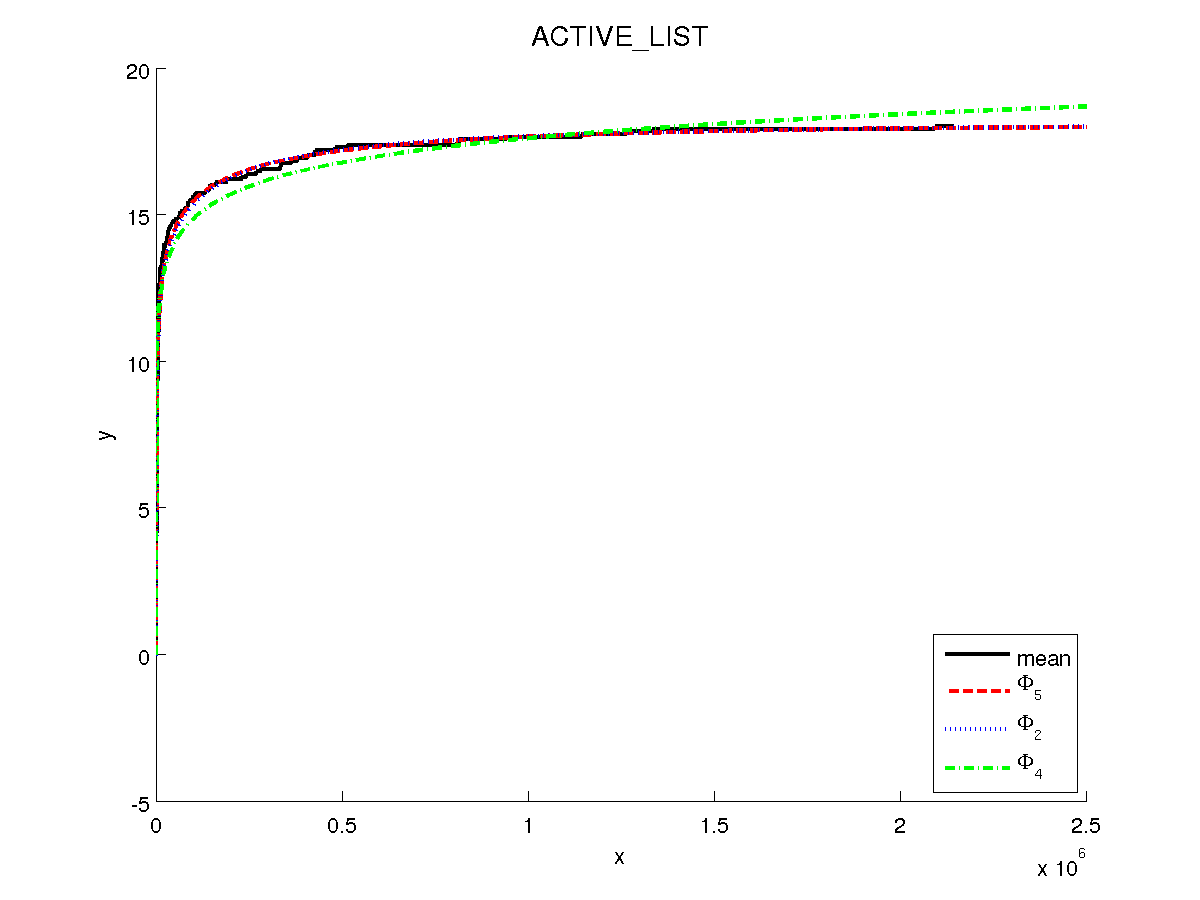} &
    \includegraphics[width={.48\textwidth}]{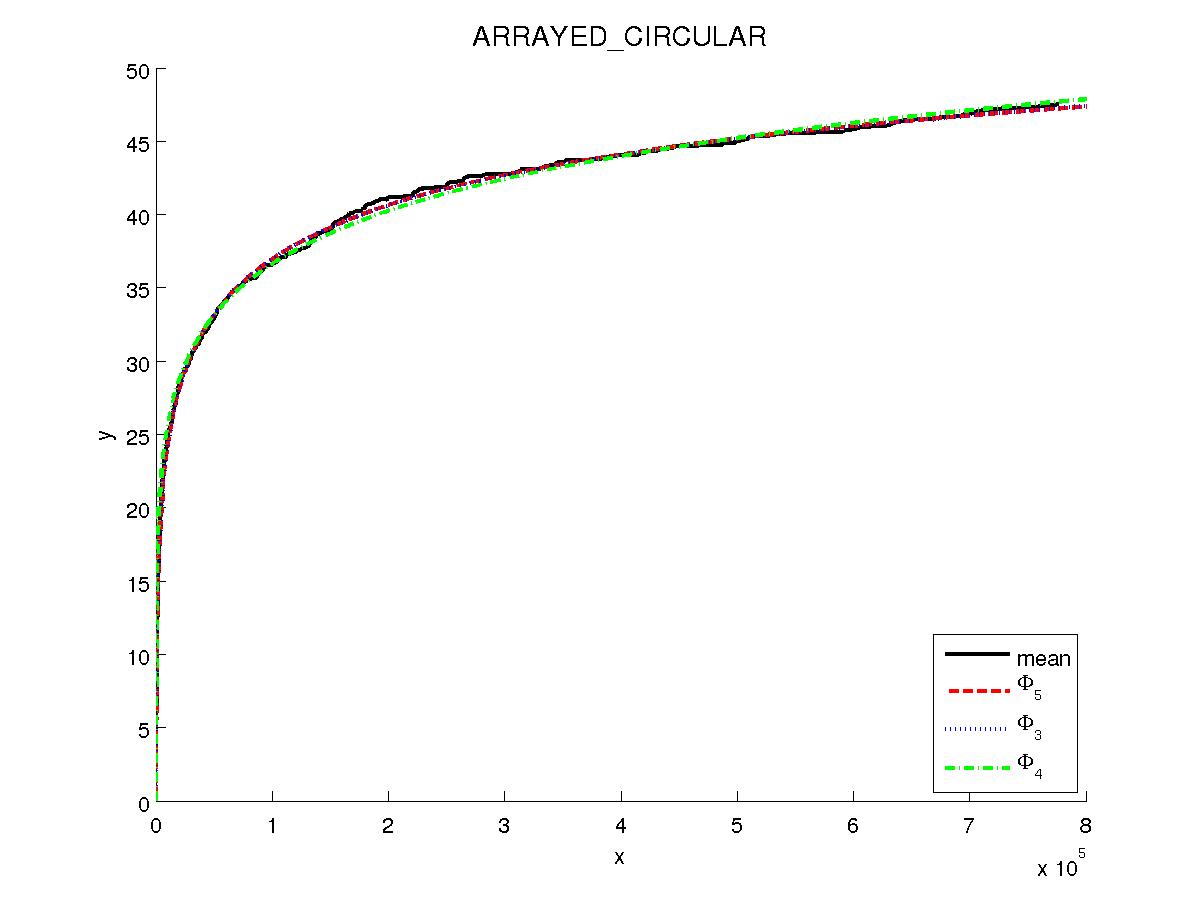} \\
    \includegraphics[width={.48\textwidth}]{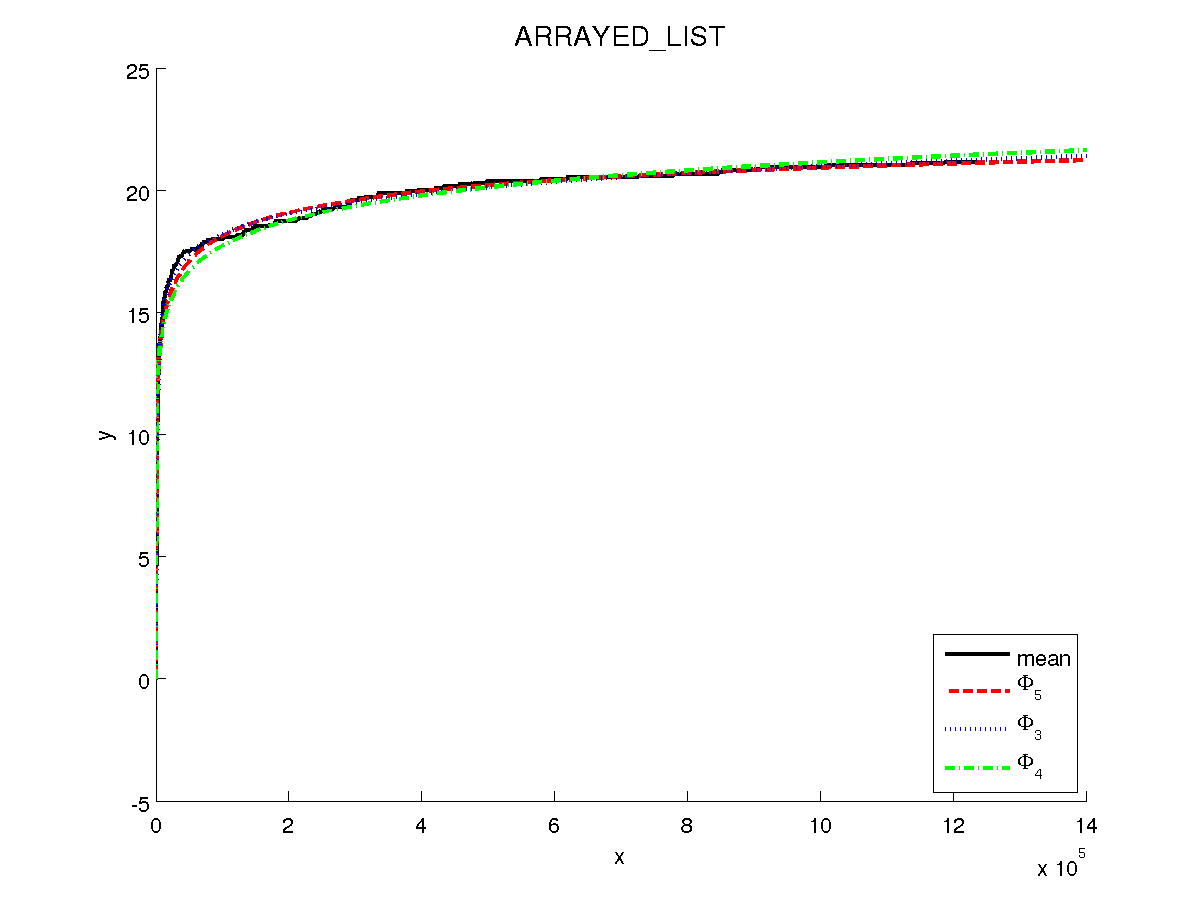} &
    \includegraphics[width={.48\textwidth}]{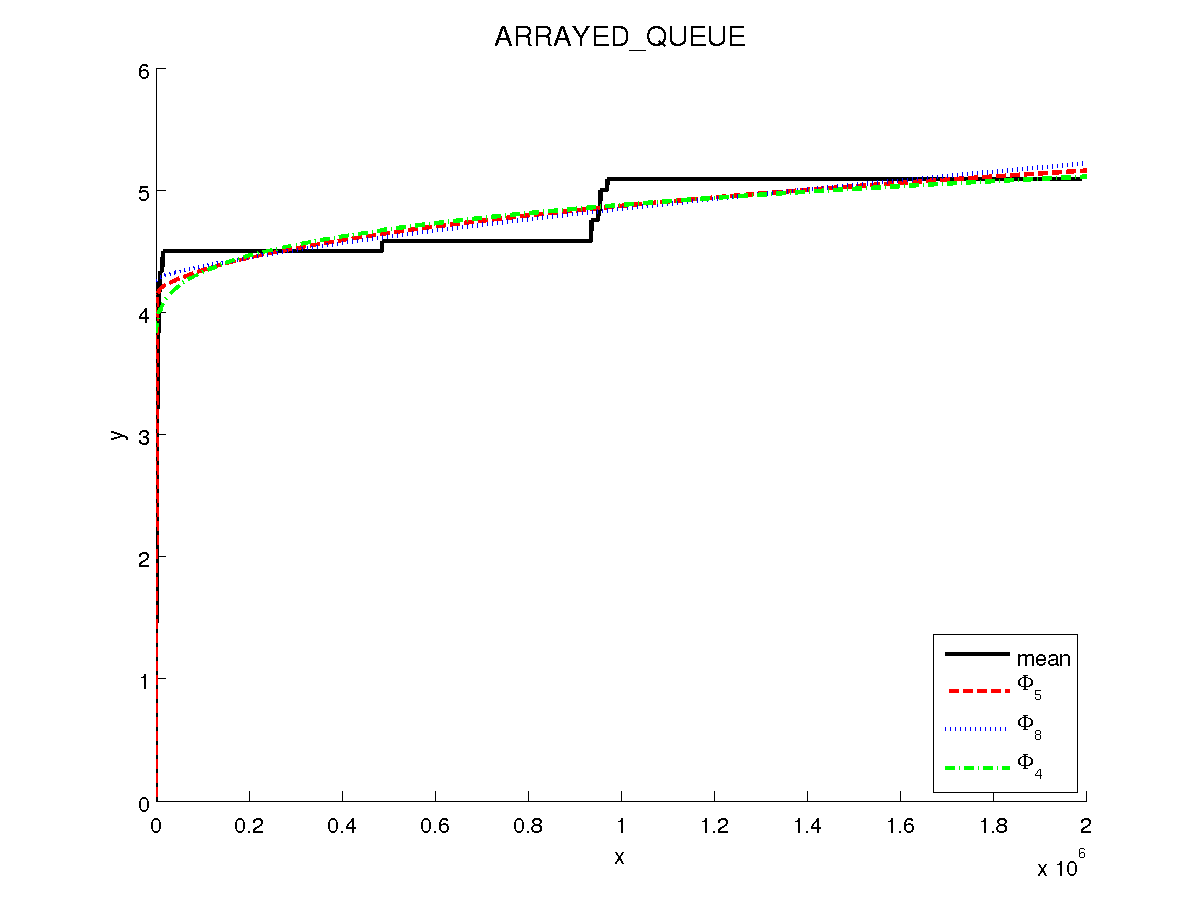} \\
    \includegraphics[width={.48\textwidth}]{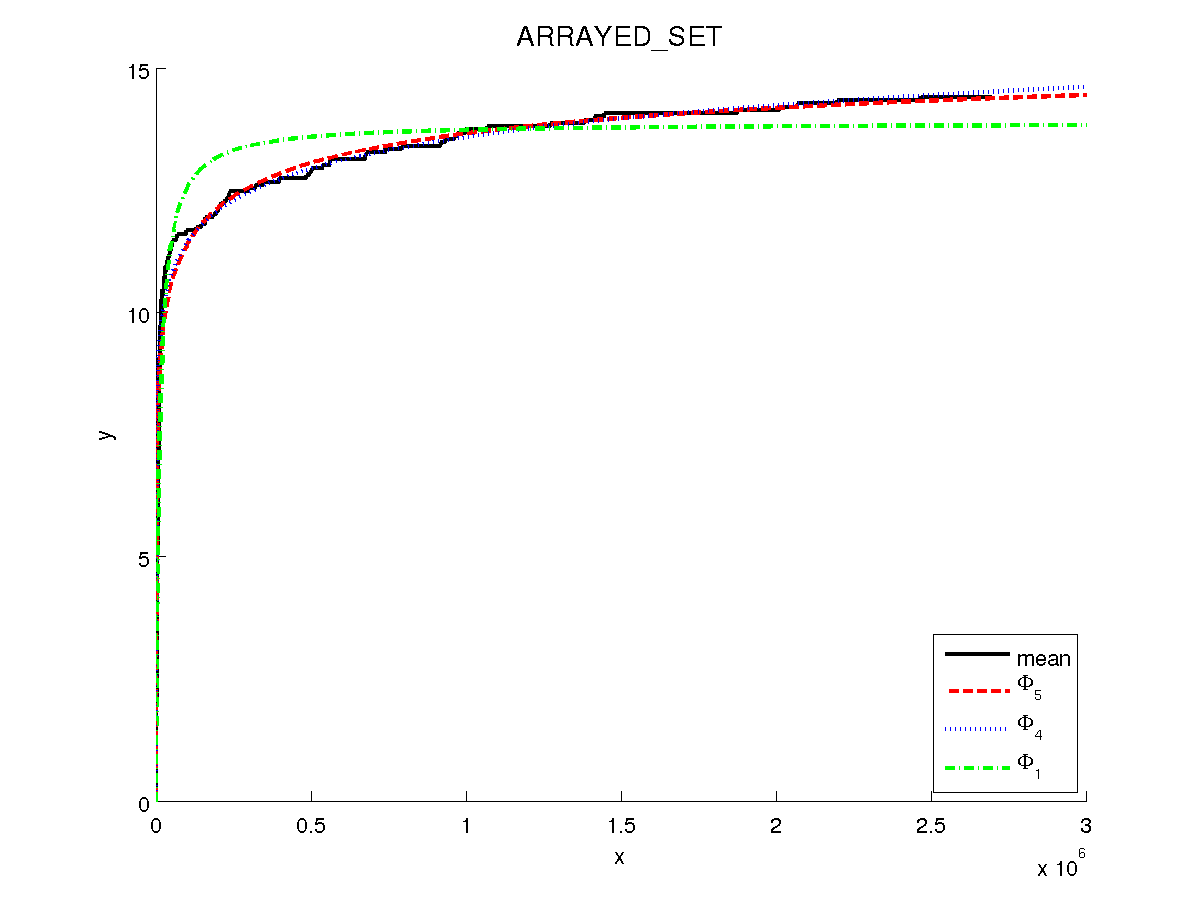} &
    \includegraphics[width={.48\textwidth}]{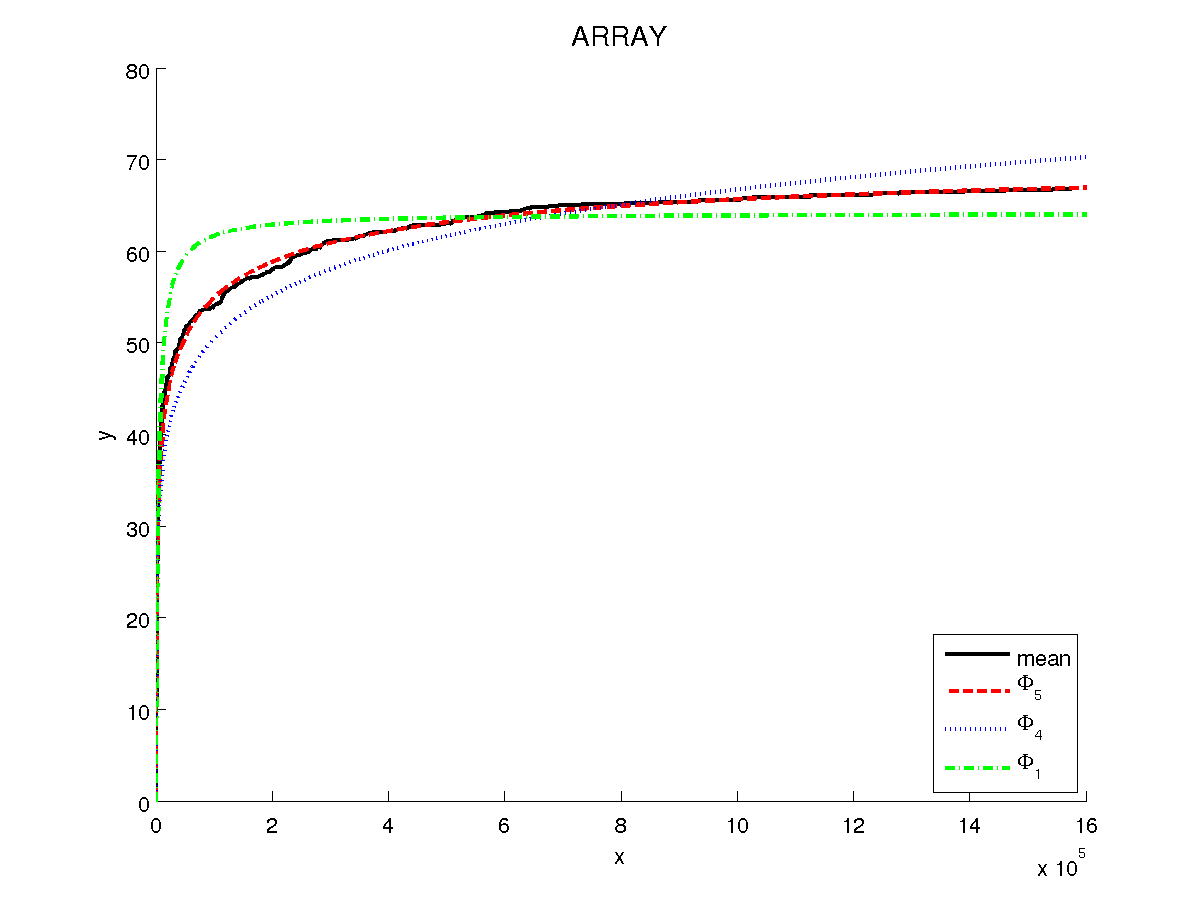} 
  \end{array}$
\end{center}
\caption{Top 3 fits with mean for six Eiffel classes.}
\label{fig:eiffel-first}
\end{figure*}

\begin{figure*}[!htb]
  \begin{center}$
  \begin{array}{cc}
    \includegraphics[width={.48\textwidth}]{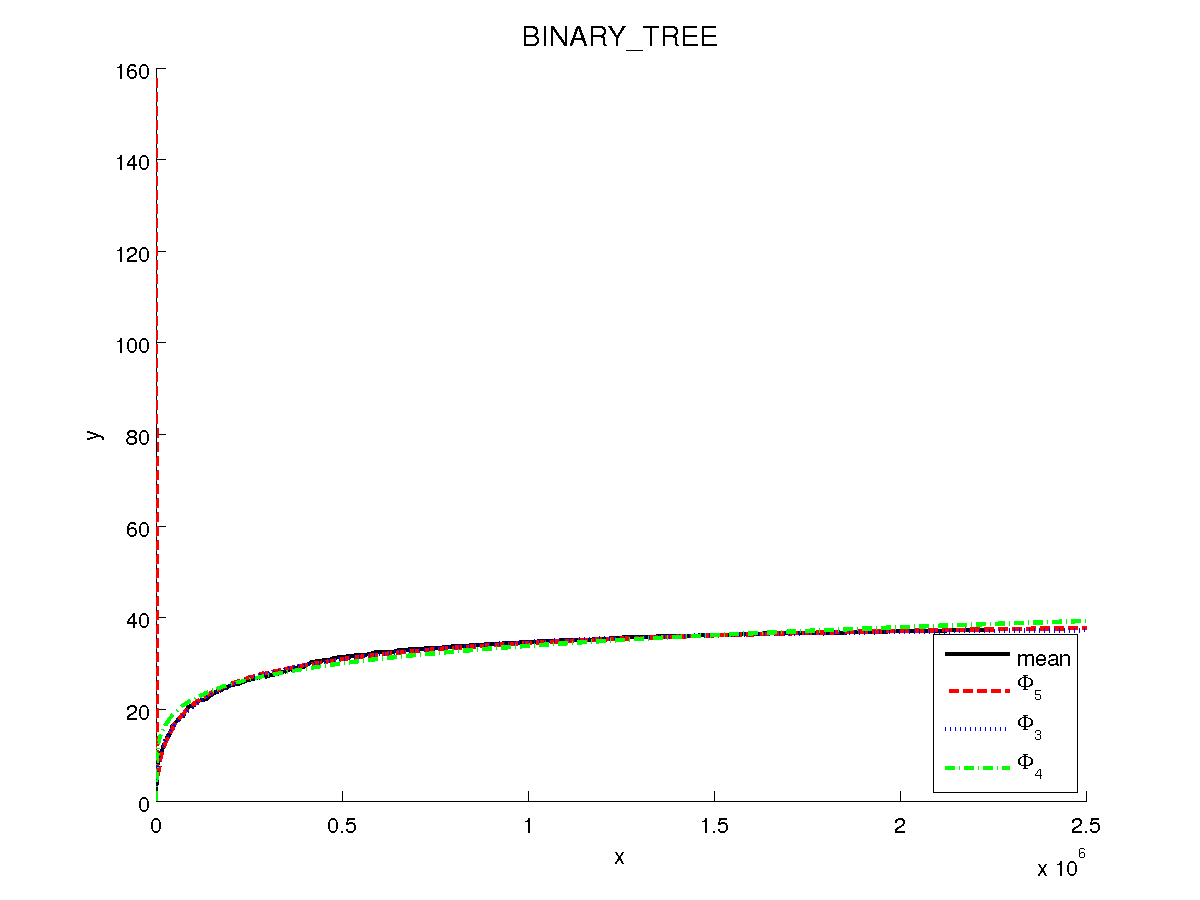} &
    \includegraphics[width={.48\textwidth}]{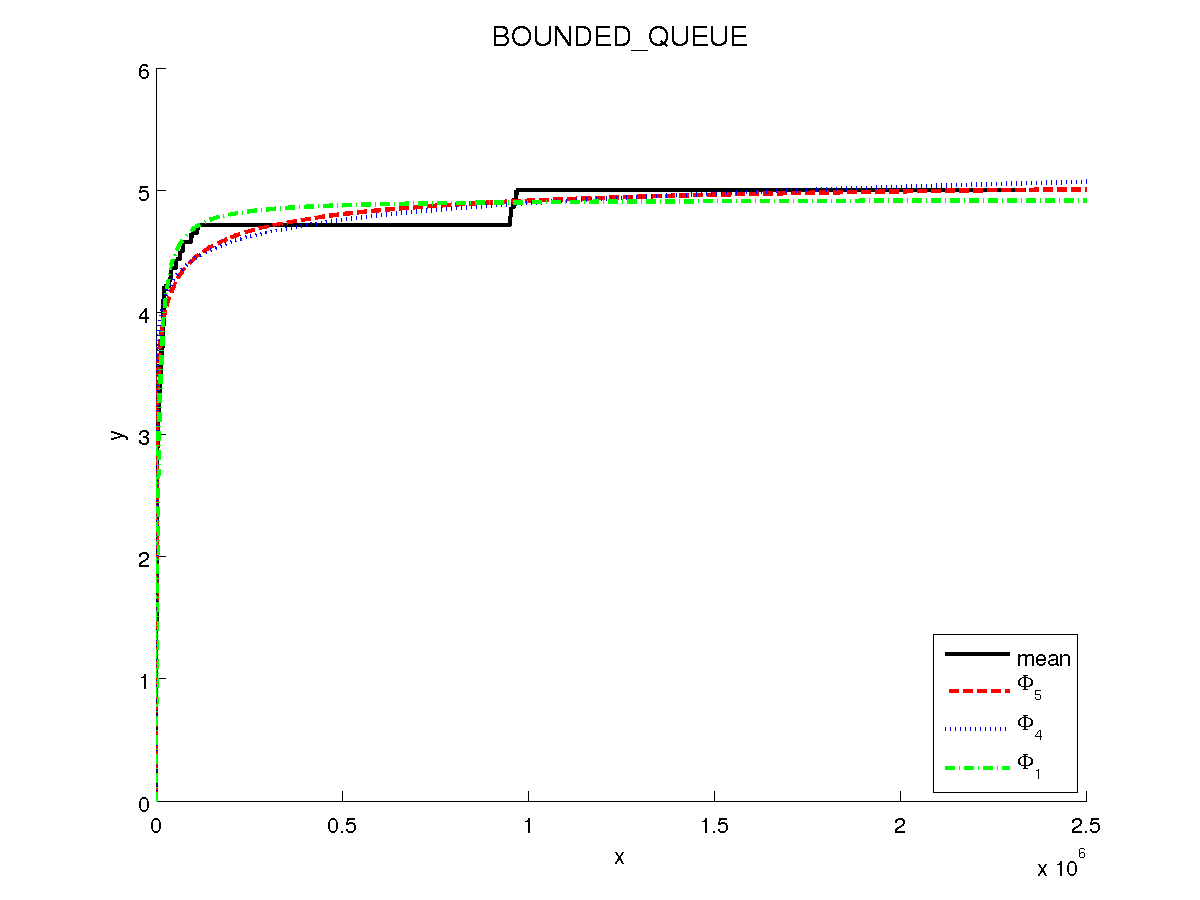} \\
    \includegraphics[width={.48\textwidth}]{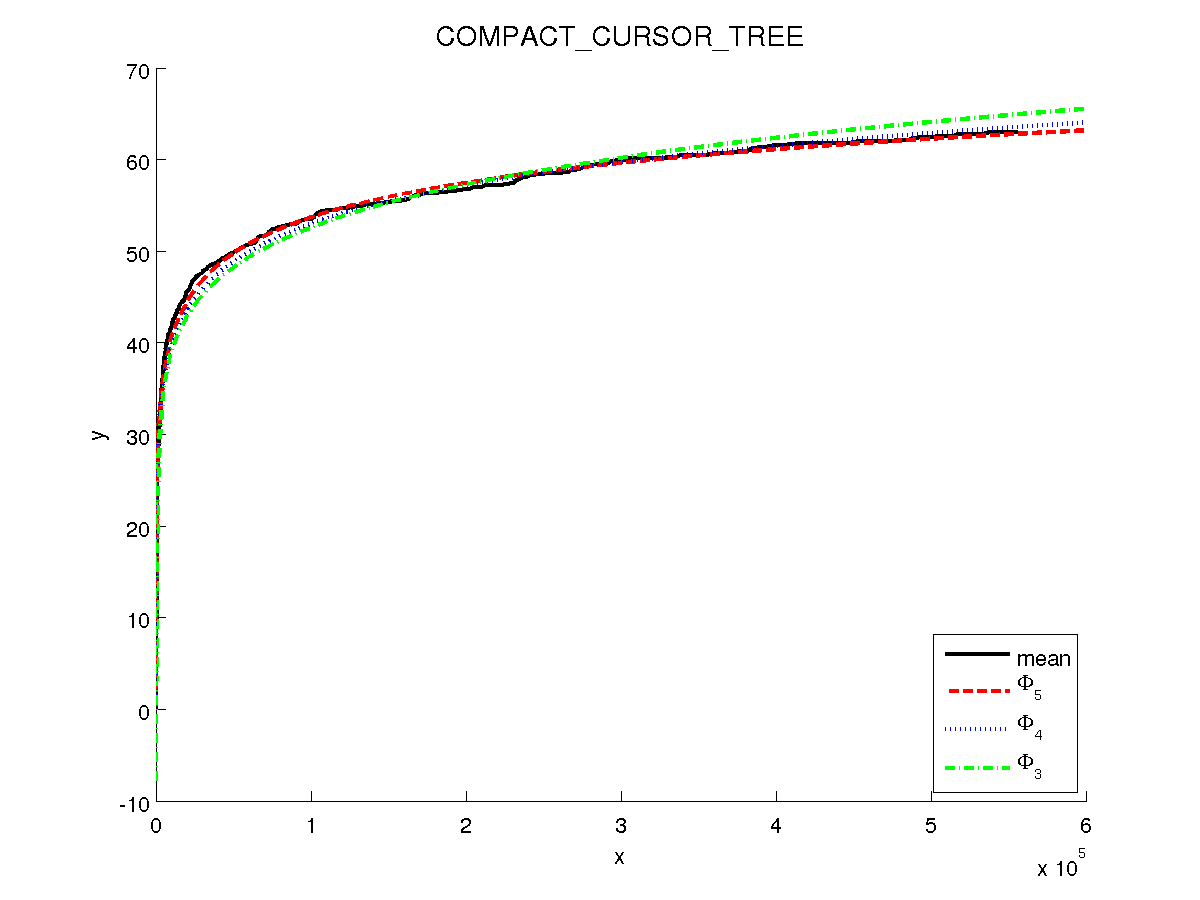} &
    \includegraphics[width={.48\textwidth}]{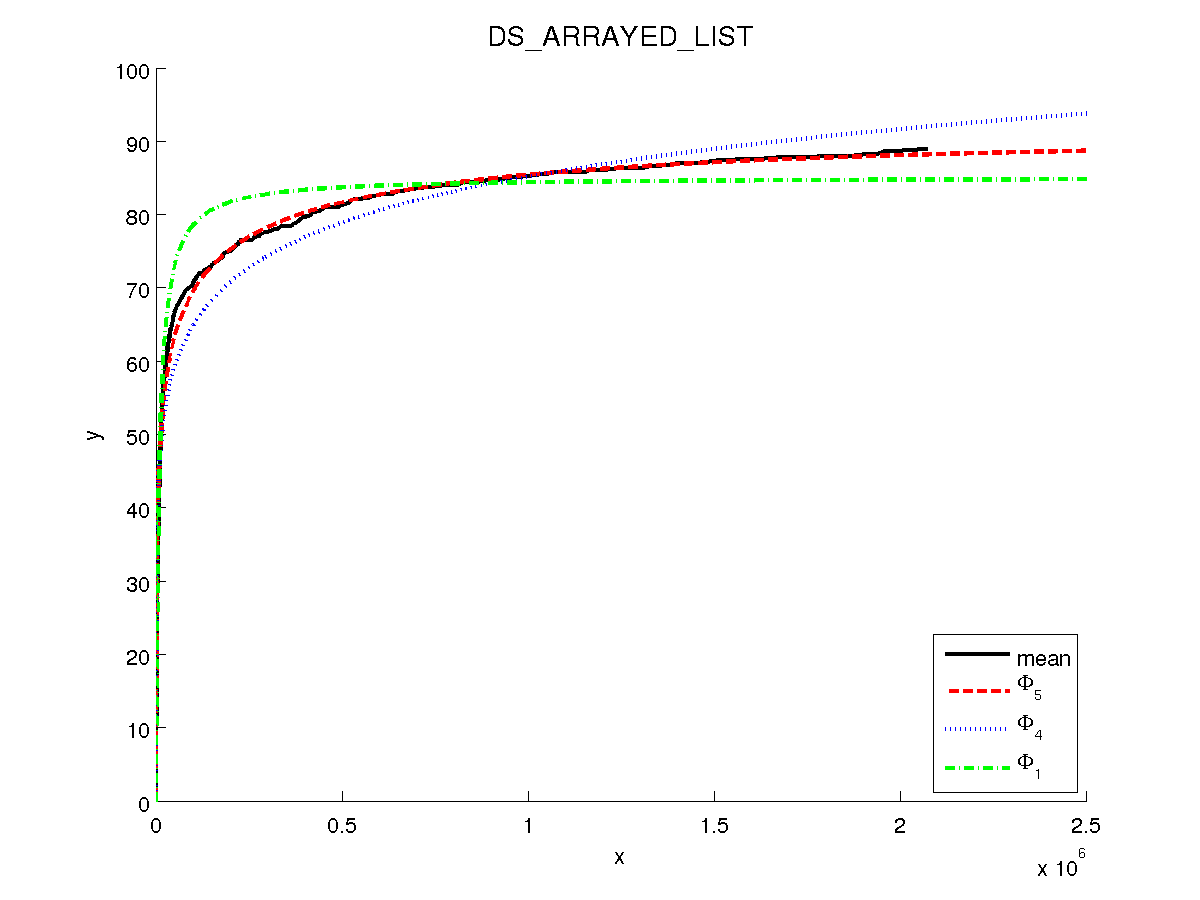} \\
    \includegraphics[width={.48\textwidth}]{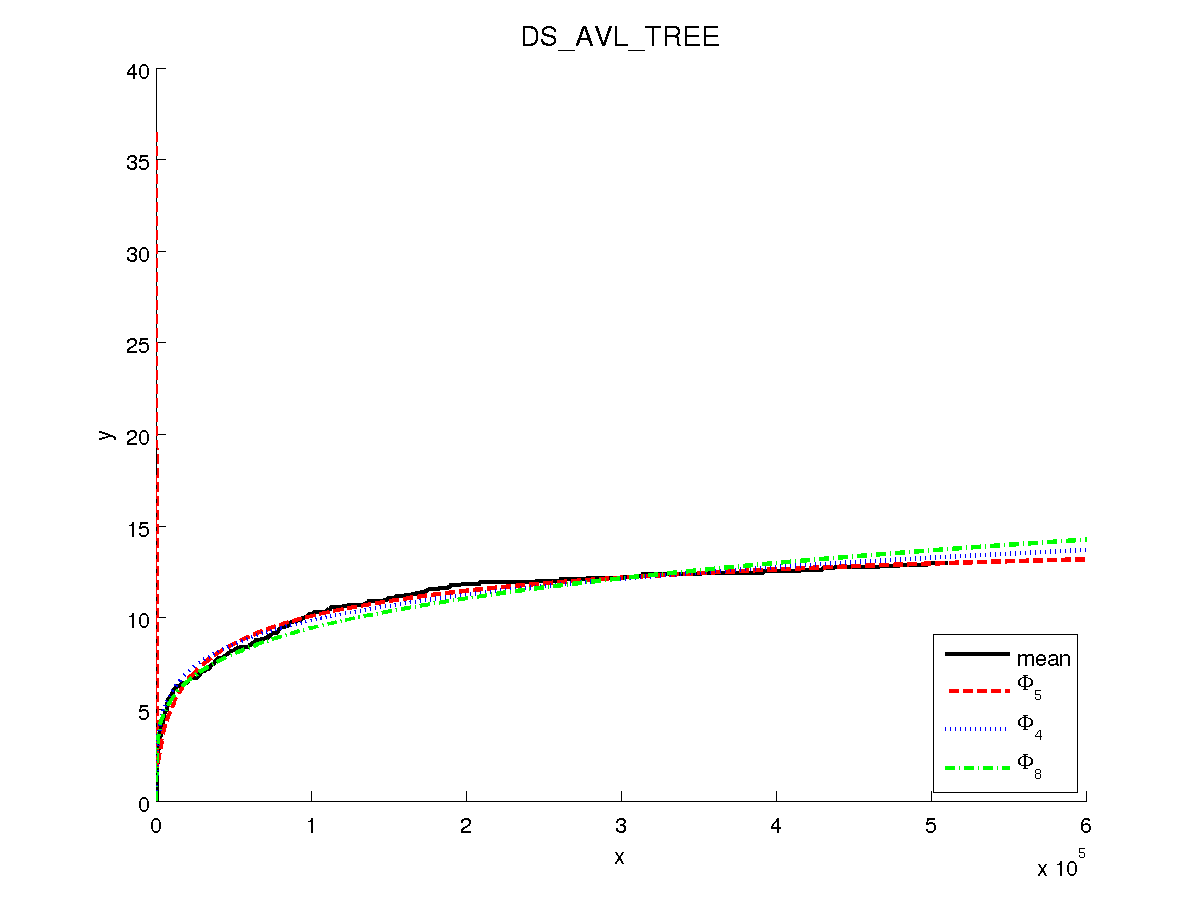} &
    \includegraphics[width={.48\textwidth}]{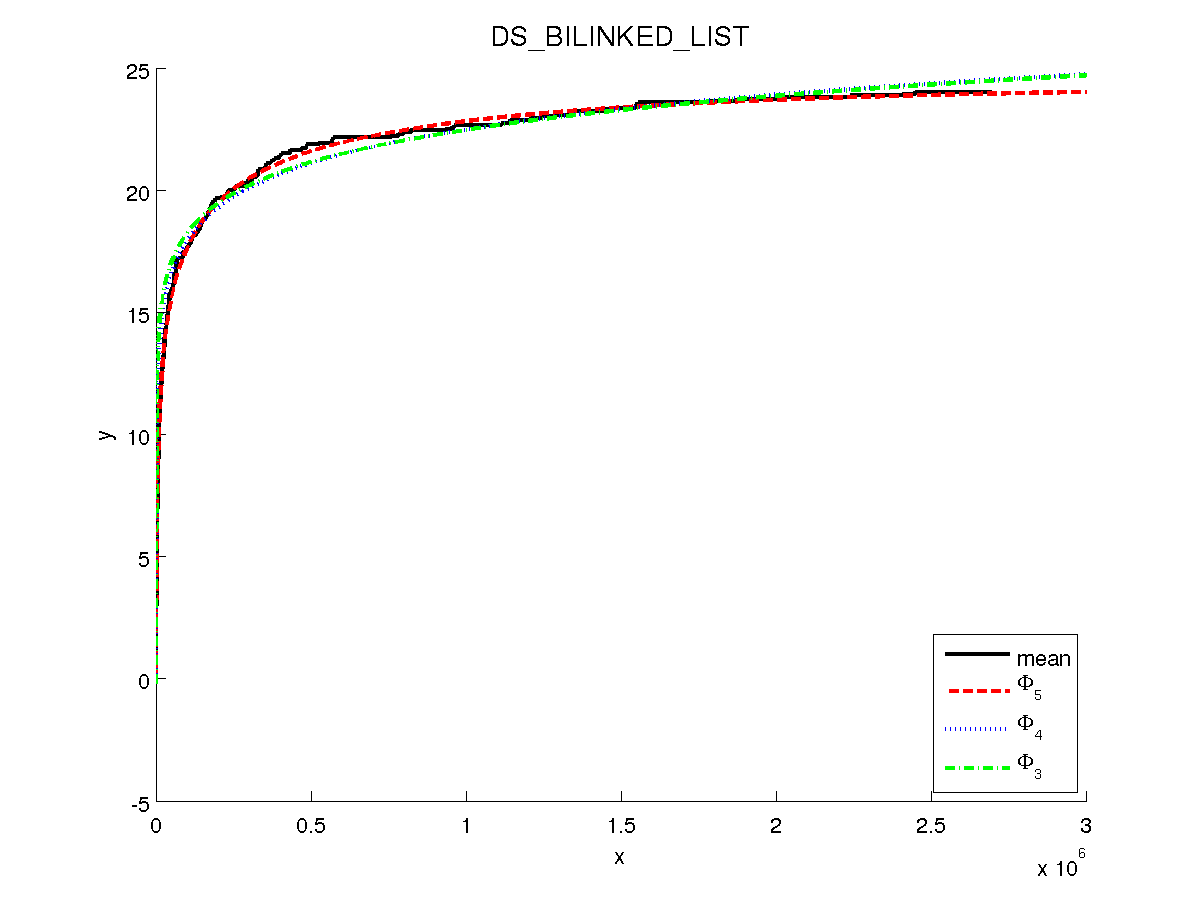} 
  \end{array}$
\end{center}
\caption{Top 3 fits with mean for six Eiffel classes.}
\end{figure*}

\begin{figure*}[!htb]
  \begin{center}$
  \begin{array}{cc}
    \includegraphics[width={.48\textwidth}]{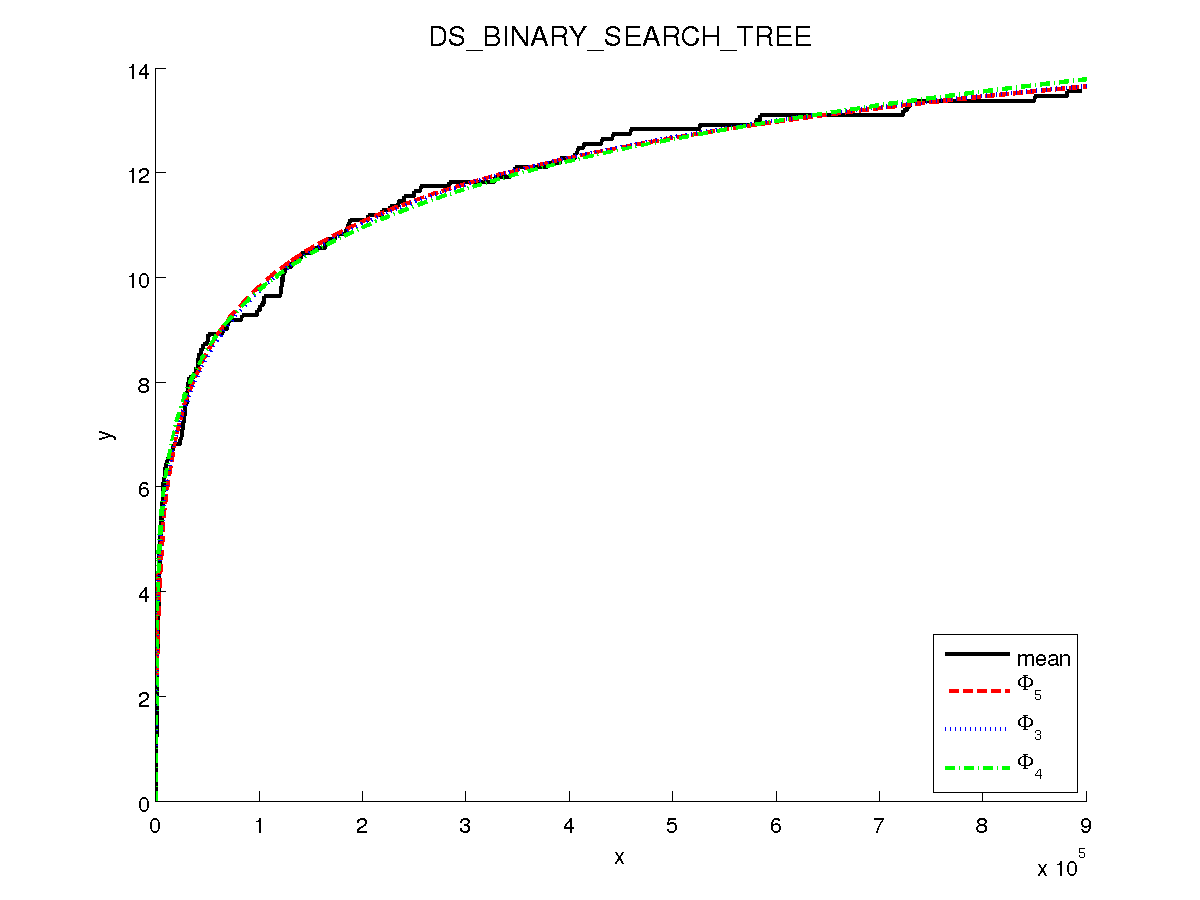} &
    \includegraphics[width={.48\textwidth}]{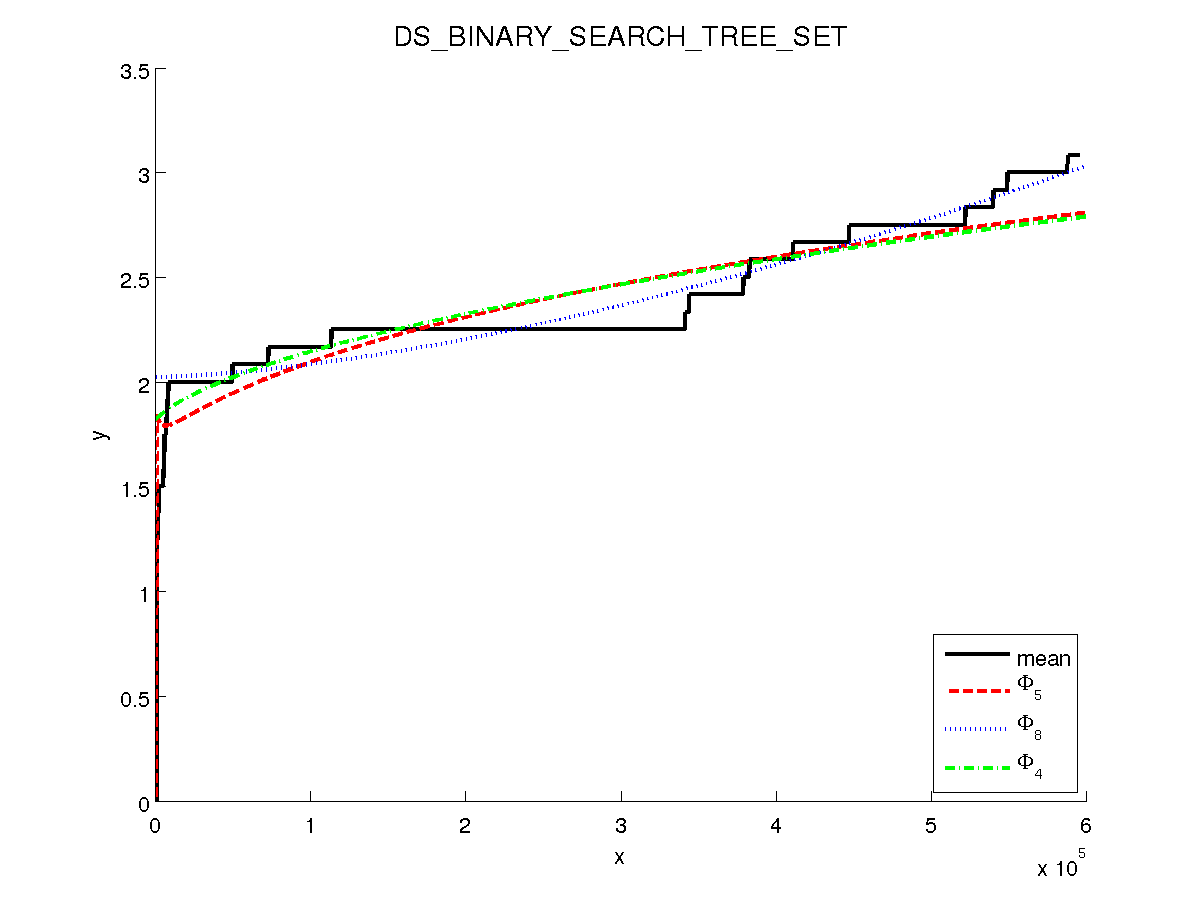} \\
    \includegraphics[width={.48\textwidth}]{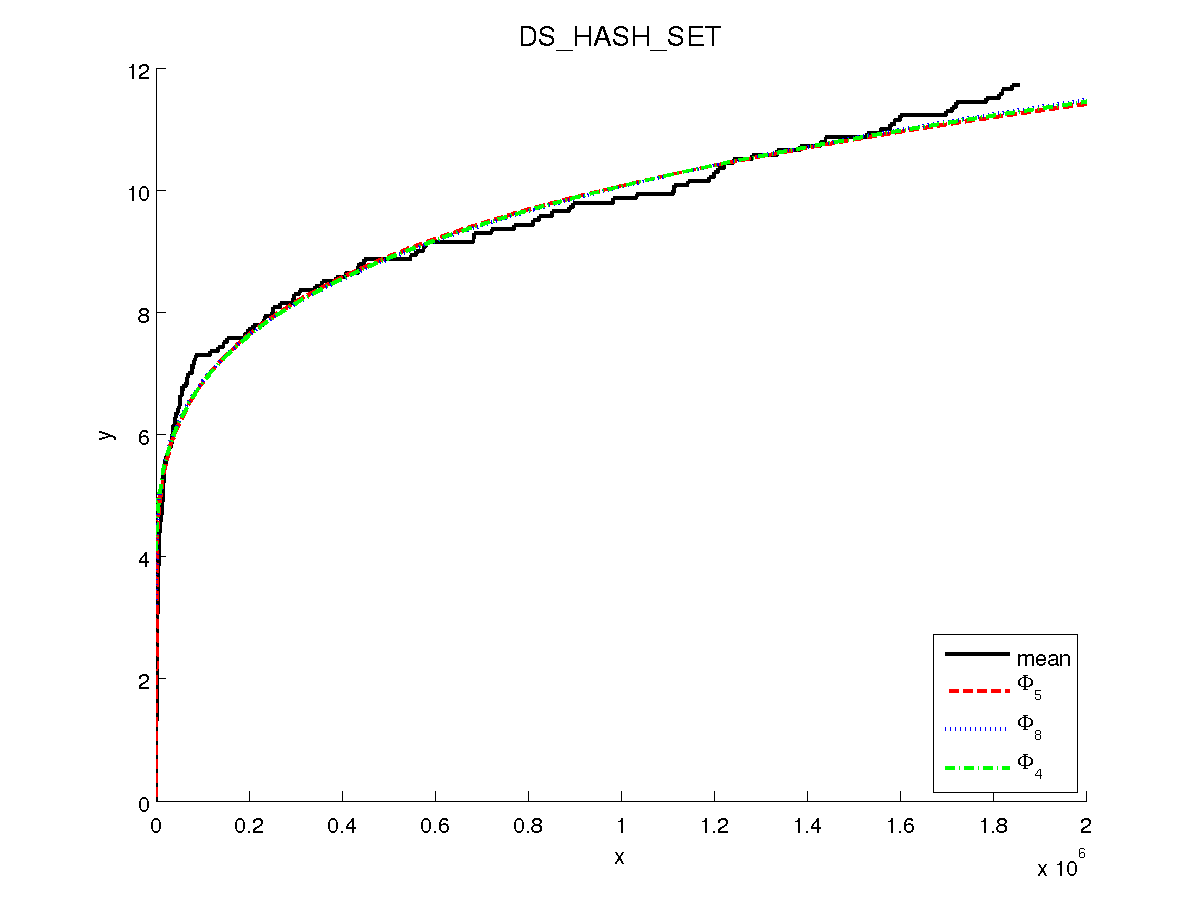} &
    \includegraphics[width={.48\textwidth}]{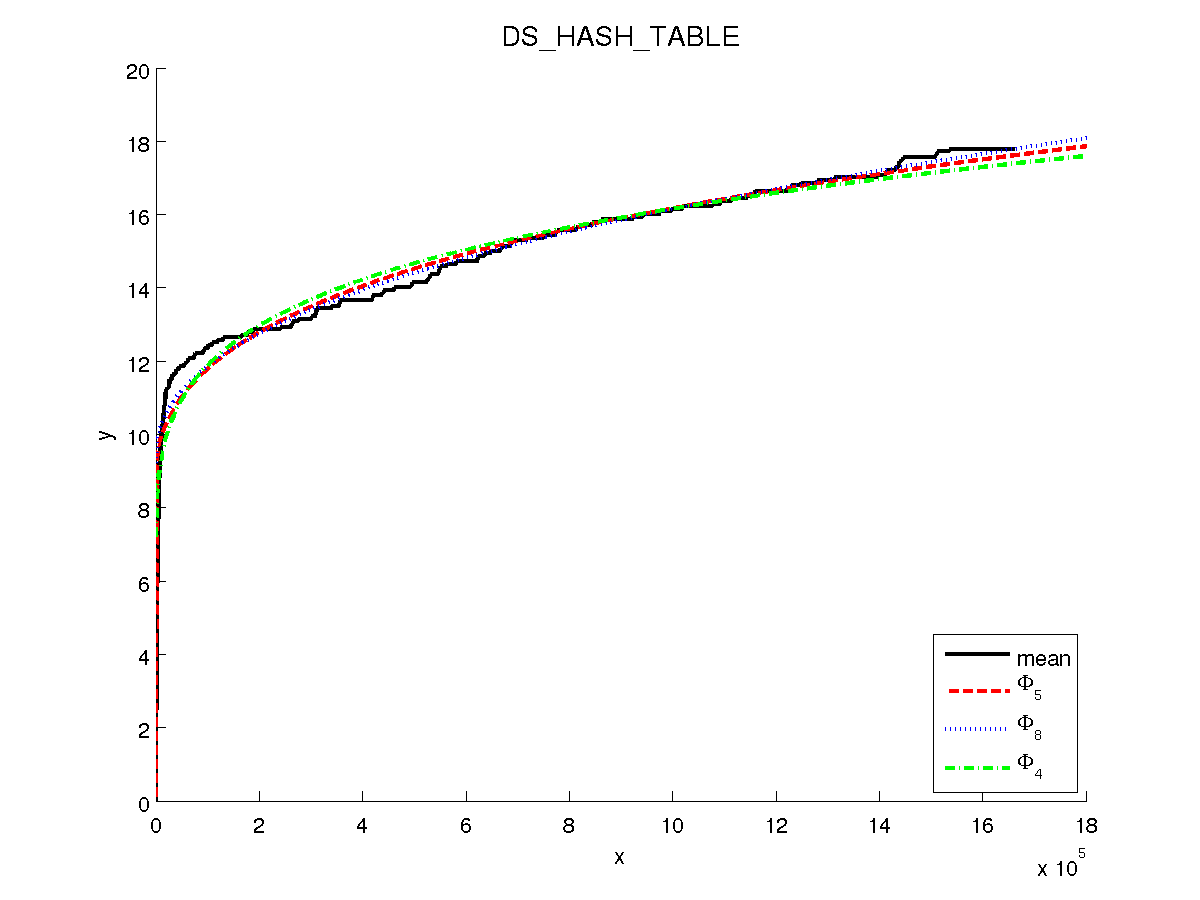} \\
    \includegraphics[width={.48\textwidth}]{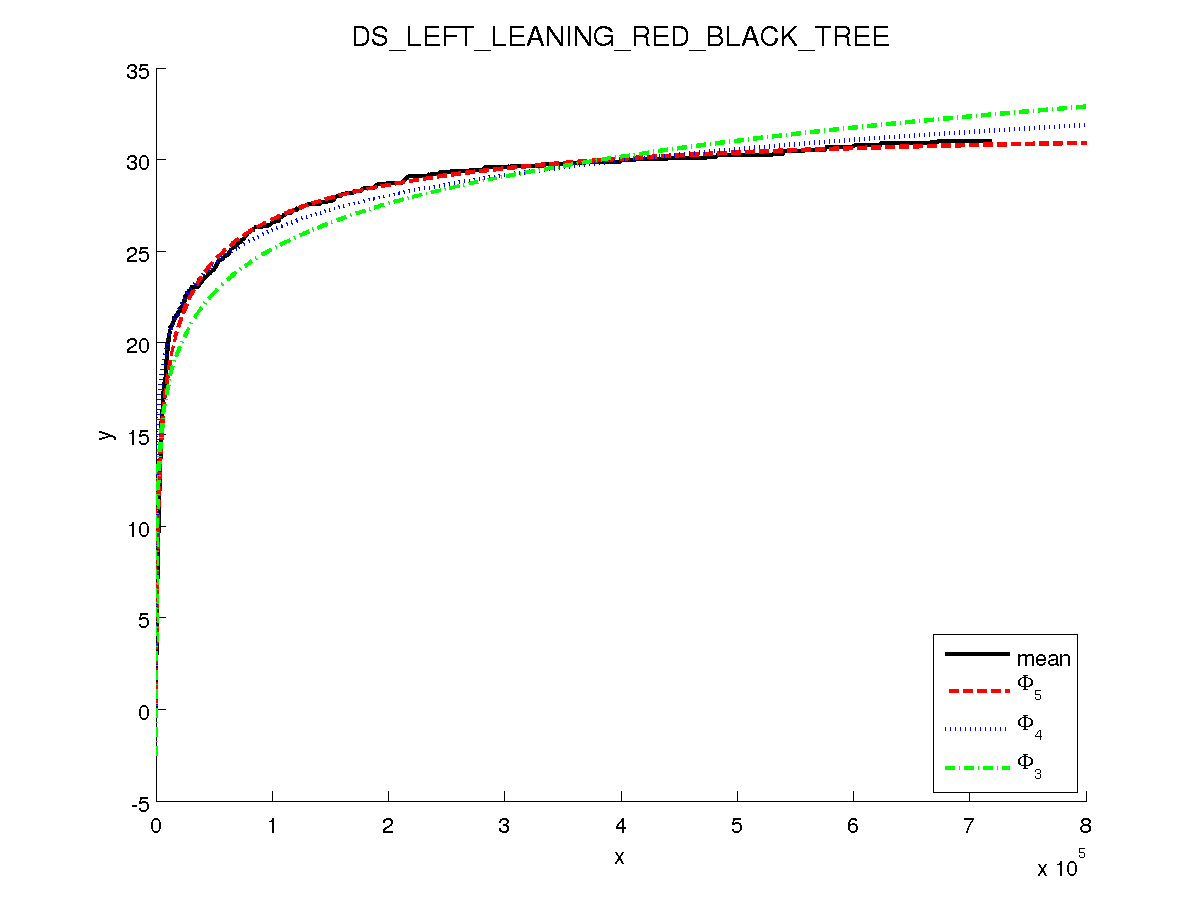} &
    \includegraphics[width={.48\textwidth}]{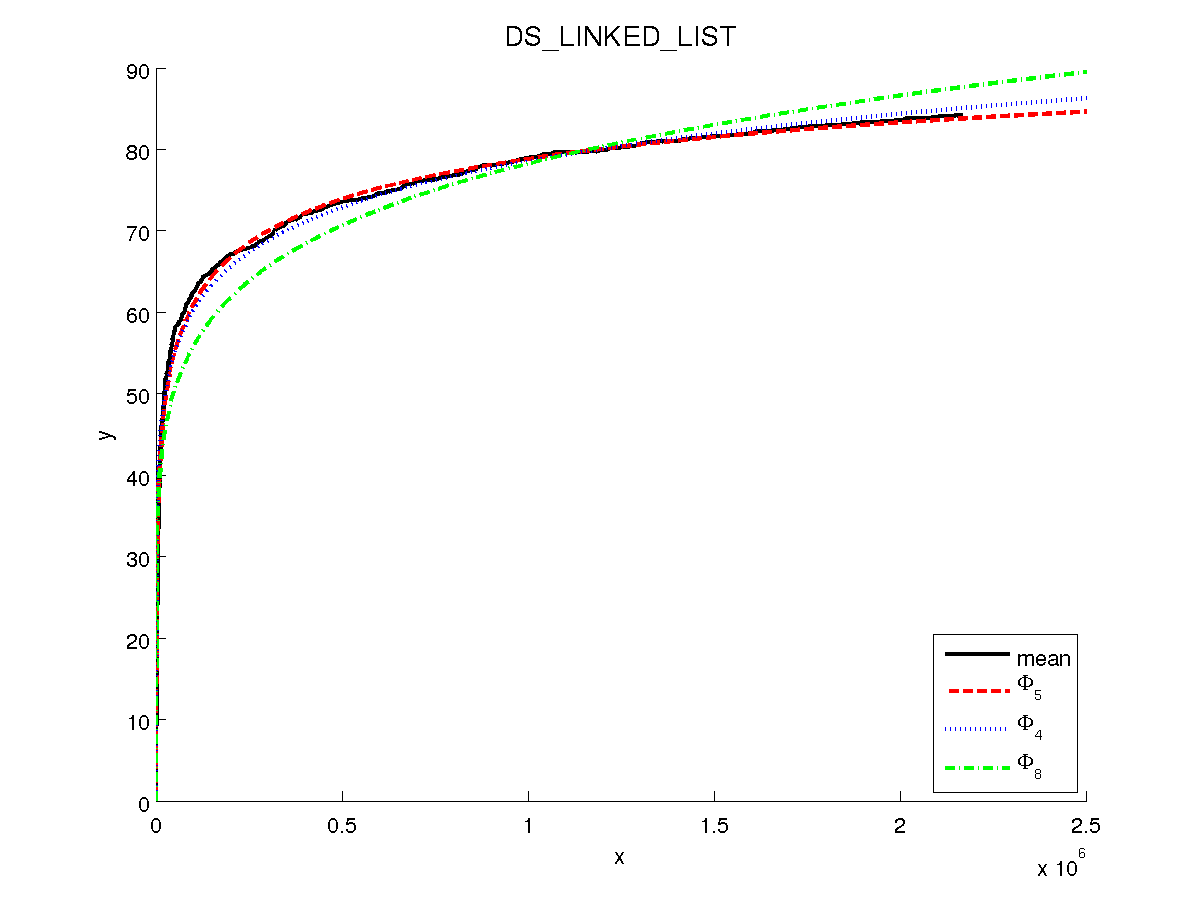} 
  \end{array}$
\end{center}
\caption{Top 3 fits with mean for six Eiffel classes.}
\end{figure*}

\begin{figure*}[!htb]
  \begin{center}$
  \begin{array}{cc}
    \includegraphics[width={.48\textwidth}]{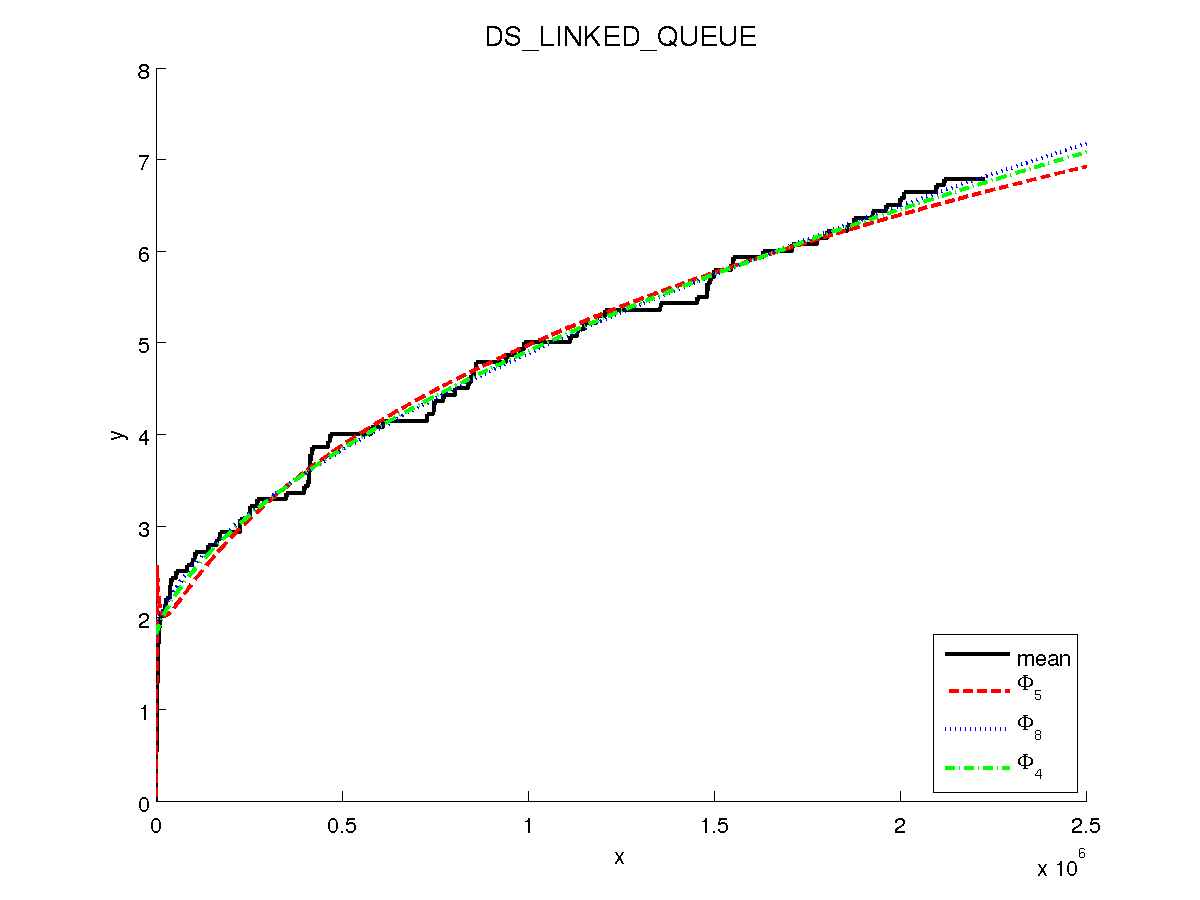} &
    \includegraphics[width={.48\textwidth}]{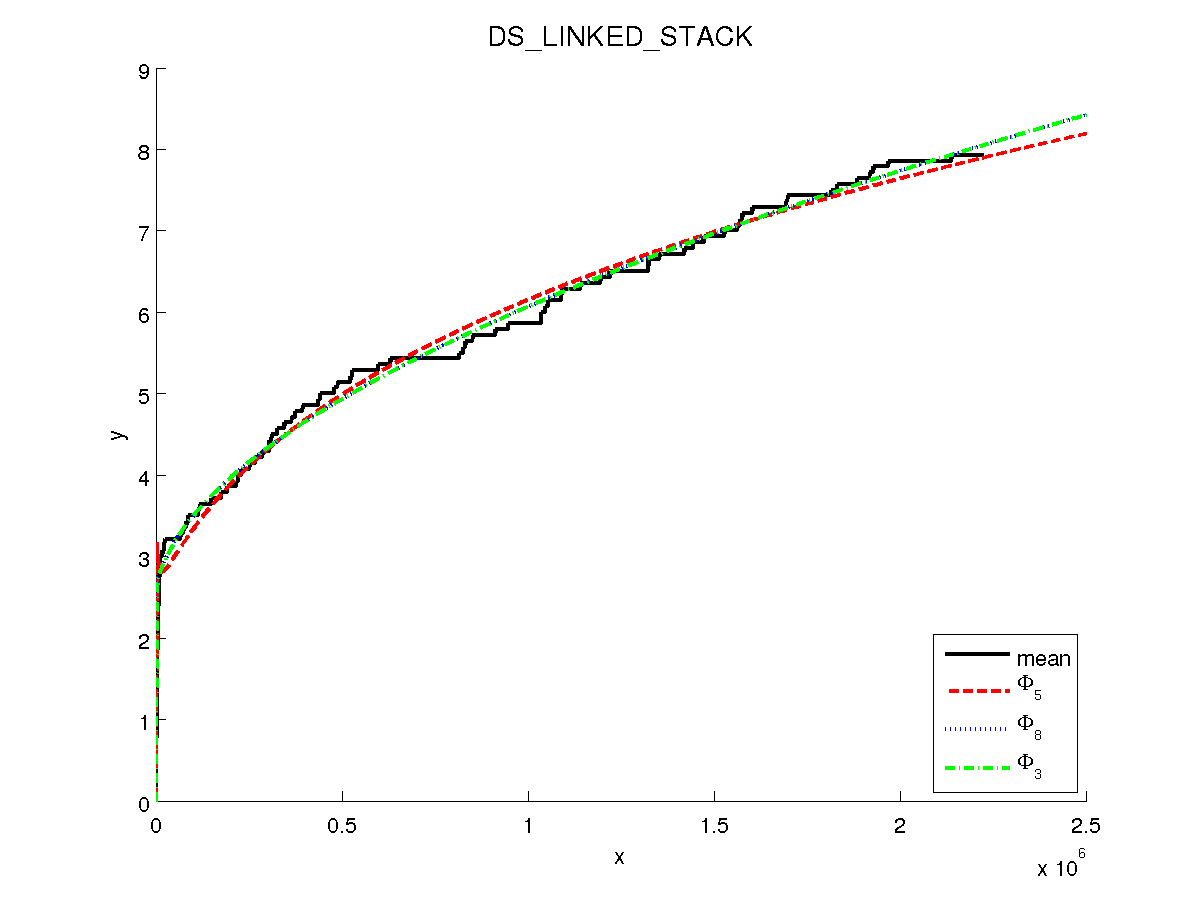} \\
    \includegraphics[width={.48\textwidth}]{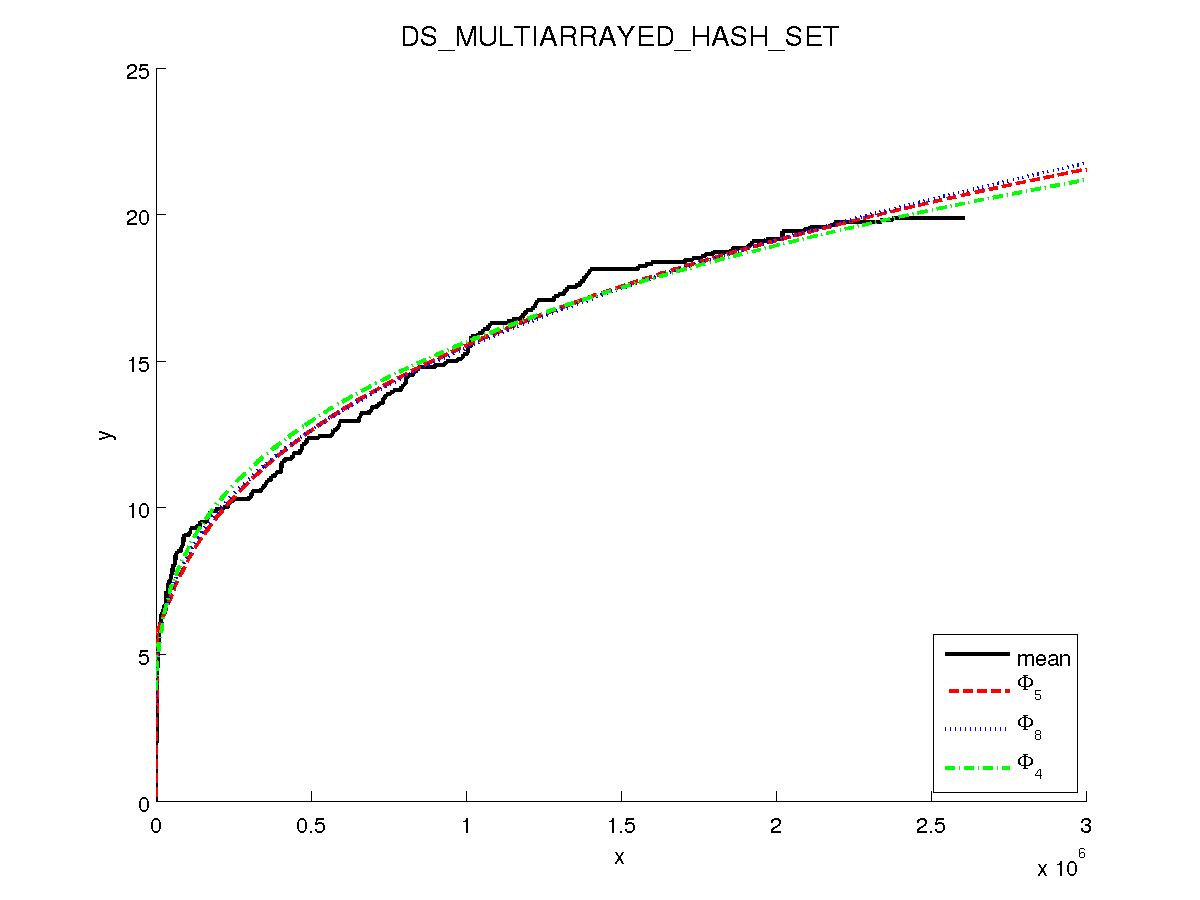} &
    \includegraphics[width={.48\textwidth}]{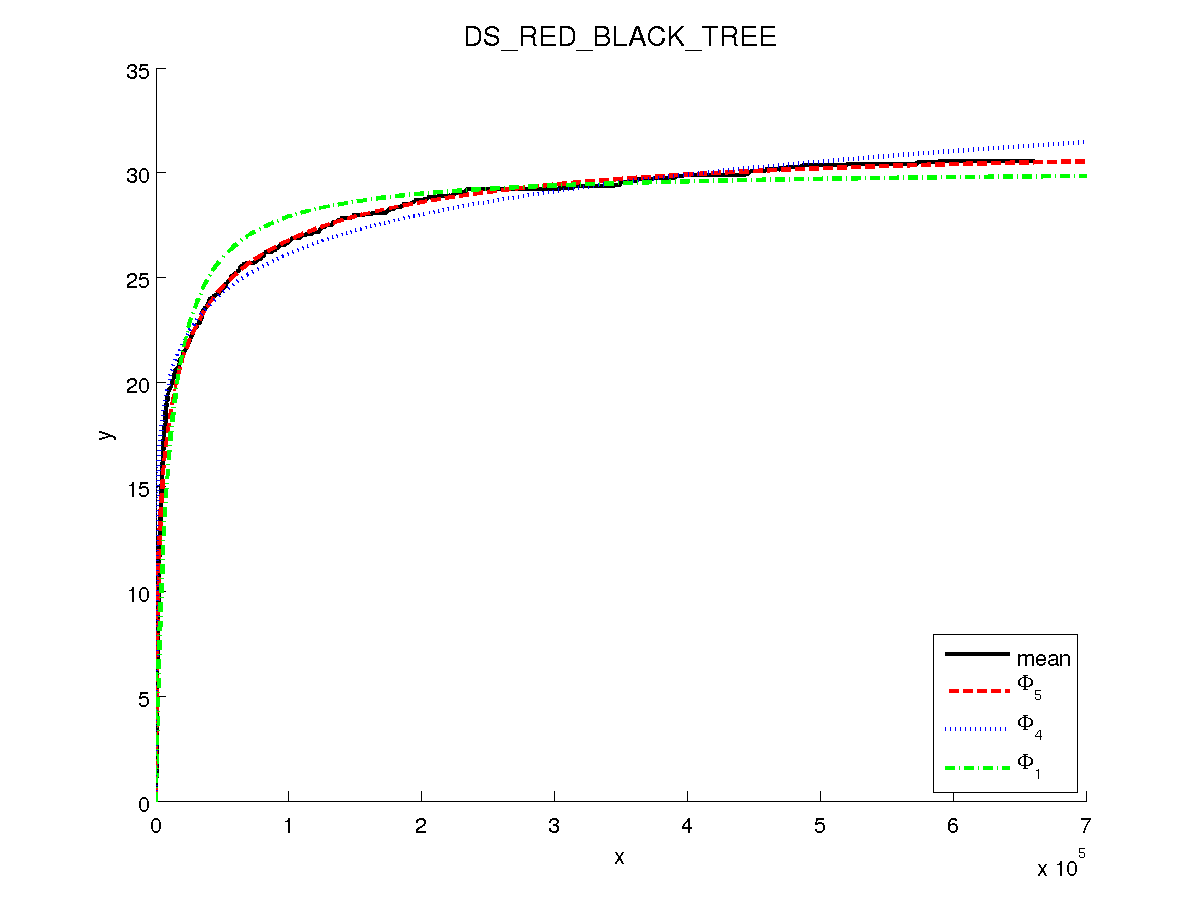} \\
    \includegraphics[width={.48\textwidth}]{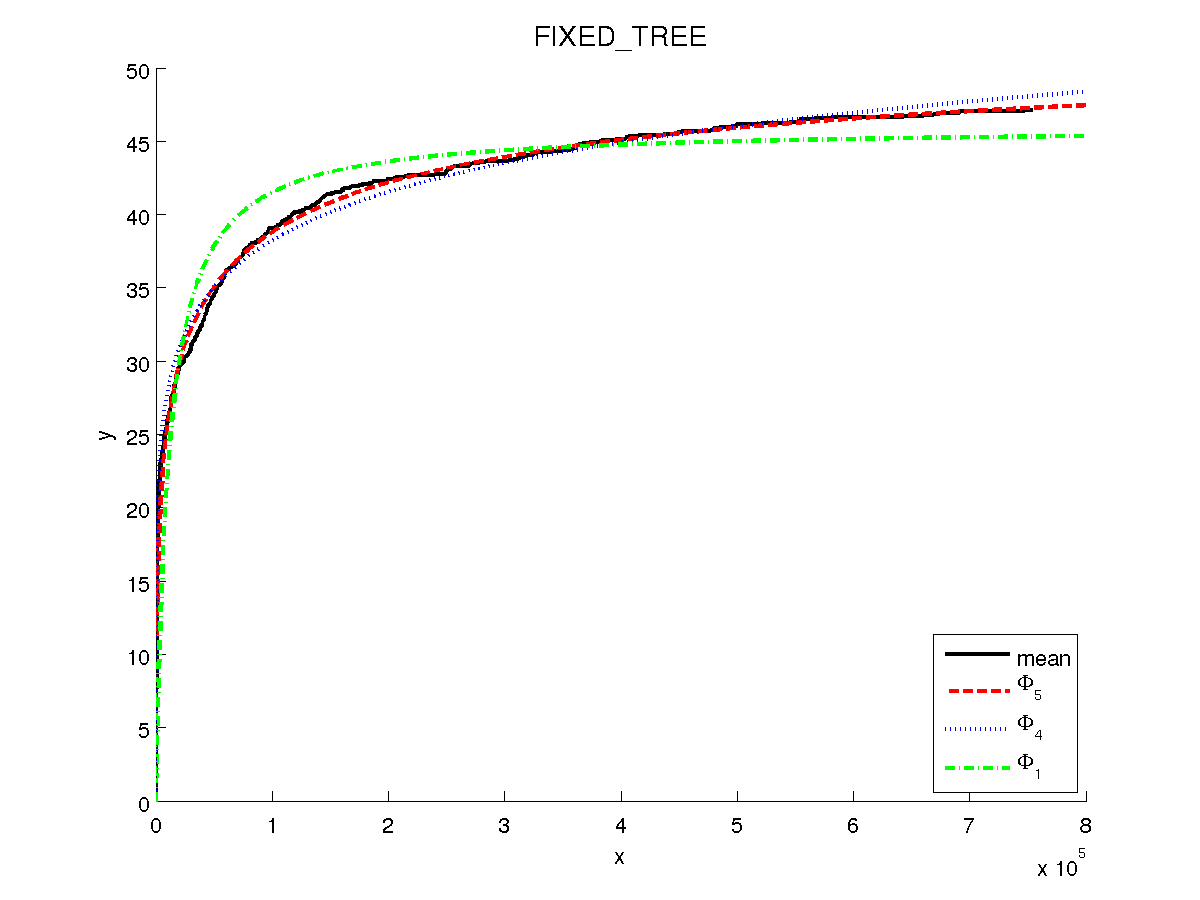} &
    \includegraphics[width={.48\textwidth}]{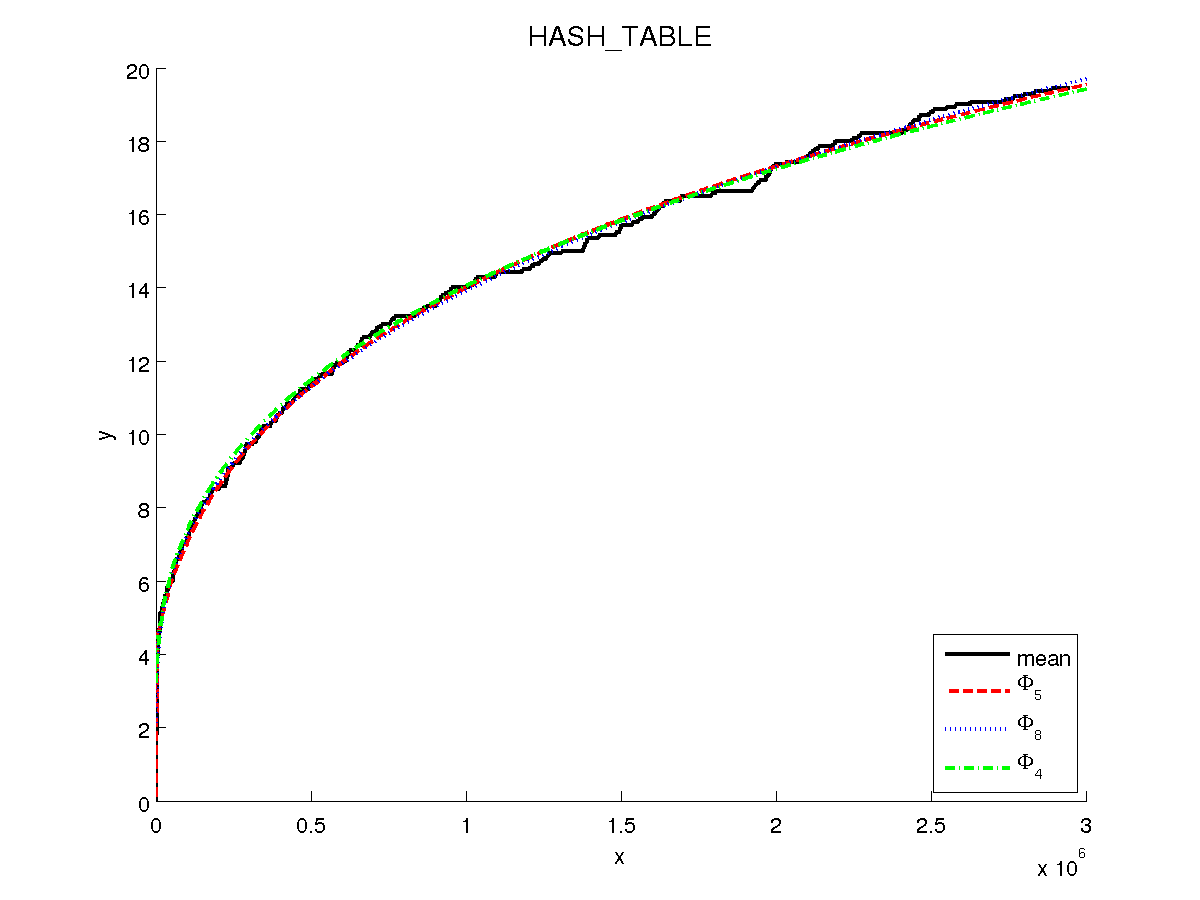}
  \end{array}$
\end{center}
\caption{Top 3 fits with mean for six Eiffel classes.}
\end{figure*}

\begin{figure*}[!htb]
  \begin{center}$
  \begin{array}{cc}
    \includegraphics[width={.48\textwidth}]{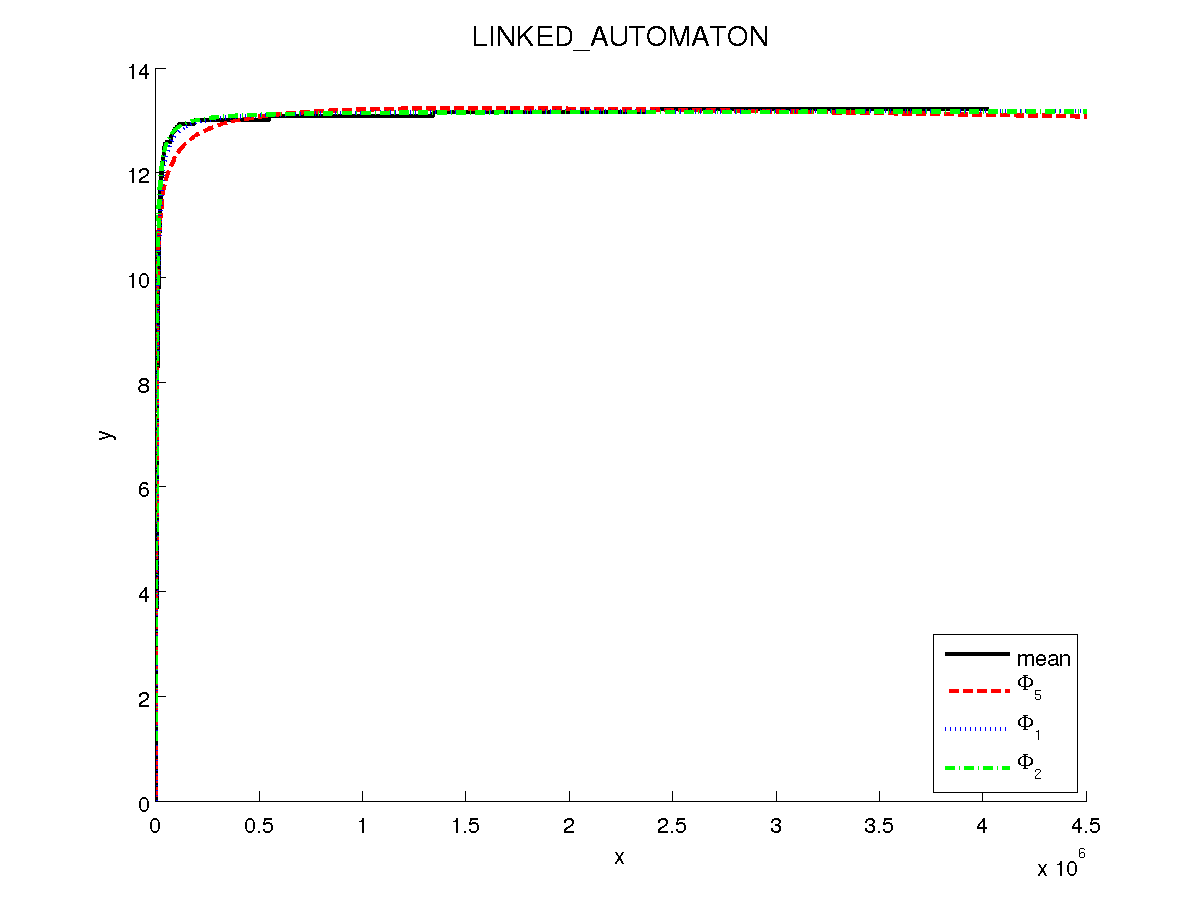} &
    \includegraphics[width={.48\textwidth}]{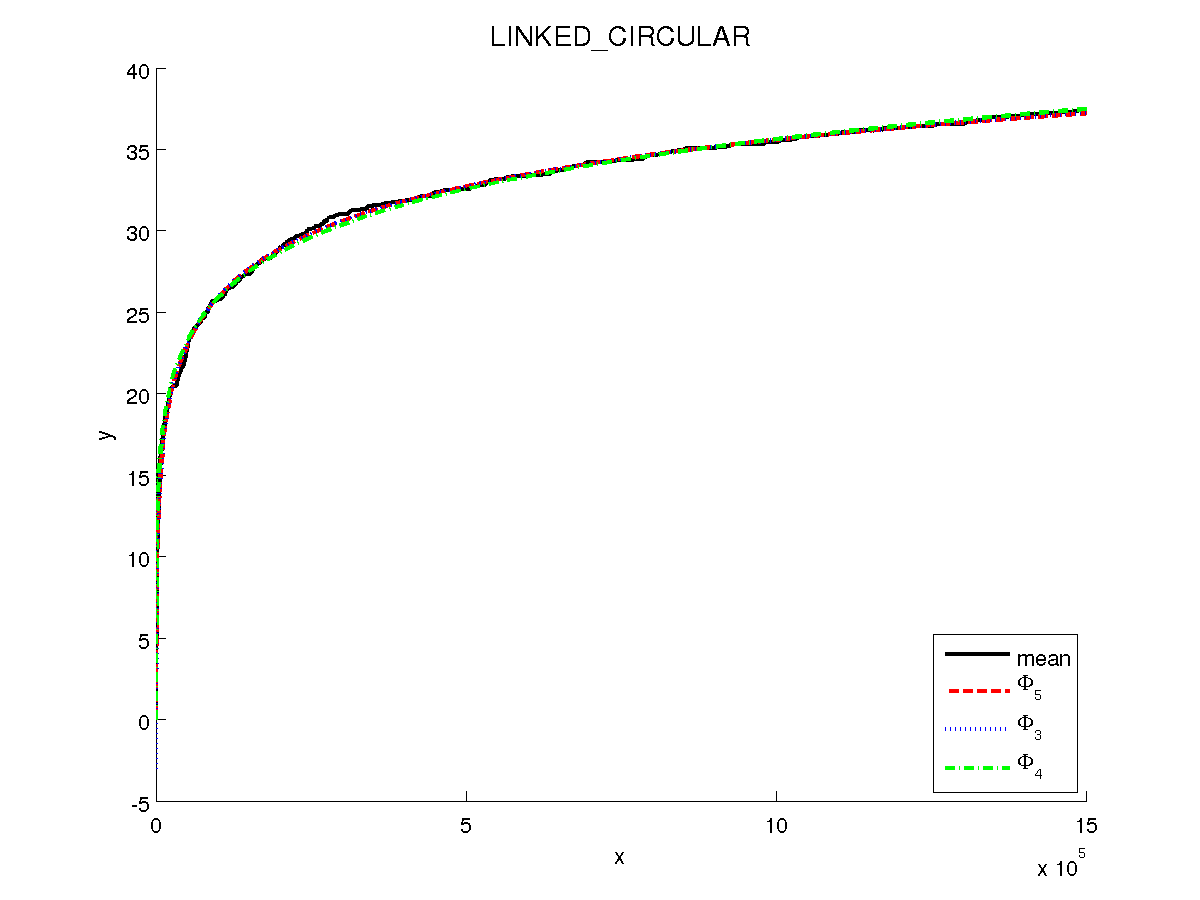} \\
    \includegraphics[width={.48\textwidth}]{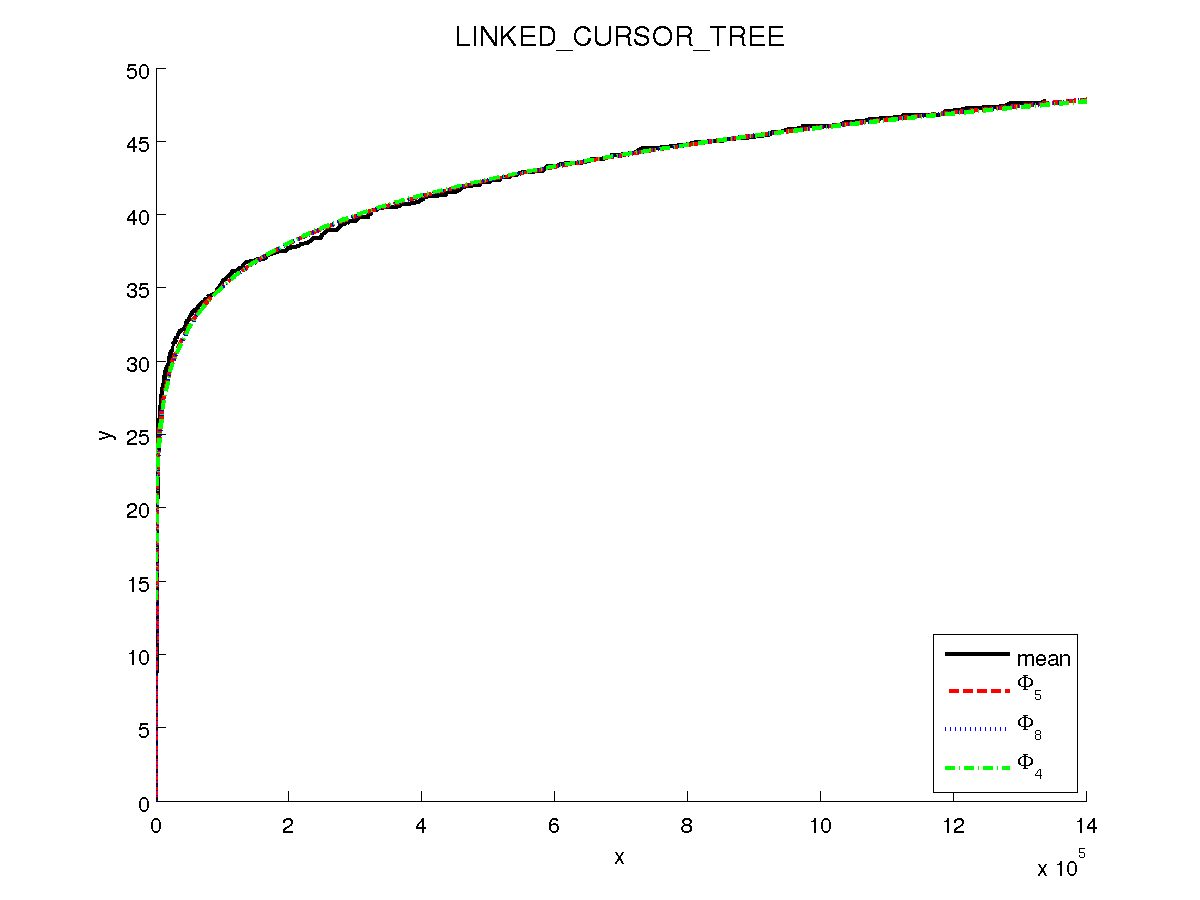} &
    \includegraphics[width={.48\textwidth}]{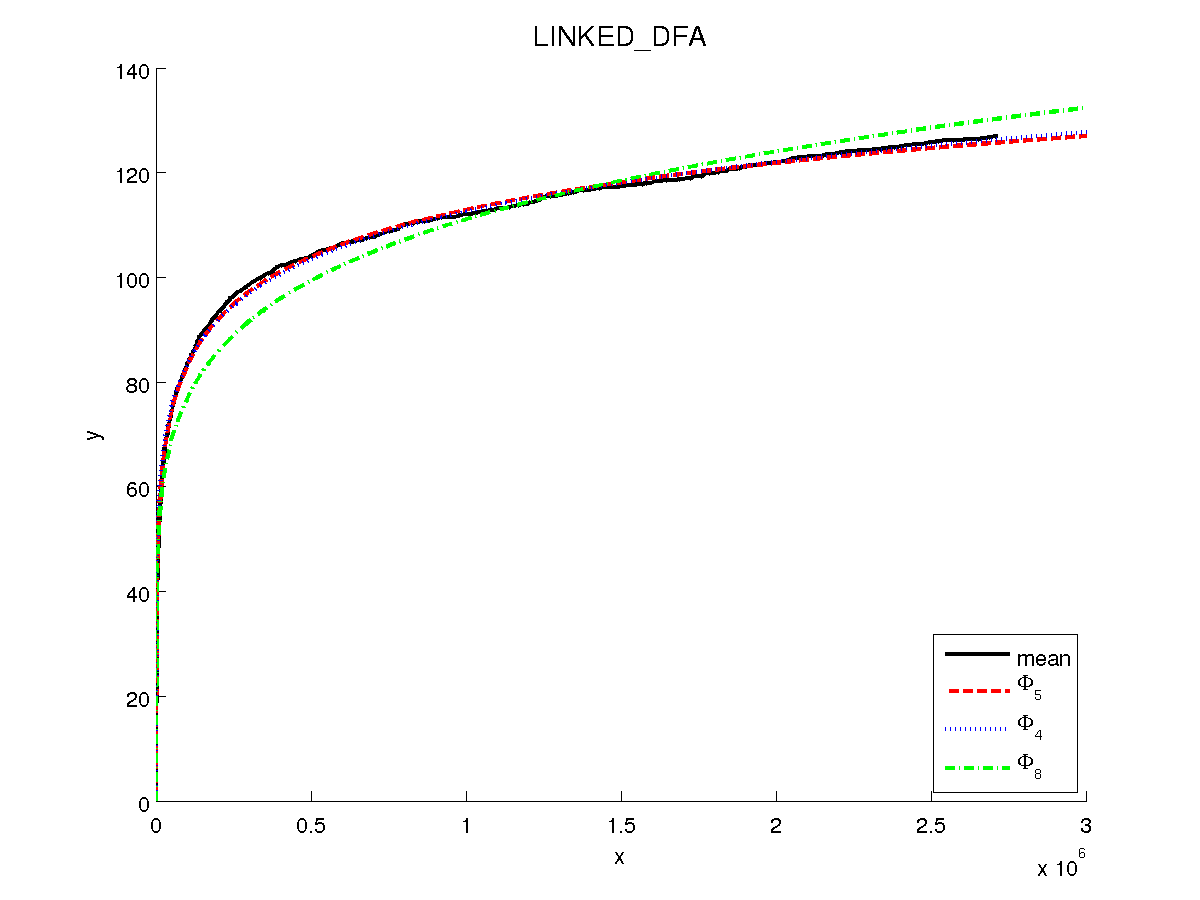} \\
    \includegraphics[width={.48\textwidth}]{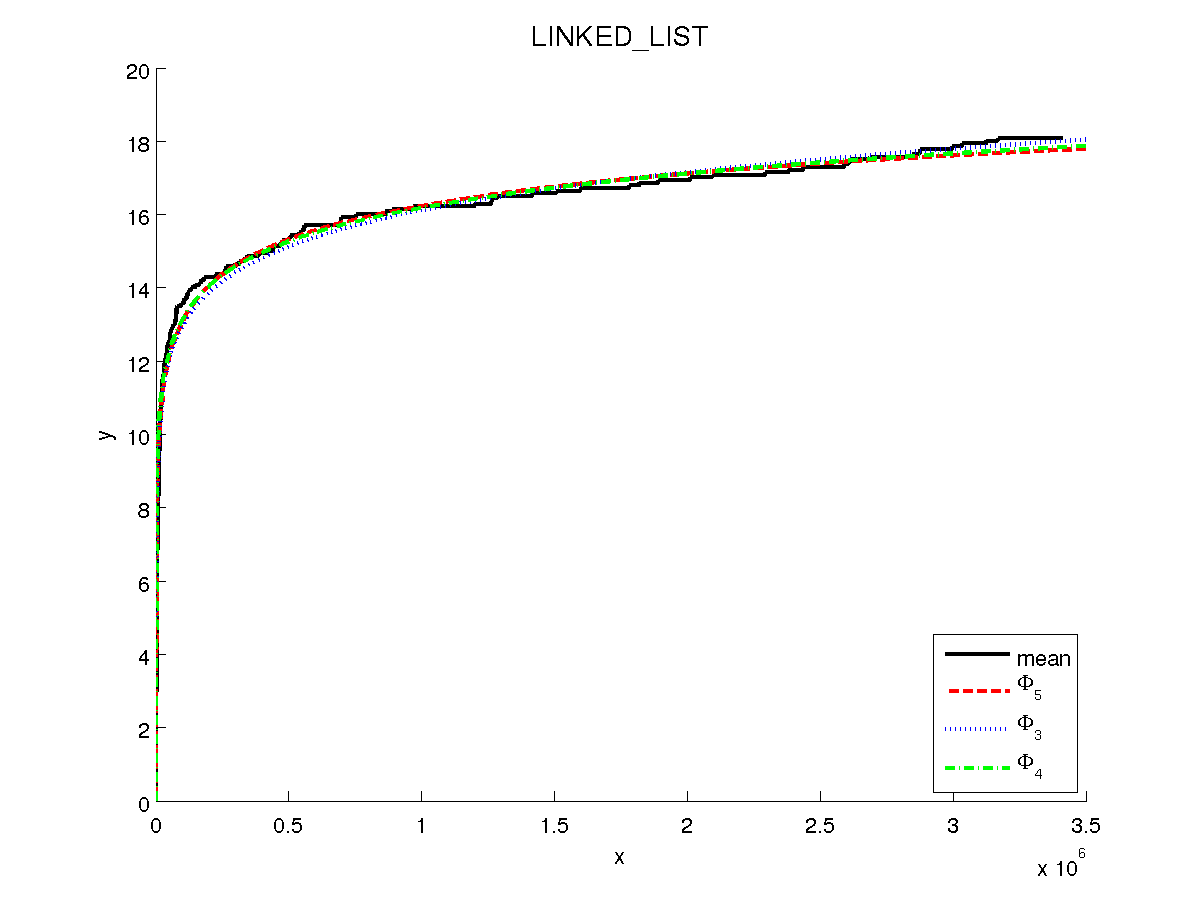} &
    \includegraphics[width={.48\textwidth}]{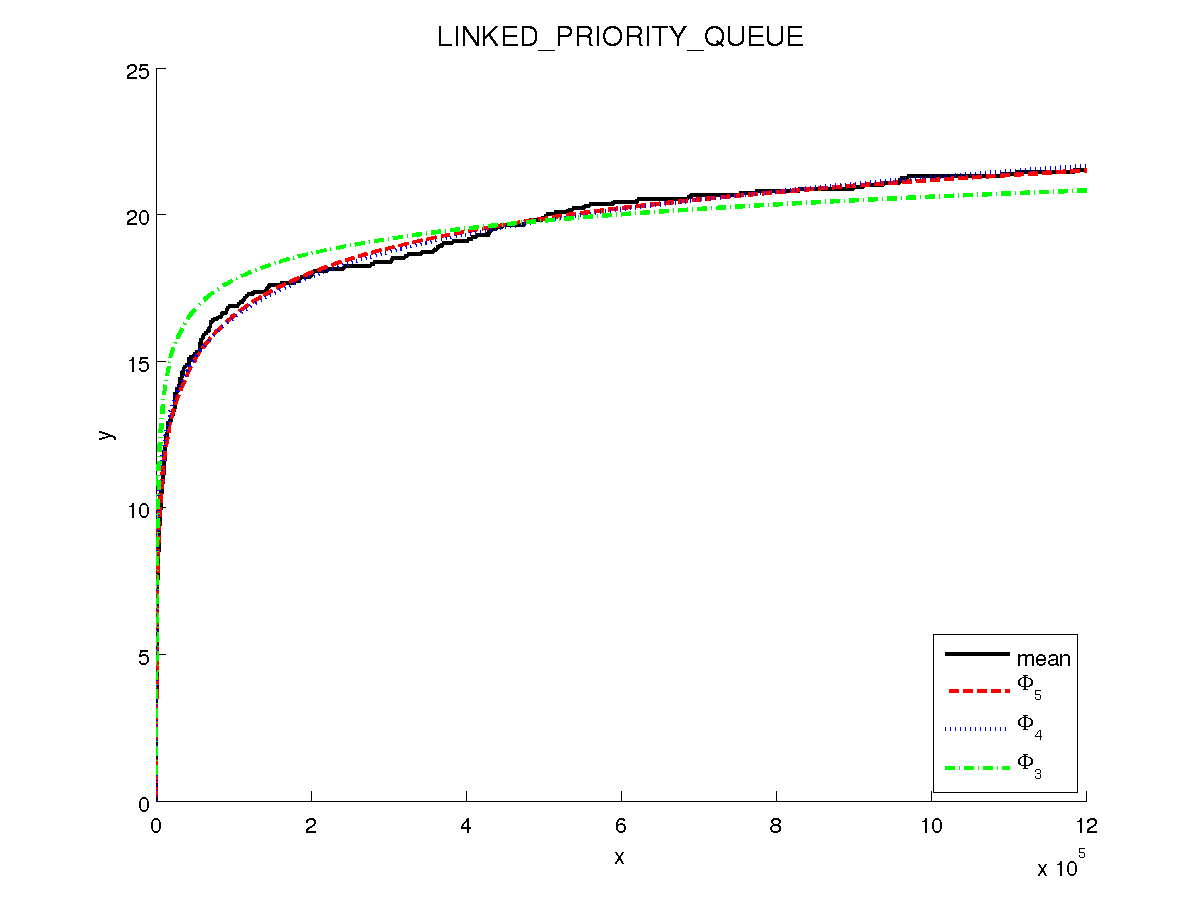}
  \end{array}$
\end{center}
\caption{Top 3 fits with mean for six Eiffel classes.}
\end{figure*}

\begin{figure*}[!htb]
  \begin{center}$
  \begin{array}{cc}
    \includegraphics[width={.48\textwidth}]{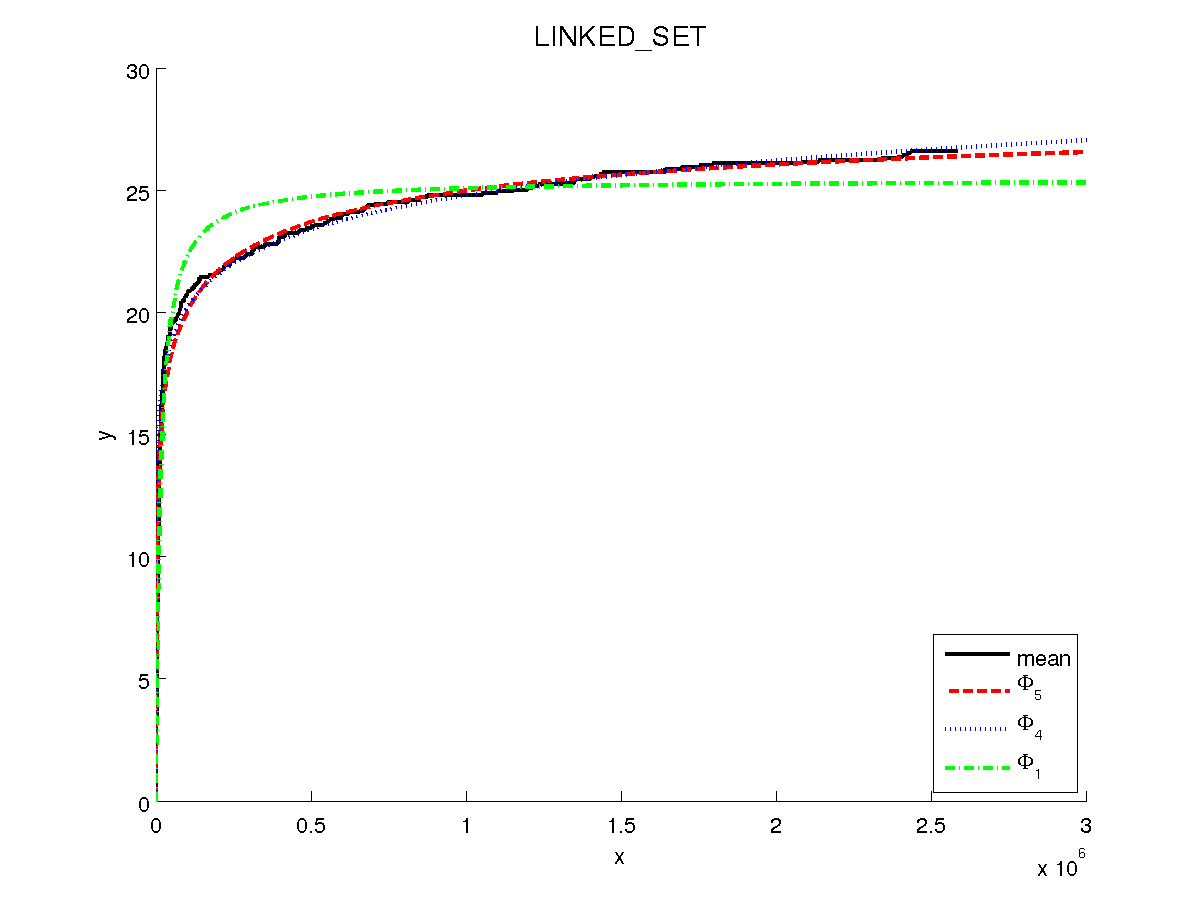} &
    \includegraphics[width={.48\textwidth}]{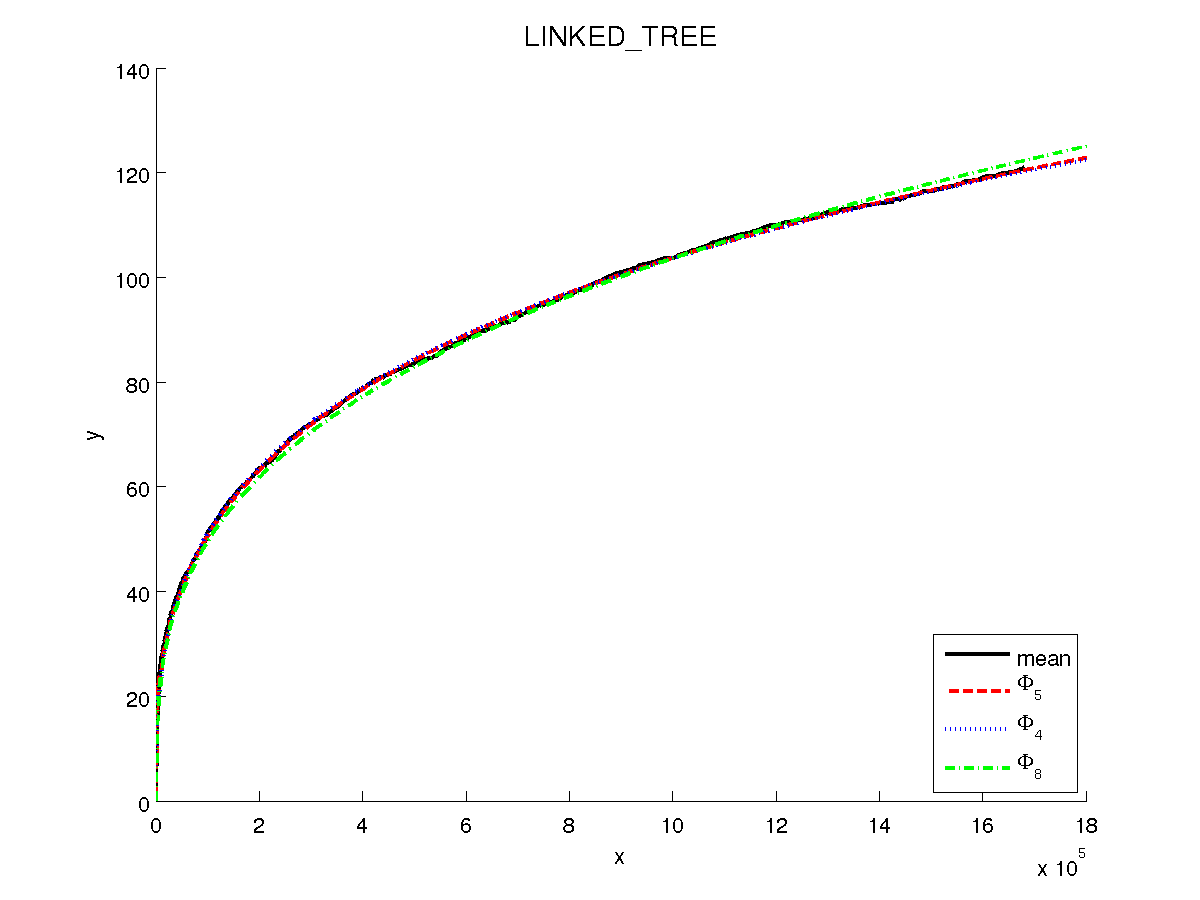} \\
    \includegraphics[width={.48\textwidth}]{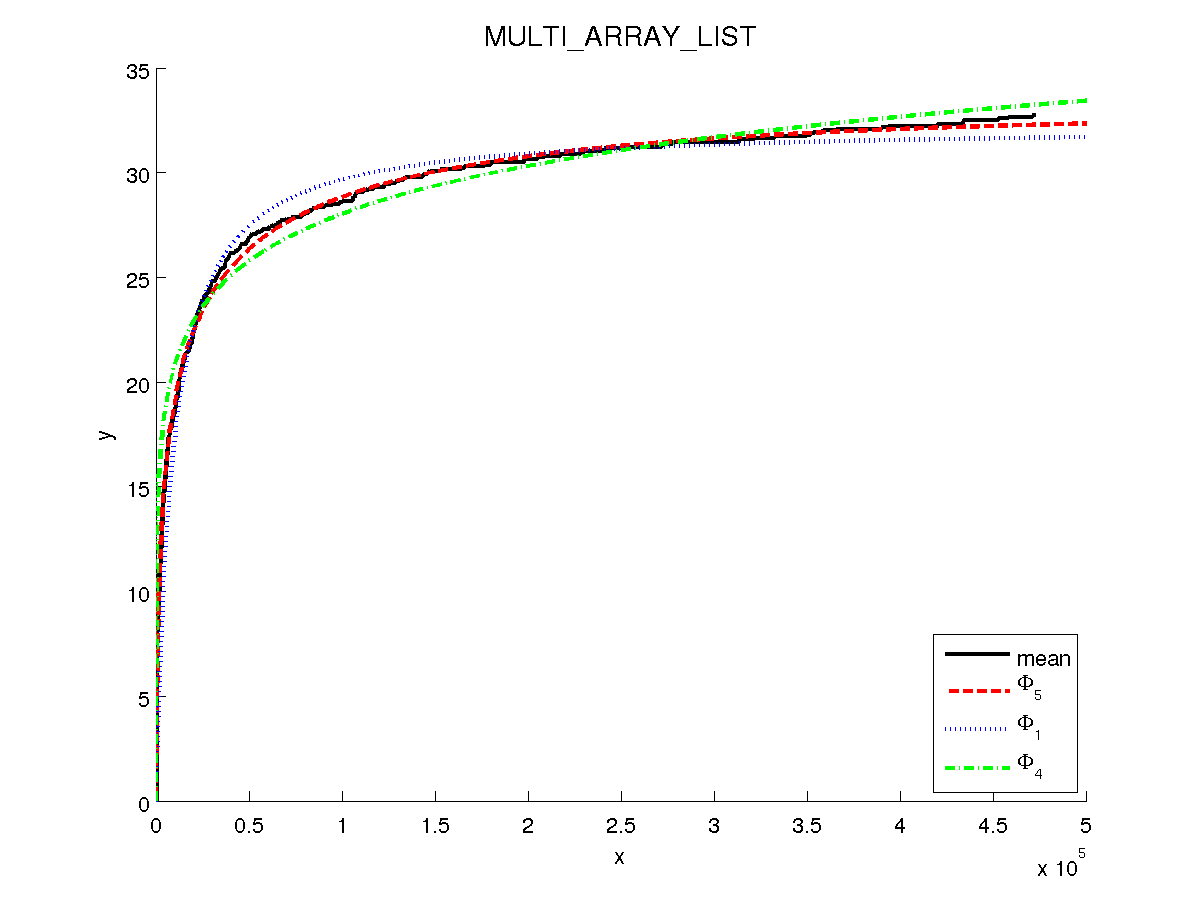} &
    \includegraphics[width={.48\textwidth}]{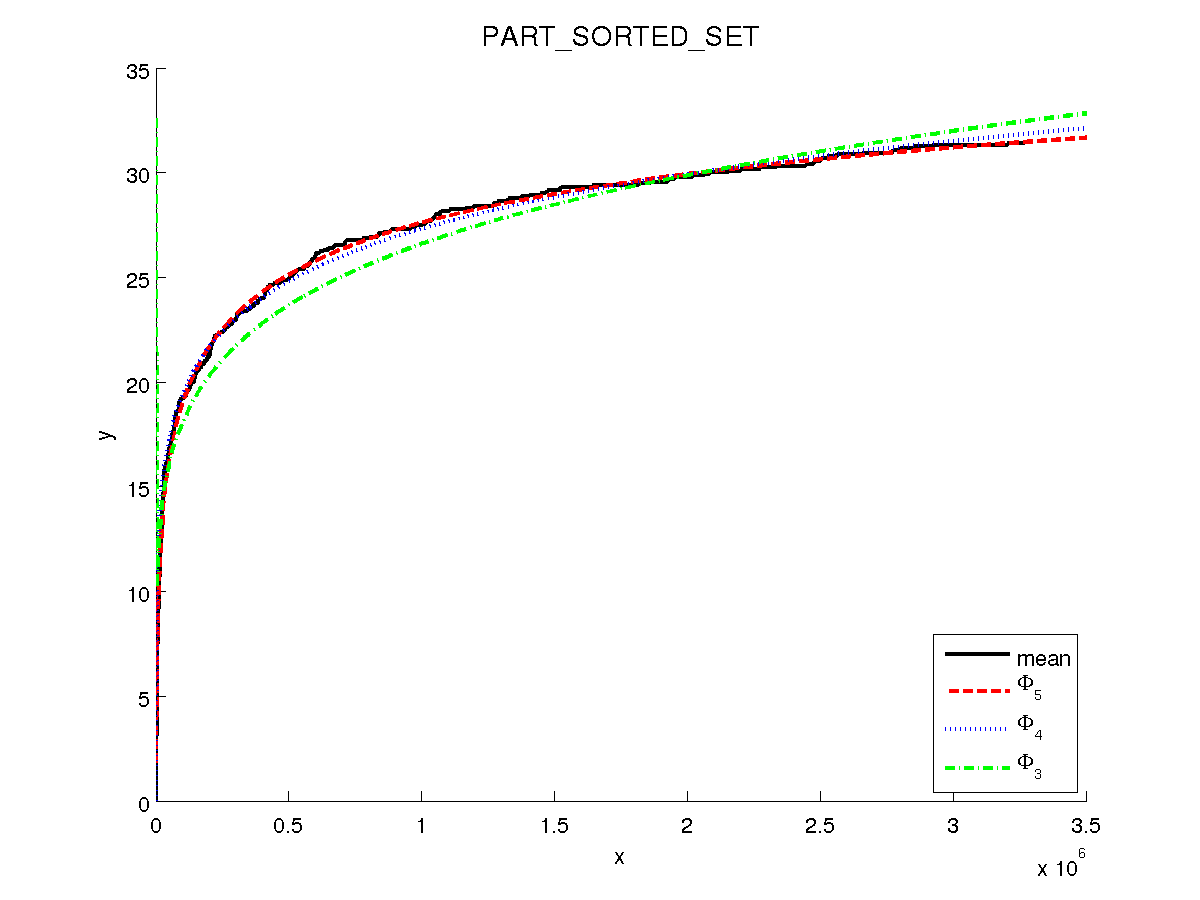} \\
    \includegraphics[width={.48\textwidth}]{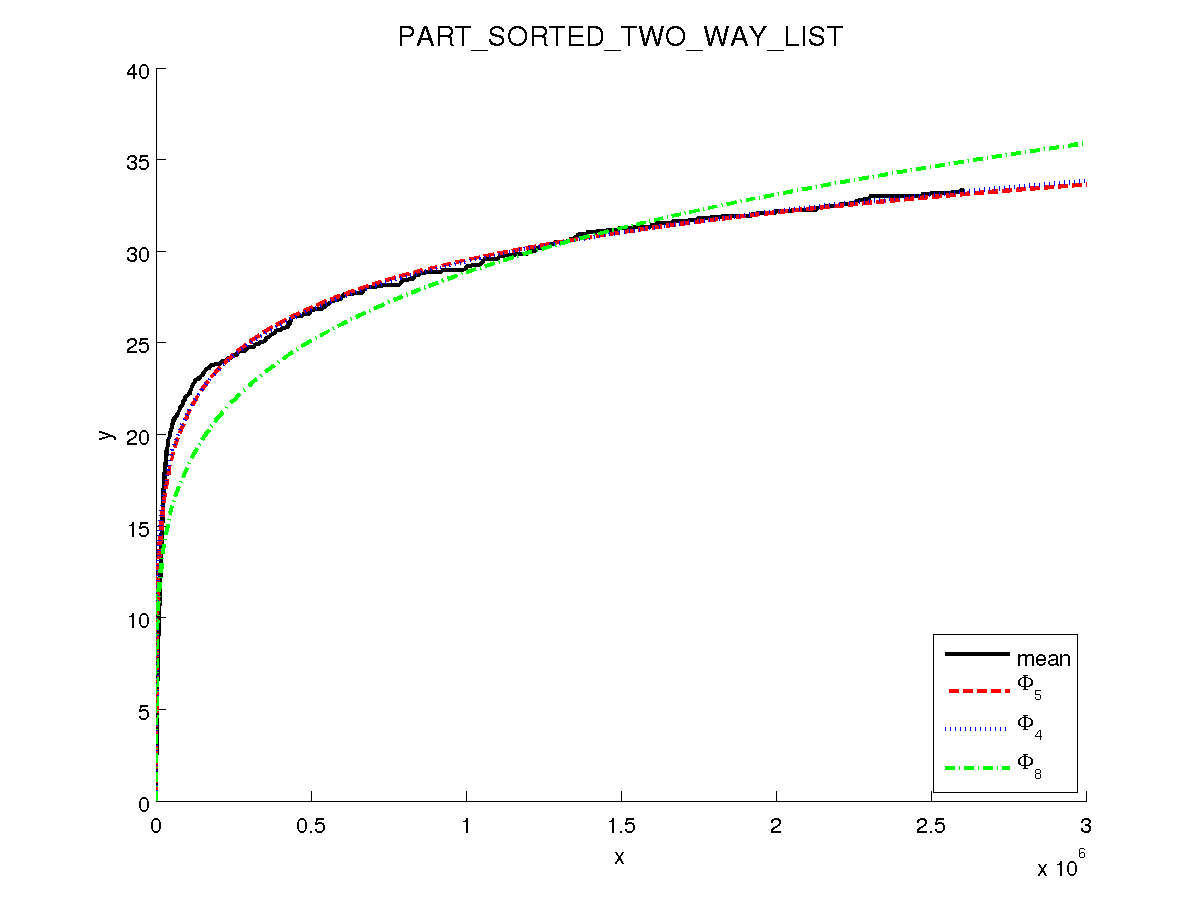} &
    \includegraphics[width={.48\textwidth}]{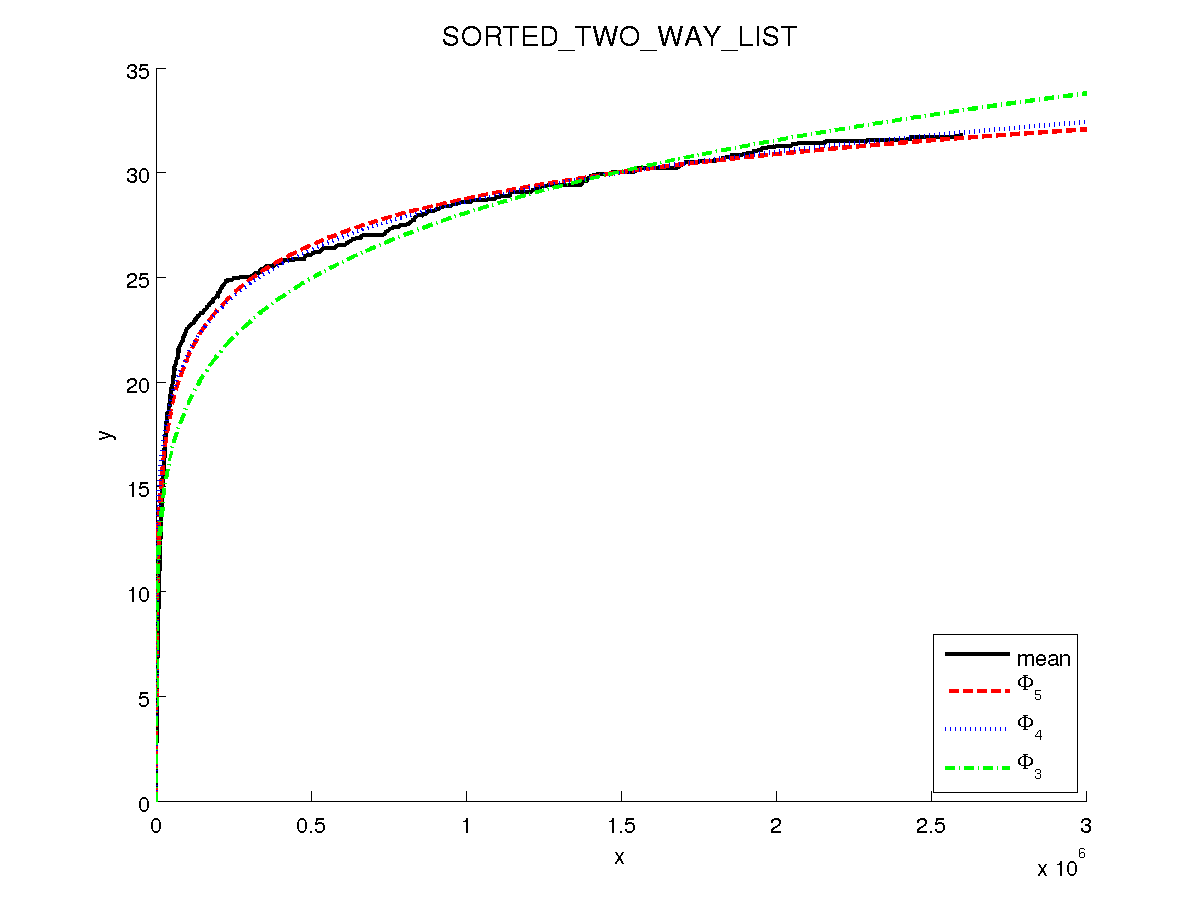}
  \end{array}$
\end{center}
\caption{Top 3 fits with mean for six Eiffel classes.}
\end{figure*}

\begin{figure*}[!htb]
  \begin{center}$
  \begin{array}{cc}
    \includegraphics[width={.48\textwidth}]{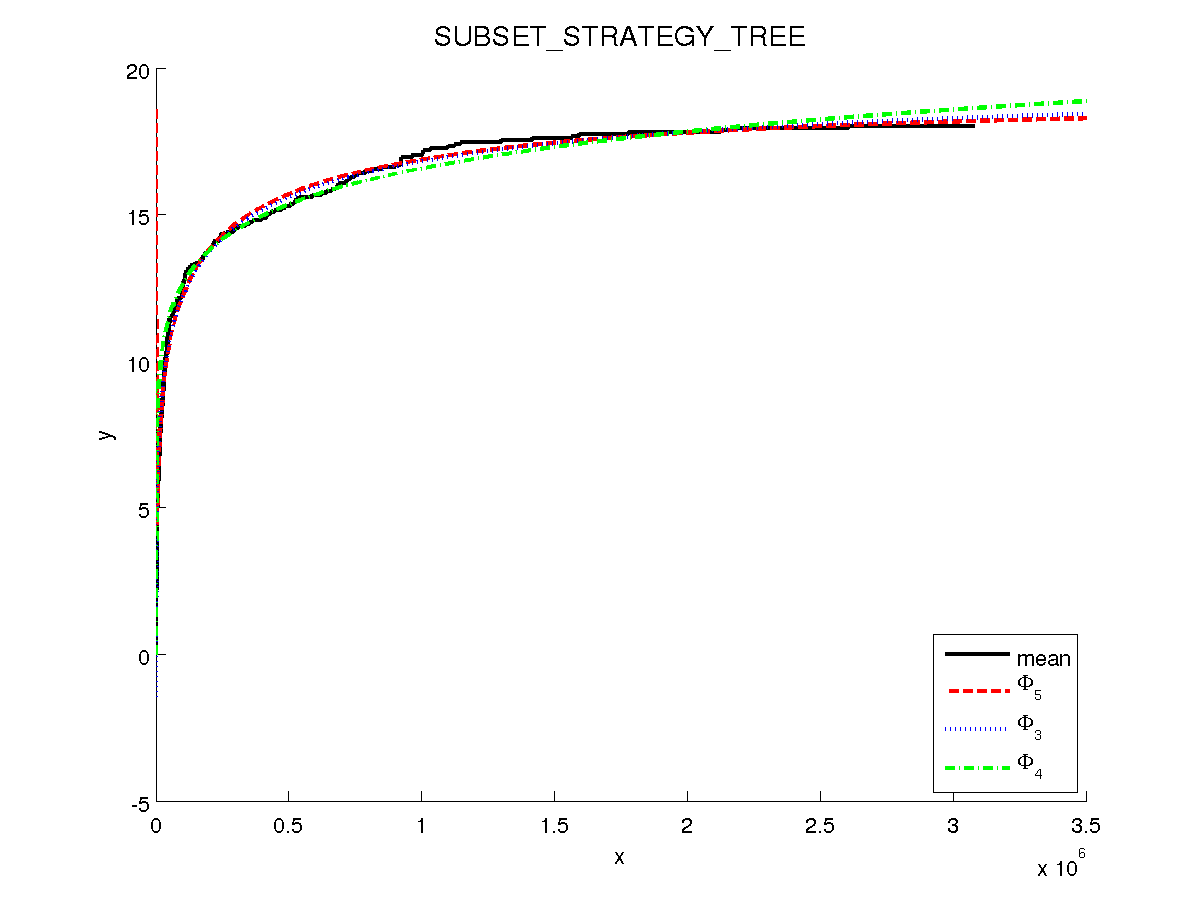} &
    \includegraphics[width={.48\textwidth}]{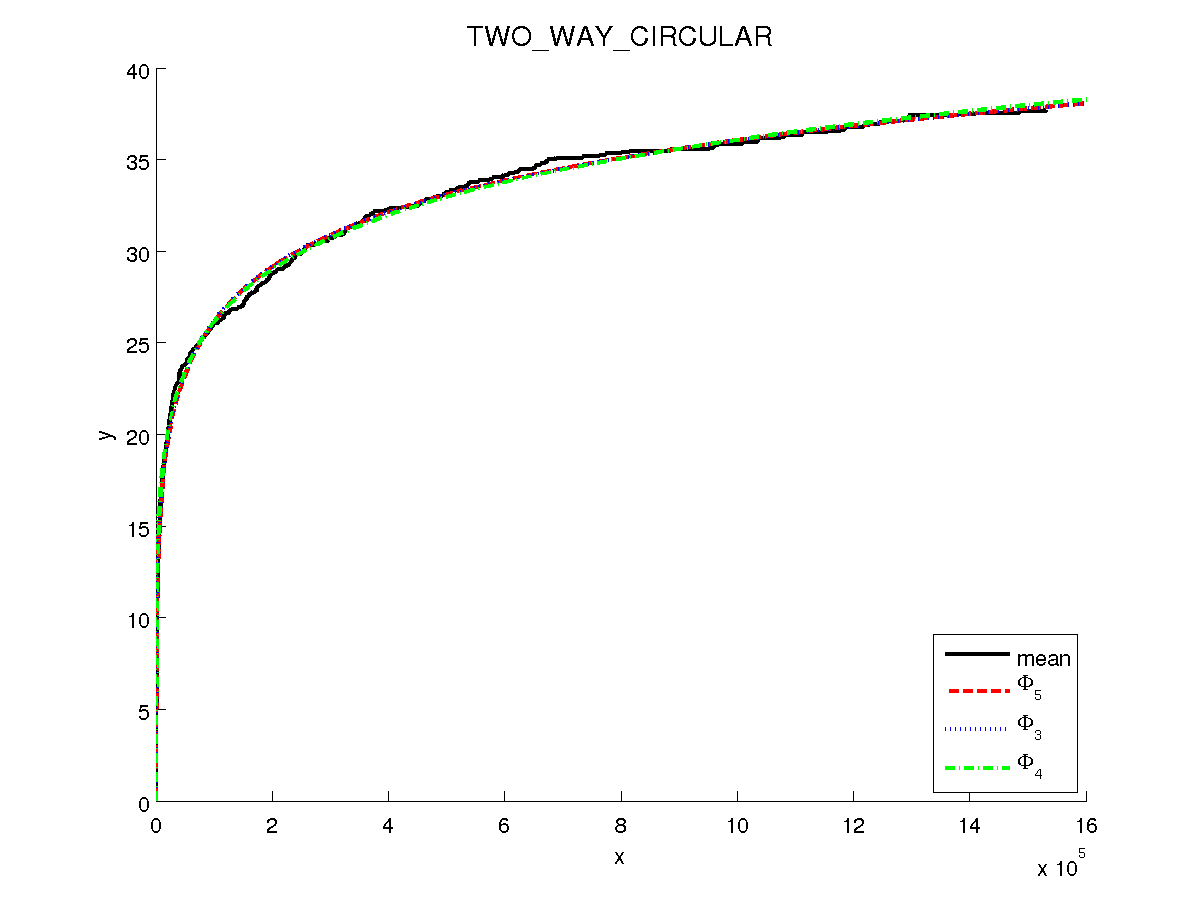} \\
    \includegraphics[width={.48\textwidth}]{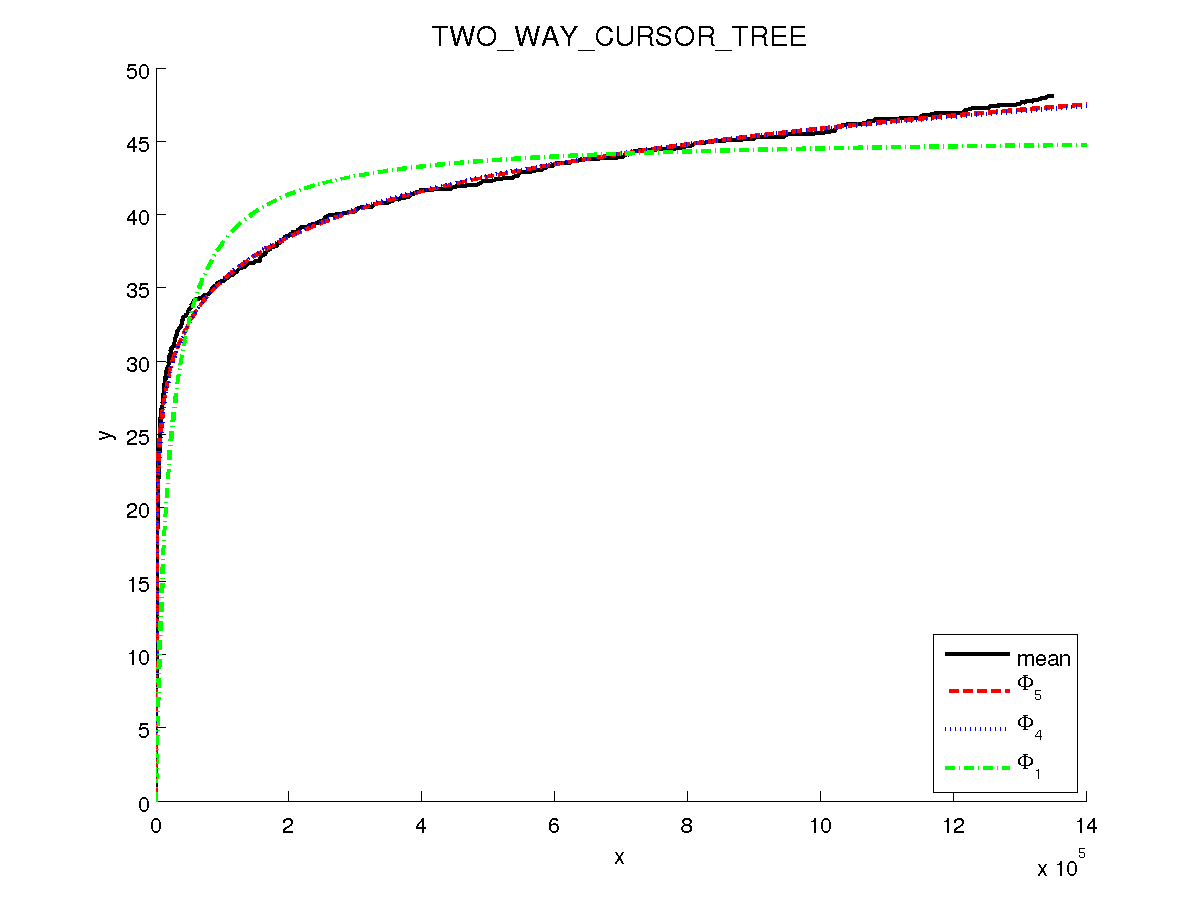} & 
    \includegraphics[width={.48\textwidth}]{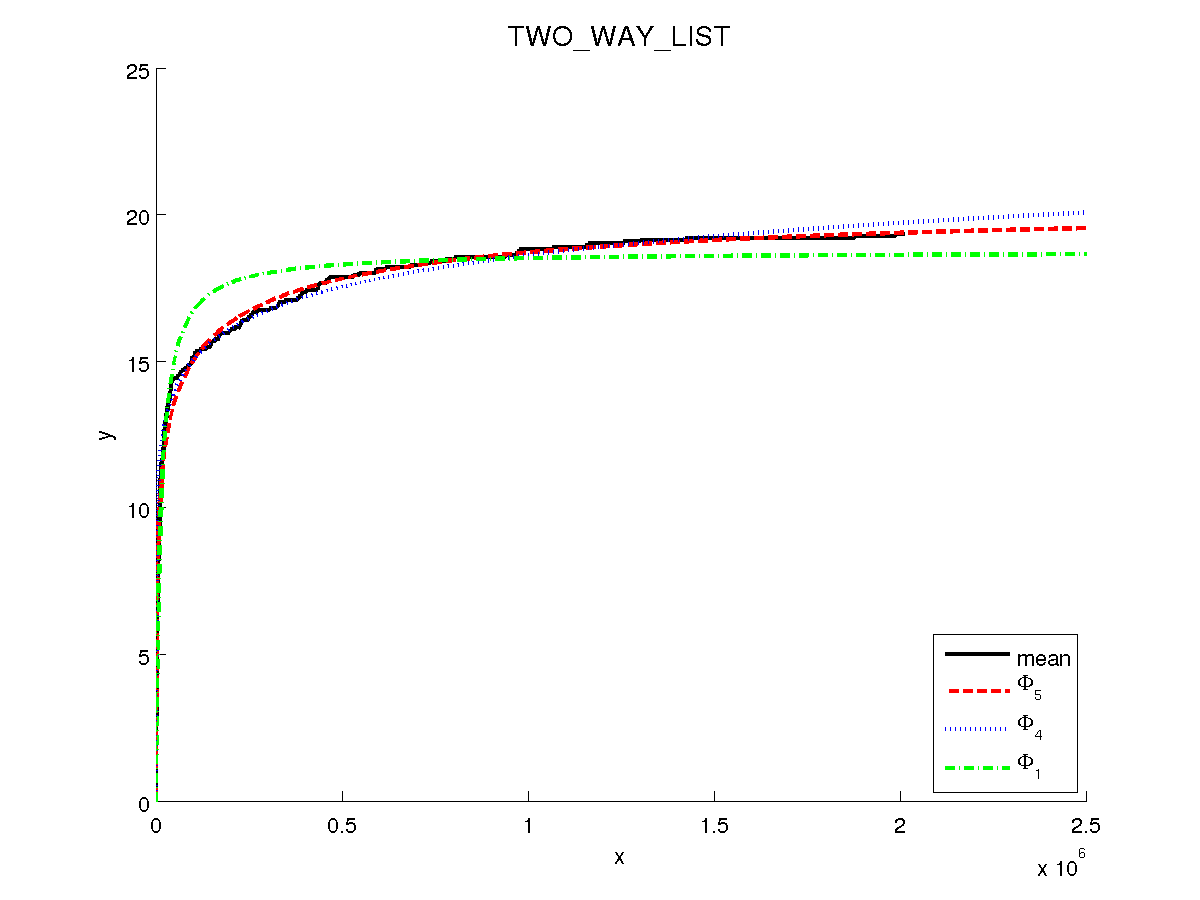} \\
    \includegraphics[width={.48\textwidth}]{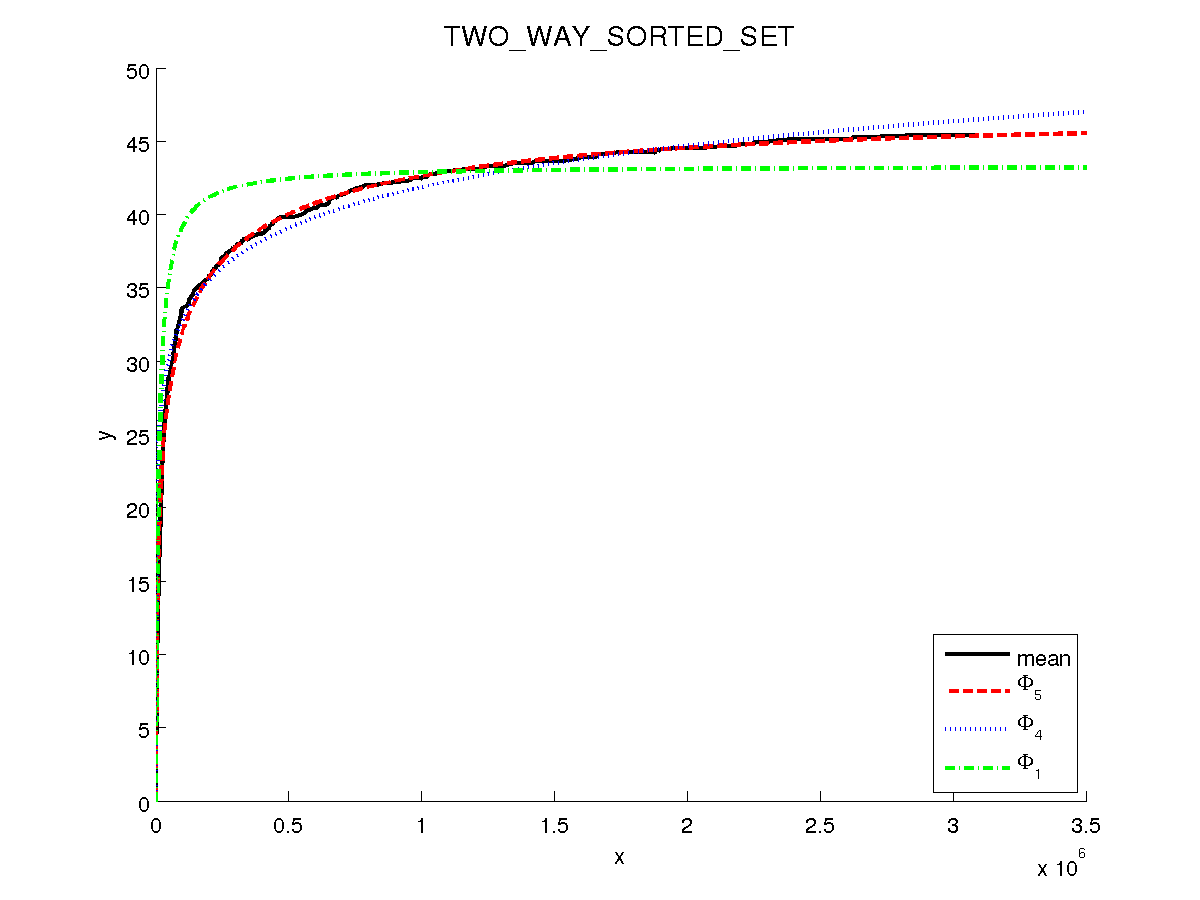} &
    \includegraphics[width={.48\textwidth}]{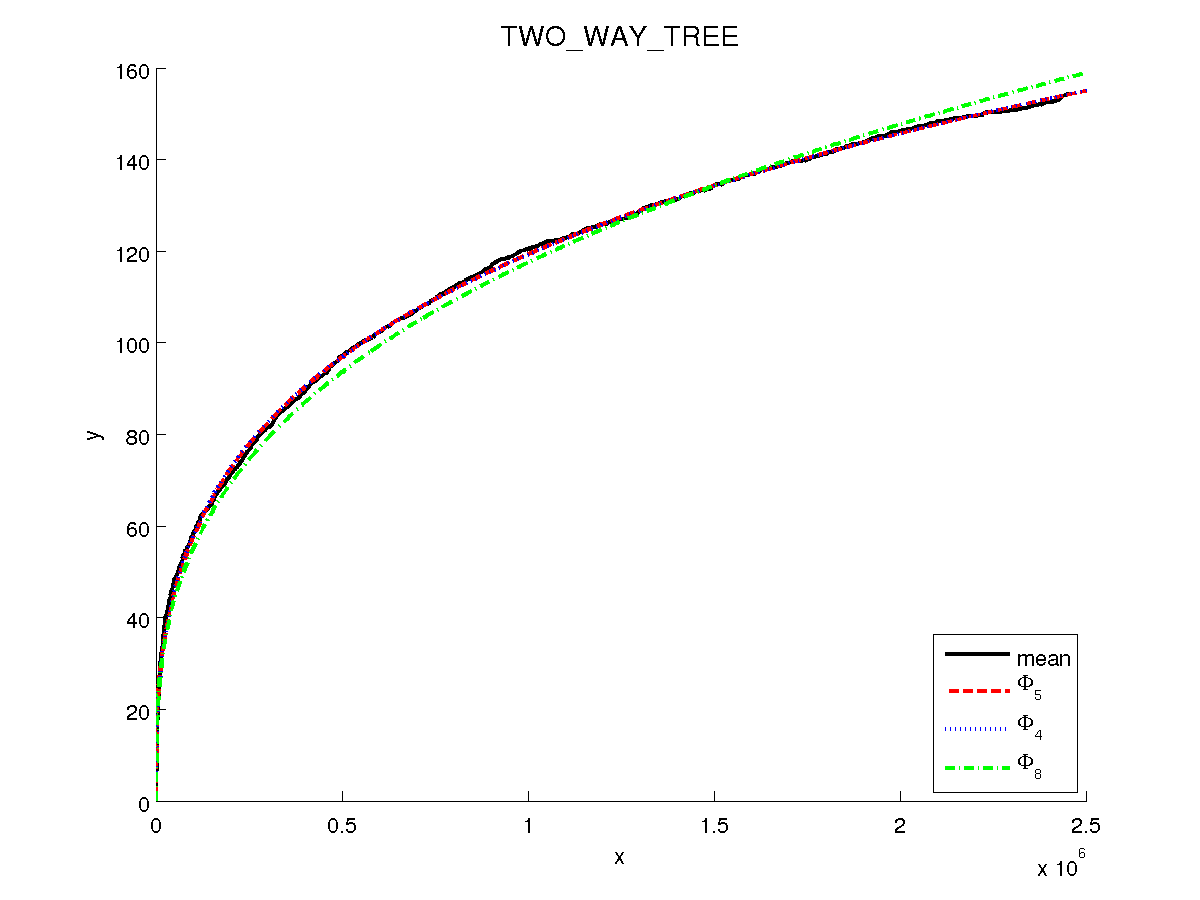} 
  \end{array}$
\end{center}
\caption{Top 3 fits with mean for six Eiffel classes.}
\label{fig:eiffel-last}
\end{figure*}

\clearpage
\subsection{Java failure experiments: all graphs}

Figures~\ref{fig:Javafailure-first}--\ref{fig:Javafailure-last}
display, for each of the 11 Java classes tested for failures, the mean
curve (in black) and the three models that fit best. Horizontal axes
are scaled by tens of thousands of test cases drawn; vertical axes by
total number of failures found.

\begin{figure*}[!htb]
  \begin{center}$
  \begin{array}{cc}
\includegraphics[width={.48\textwidth}]{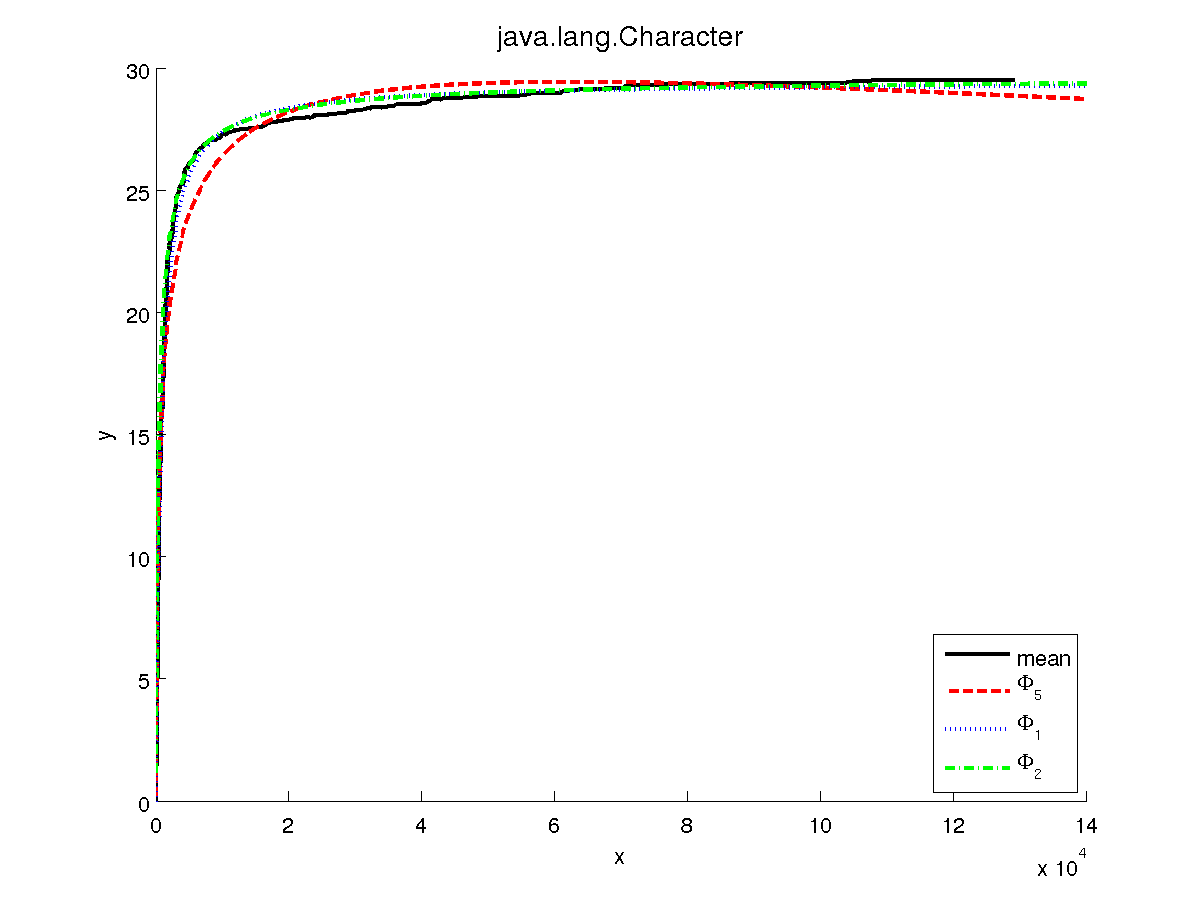} &
\includegraphics[width={.48\textwidth}]{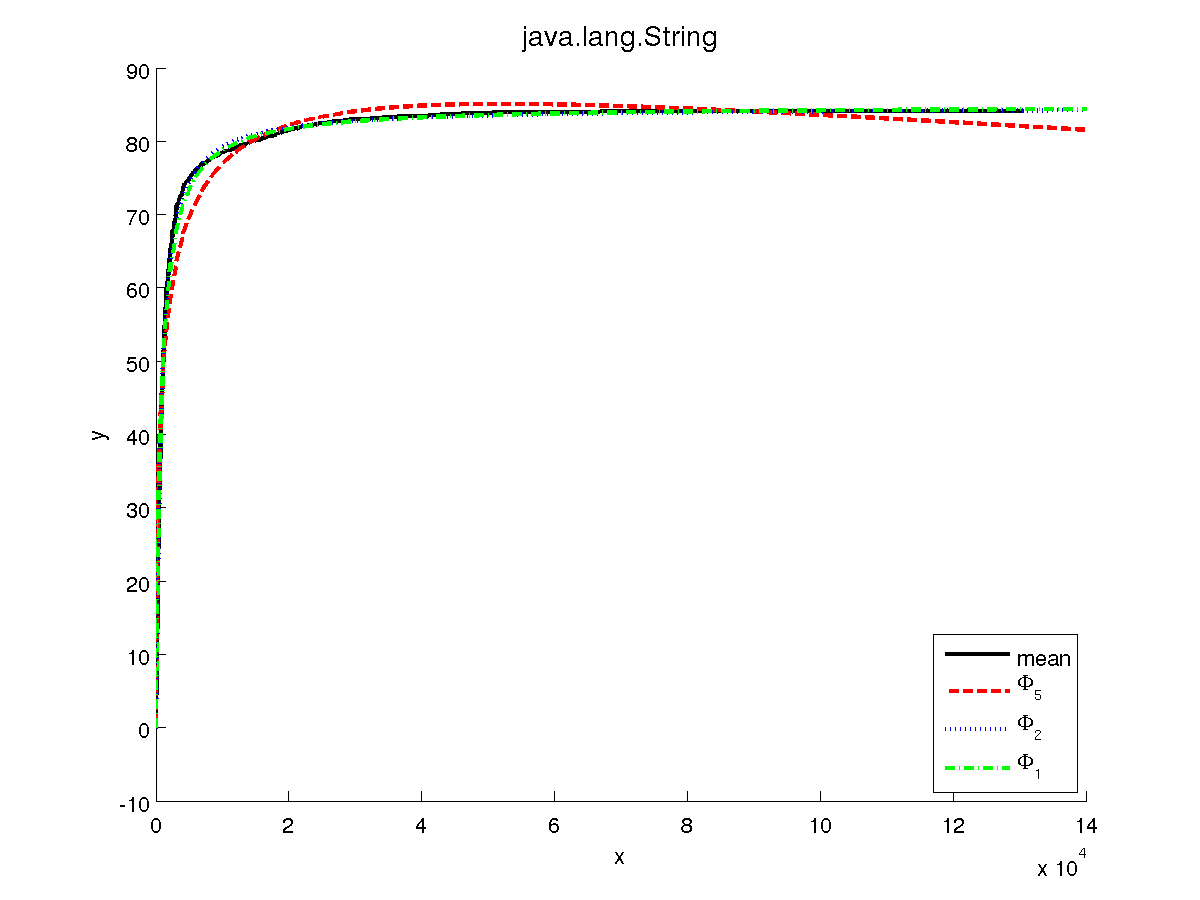} \\
\includegraphics[width={.48\textwidth}]{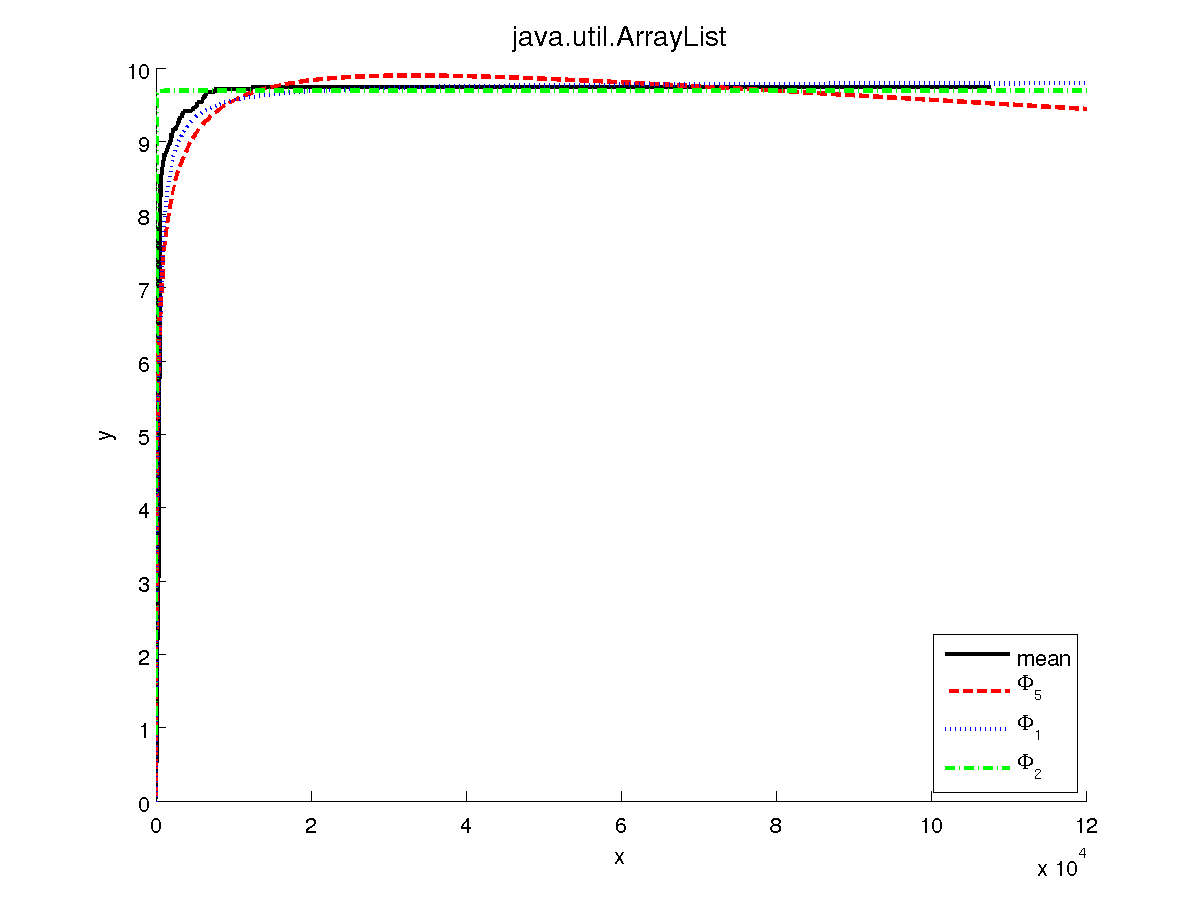} &
\includegraphics[width={.48\textwidth}]{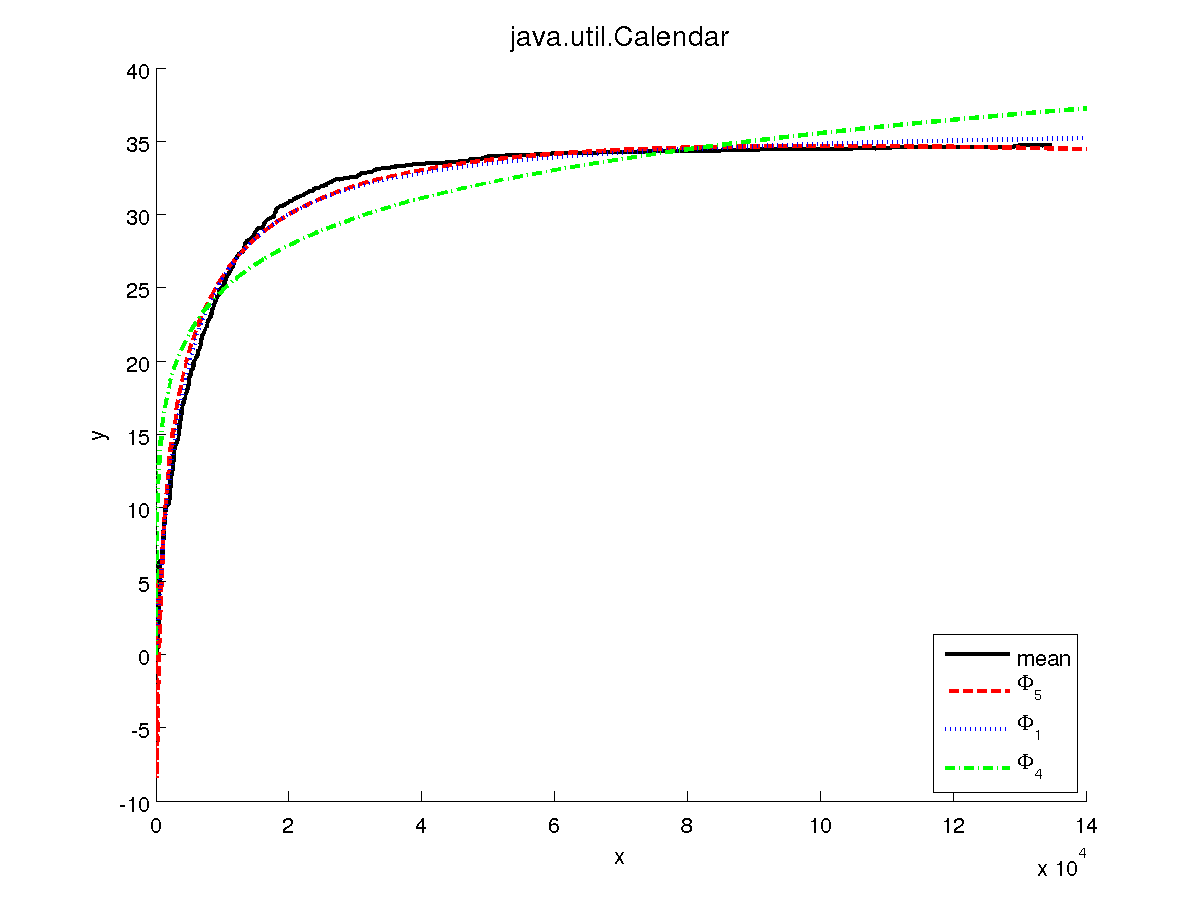} \\ 
\includegraphics[width={.48\textwidth}]{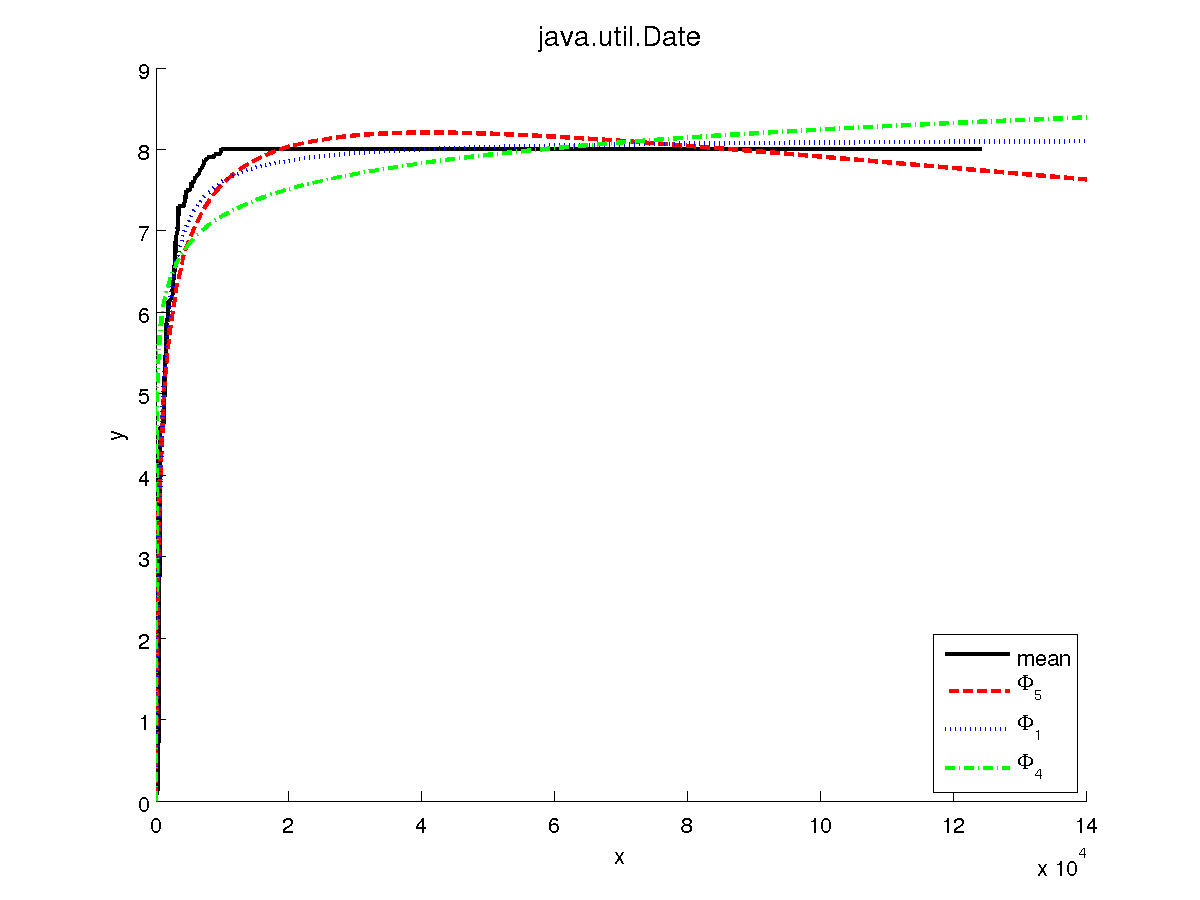} &
\includegraphics[width={.48\textwidth}]{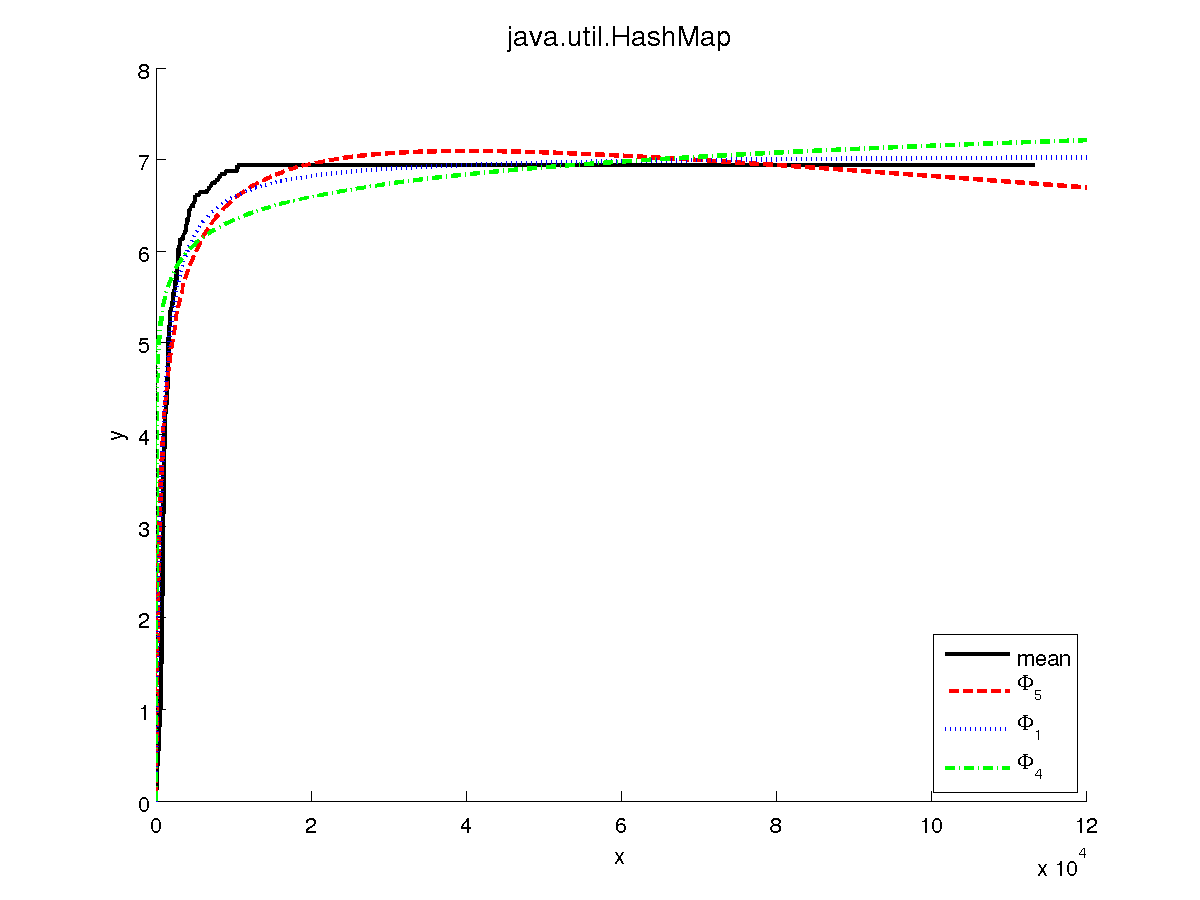}   \end{array}$
\end{center}
\caption{Top 3 fits with mean for six Java classes (failures).}
\label{fig:Javafailure-first}
\end{figure*}

\begin{figure*}[!p]
  \begin{center}$
  \begin{array}{cc}
\includegraphics[width={.48\textwidth}]{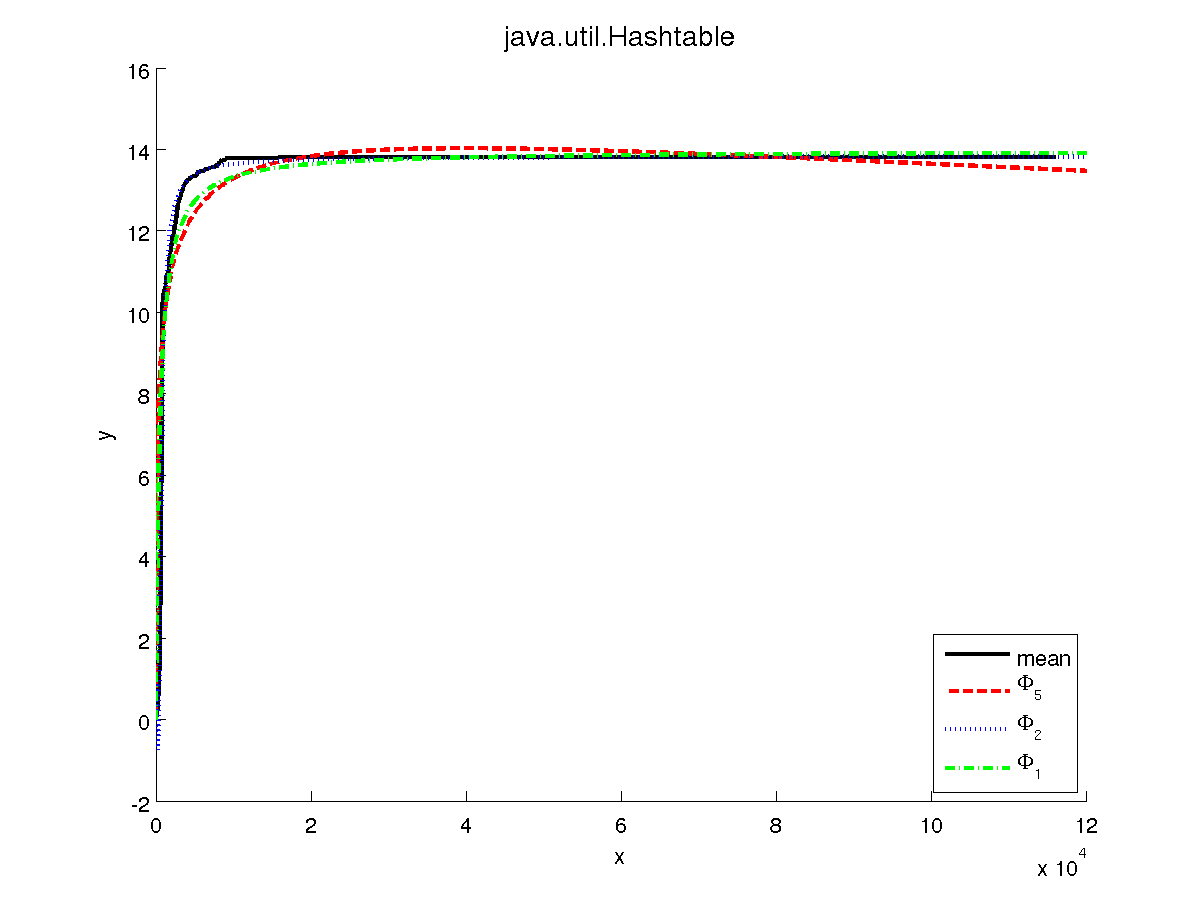} &
\includegraphics[width={.48\textwidth}]{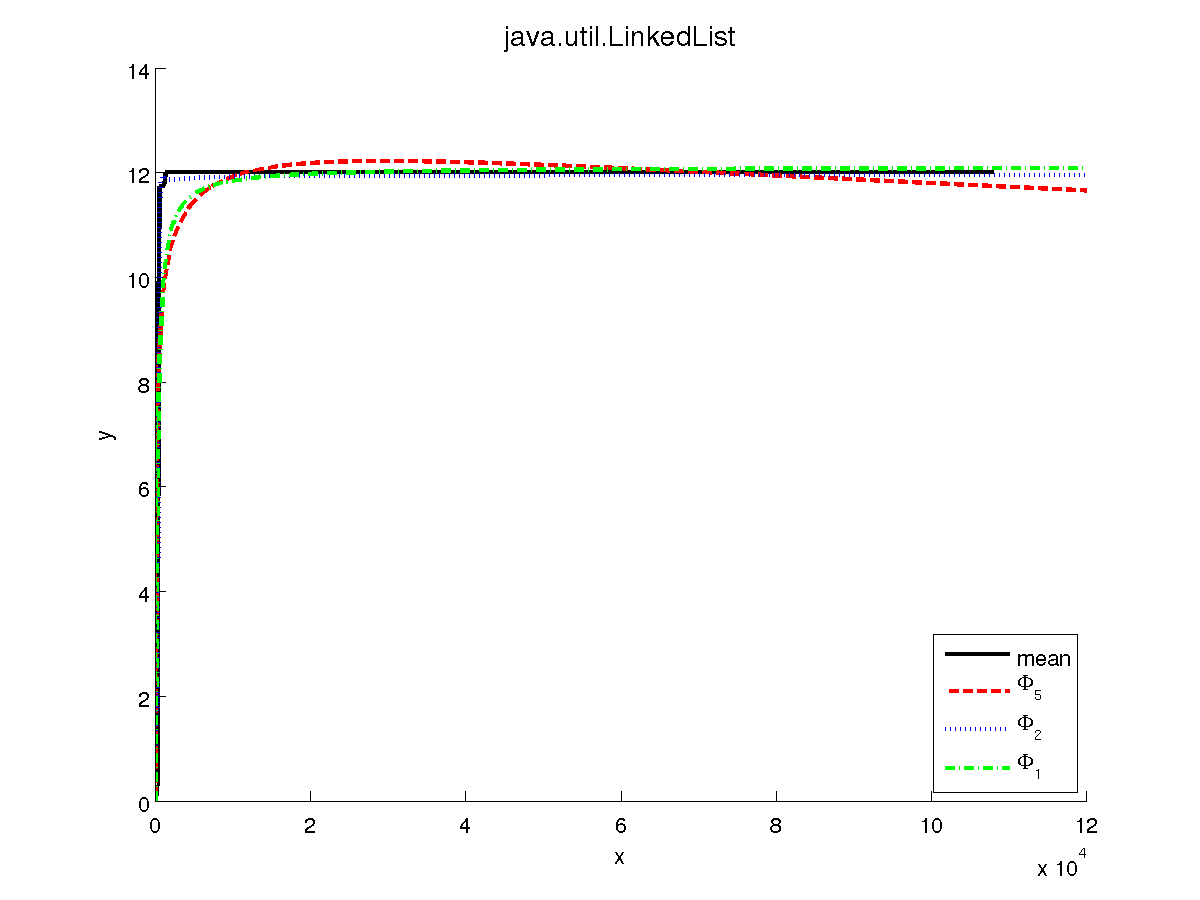} \\
\includegraphics[width={.48\textwidth}]{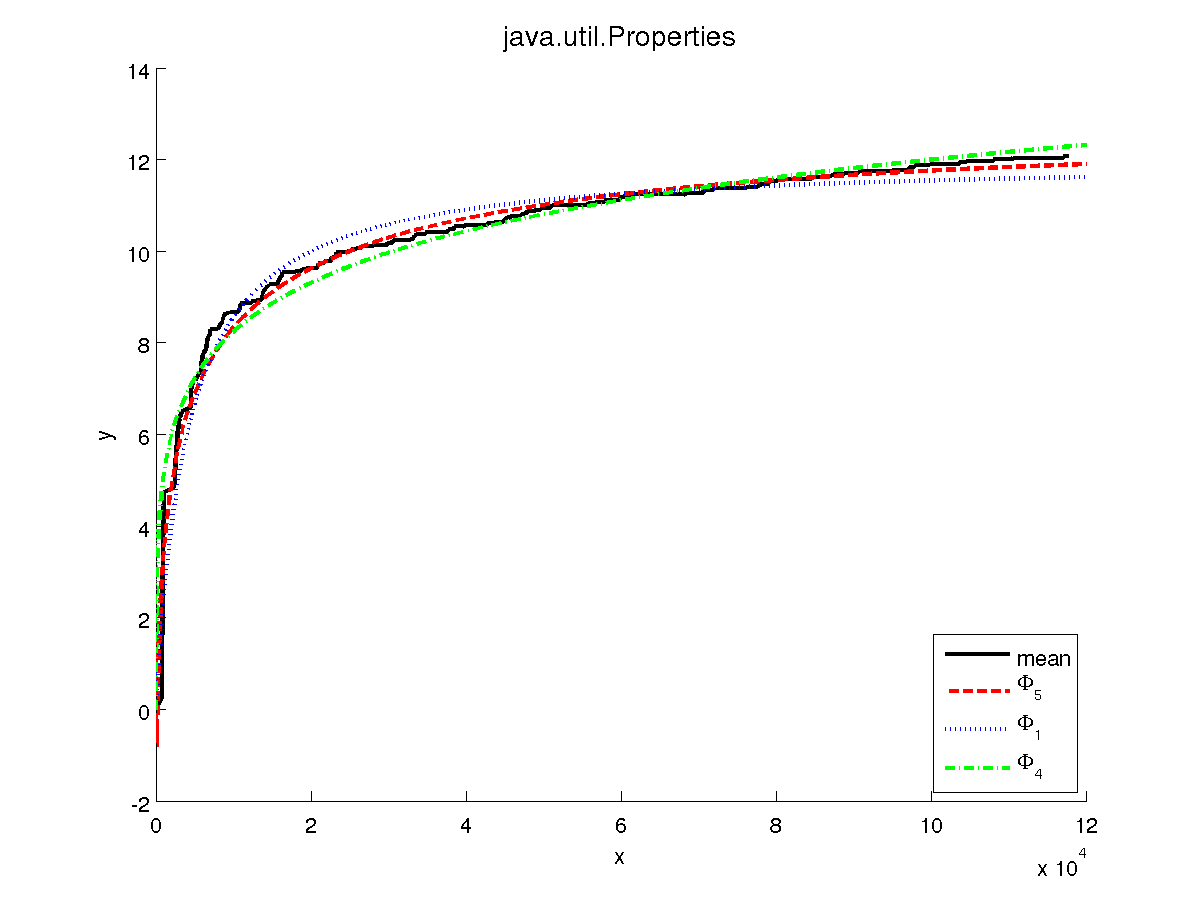} &
\includegraphics[width={.48\textwidth}]{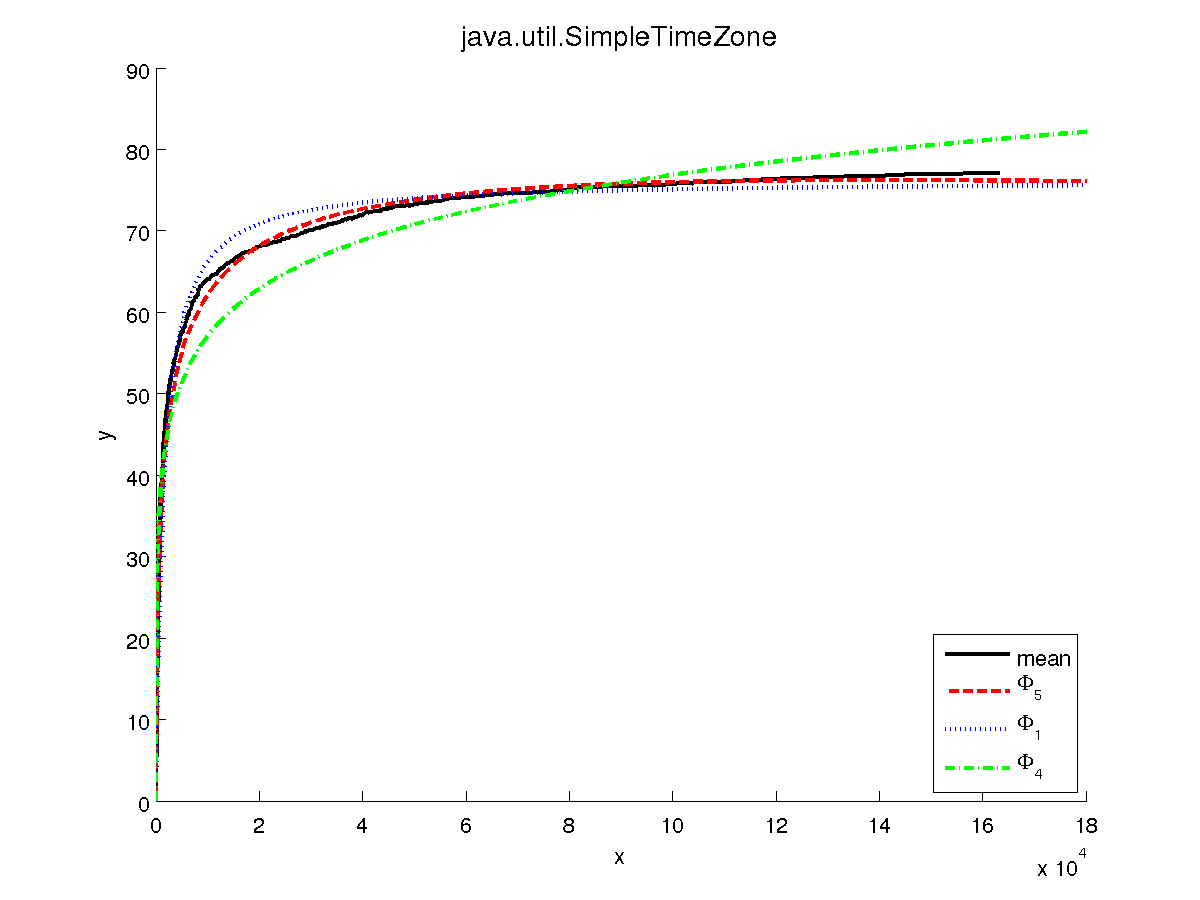} \\
\includegraphics[width={.48\textwidth}]{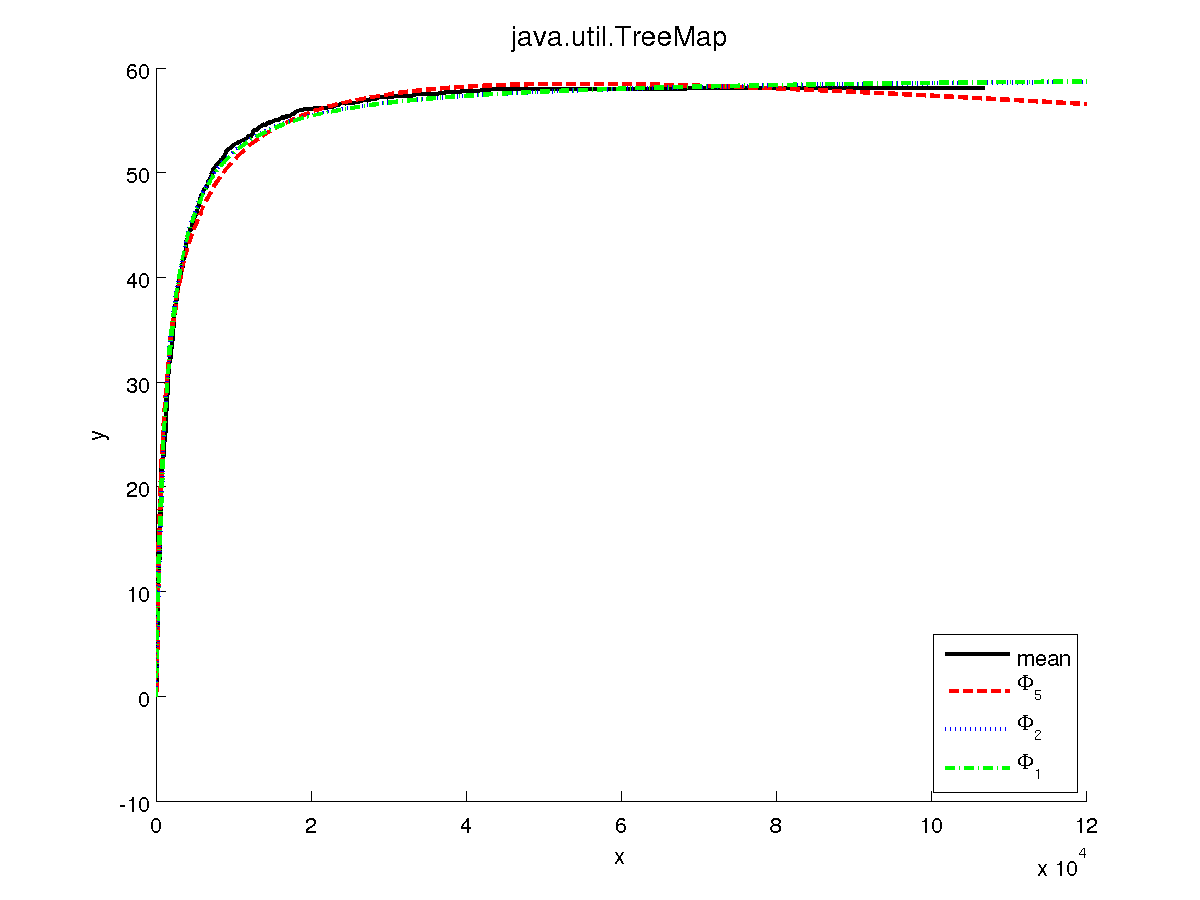}
  \end{array}$
\end{center}
\caption{Top 3 fits with mean for five Java classes (failures).}
\label{fig:Javafailure-last}
\end{figure*}

\clearpage
\subsection{Java fault experiments: all graphs}

Figures~\ref{fig:Javafault-first}--\ref{fig:Javafault-last} display,
for each of the 29 Java classes tested for faults where at least one
fault was found, the mean curve (in black) and the three models that
fit best. Horizontal axes are scaled by hundreds of thousands of test
cases drawn (except for class \lstinline|java.lang.StringBuilder|
which is not scaled); vertical axes by total number of faults found.
The last values of the mean curve for \lstinline|StringBuffer| and
\lstinline|StringBuilder| are measurement errors that should be ignored.

\begin{figure*}[!htb]
  \begin{center}$
    \begin{array}{cc}
\includegraphics[width={.48\textwidth}]{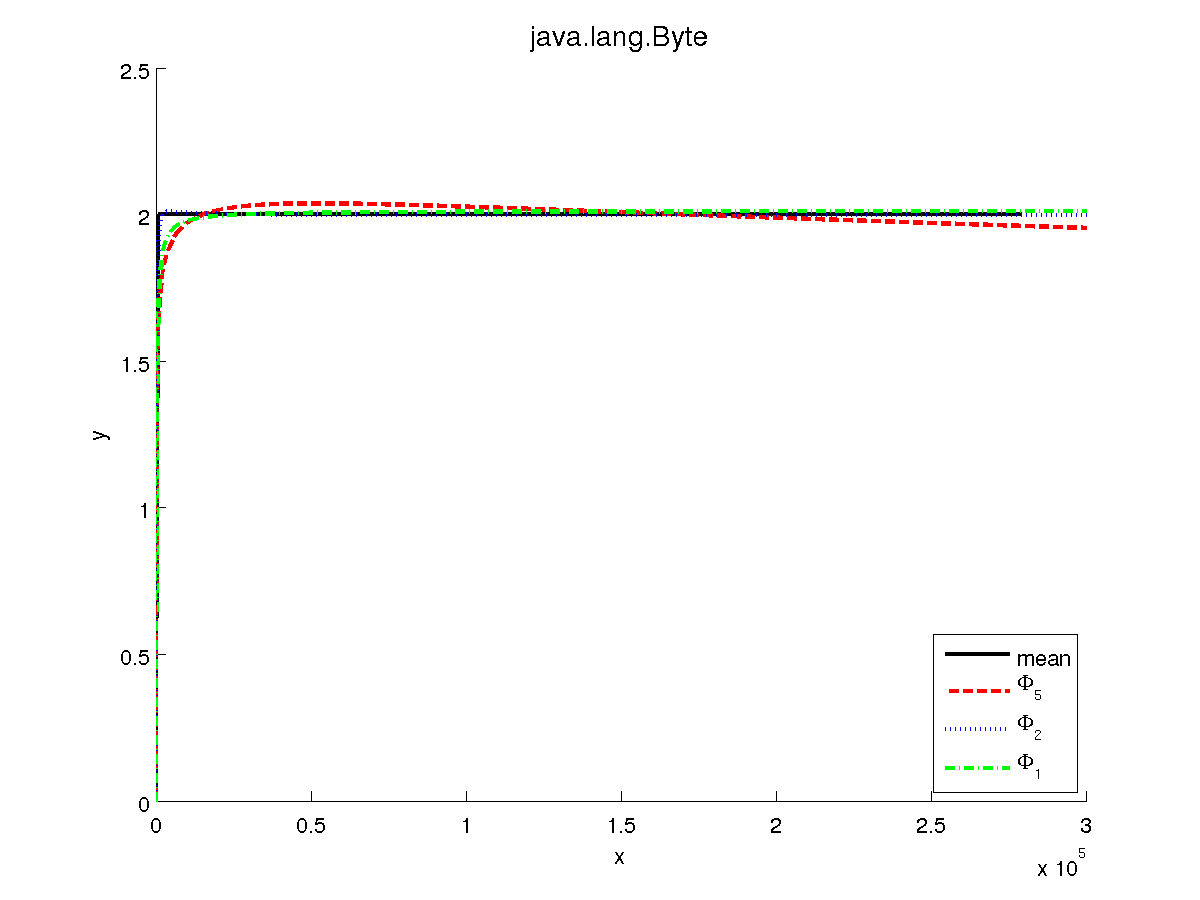} &
\includegraphics[width={.48\textwidth}]{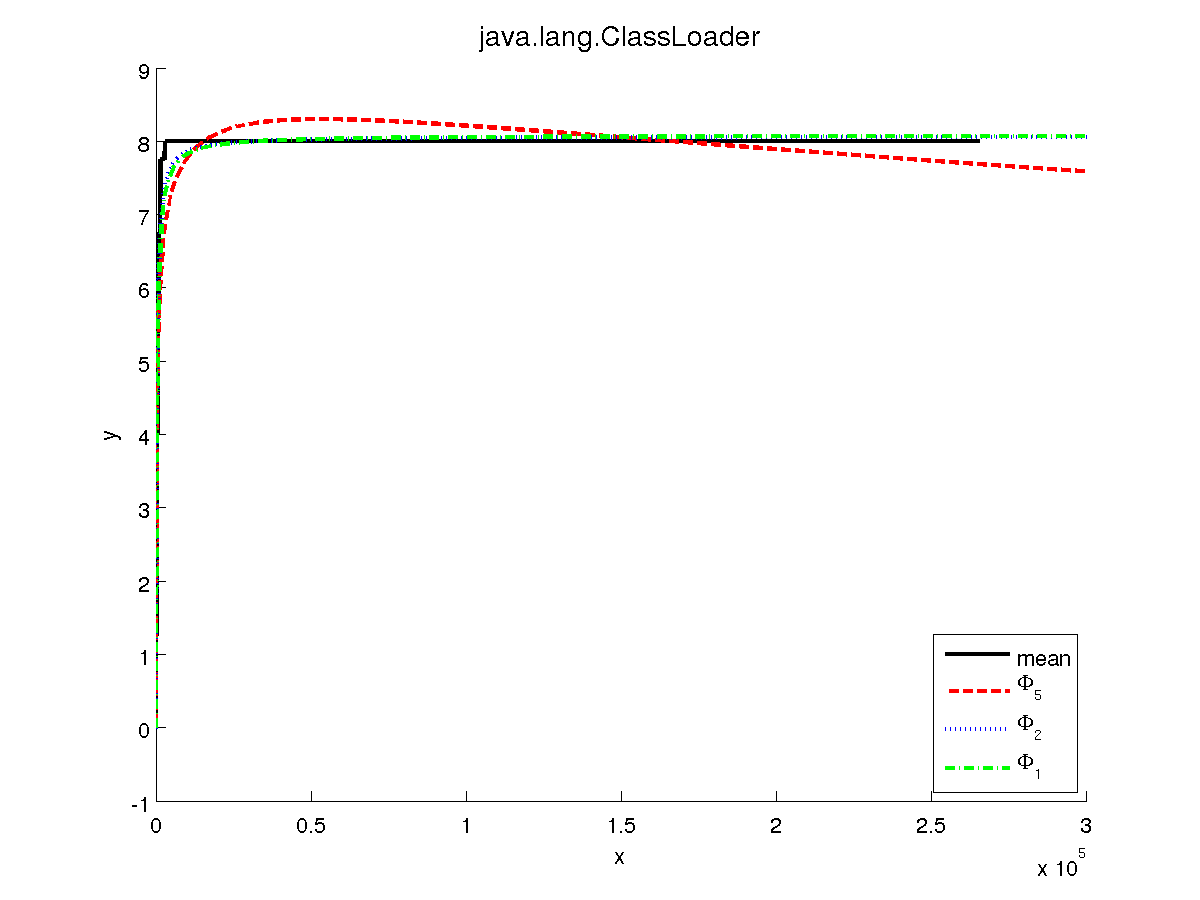} \\
\includegraphics[width={.48\textwidth}]{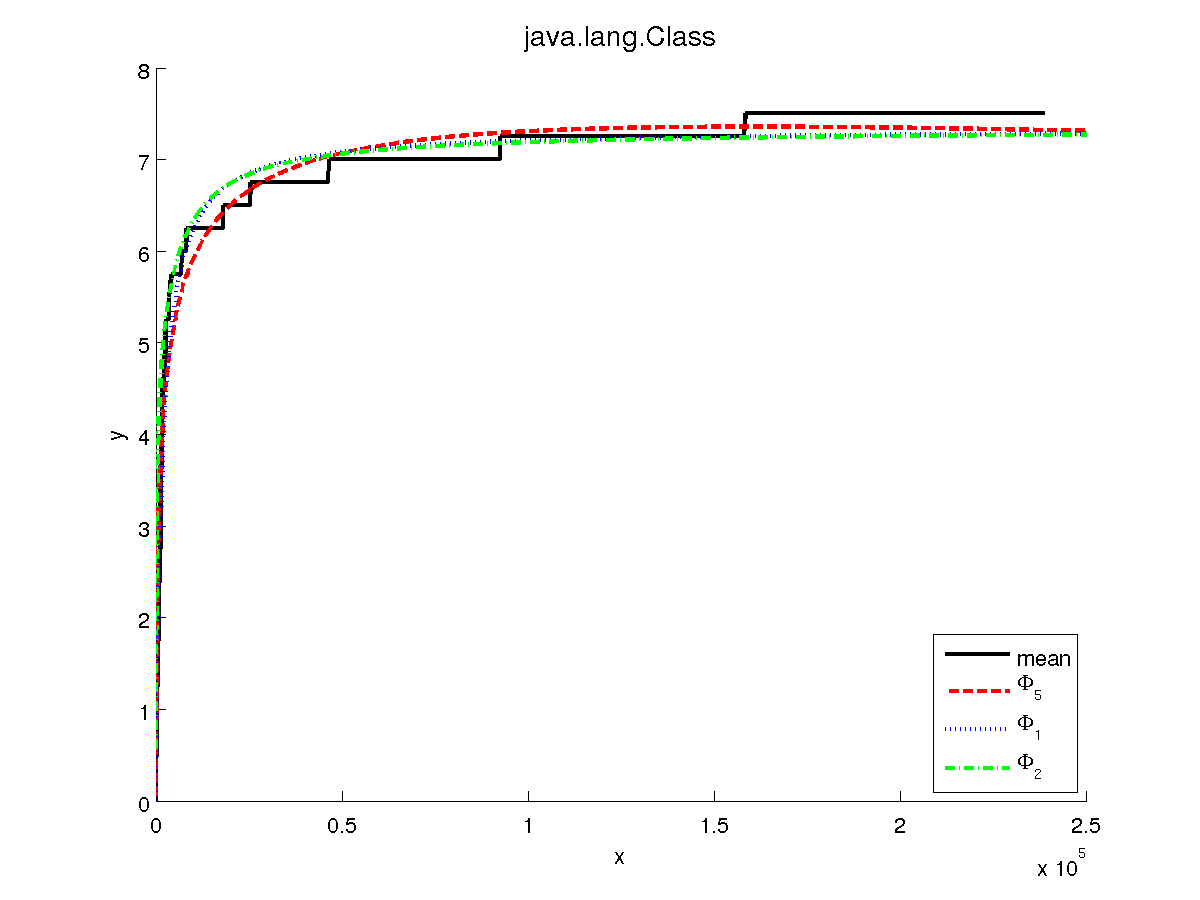} &
\includegraphics[width={.48\textwidth}]{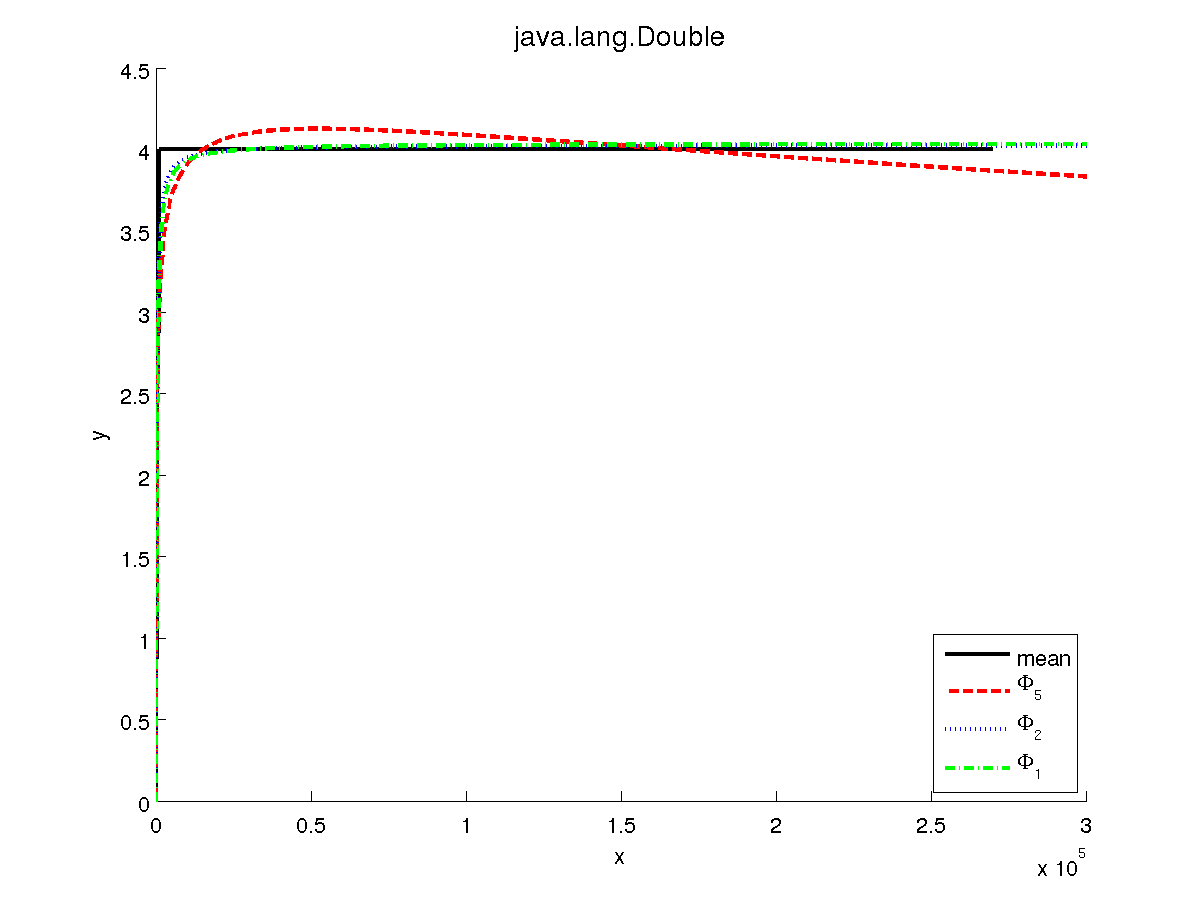} \\
\includegraphics[width={.48\textwidth}]{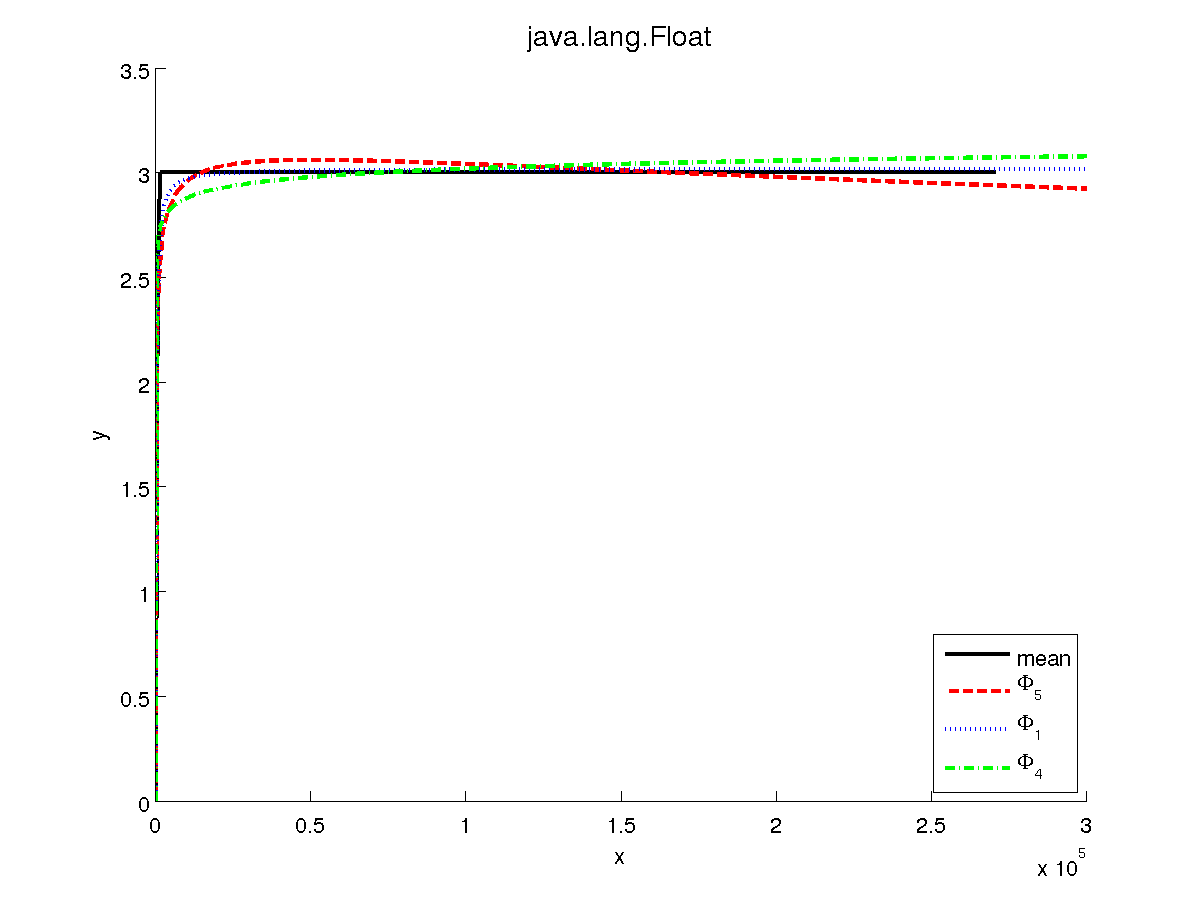} &
\includegraphics[width={.48\textwidth}]{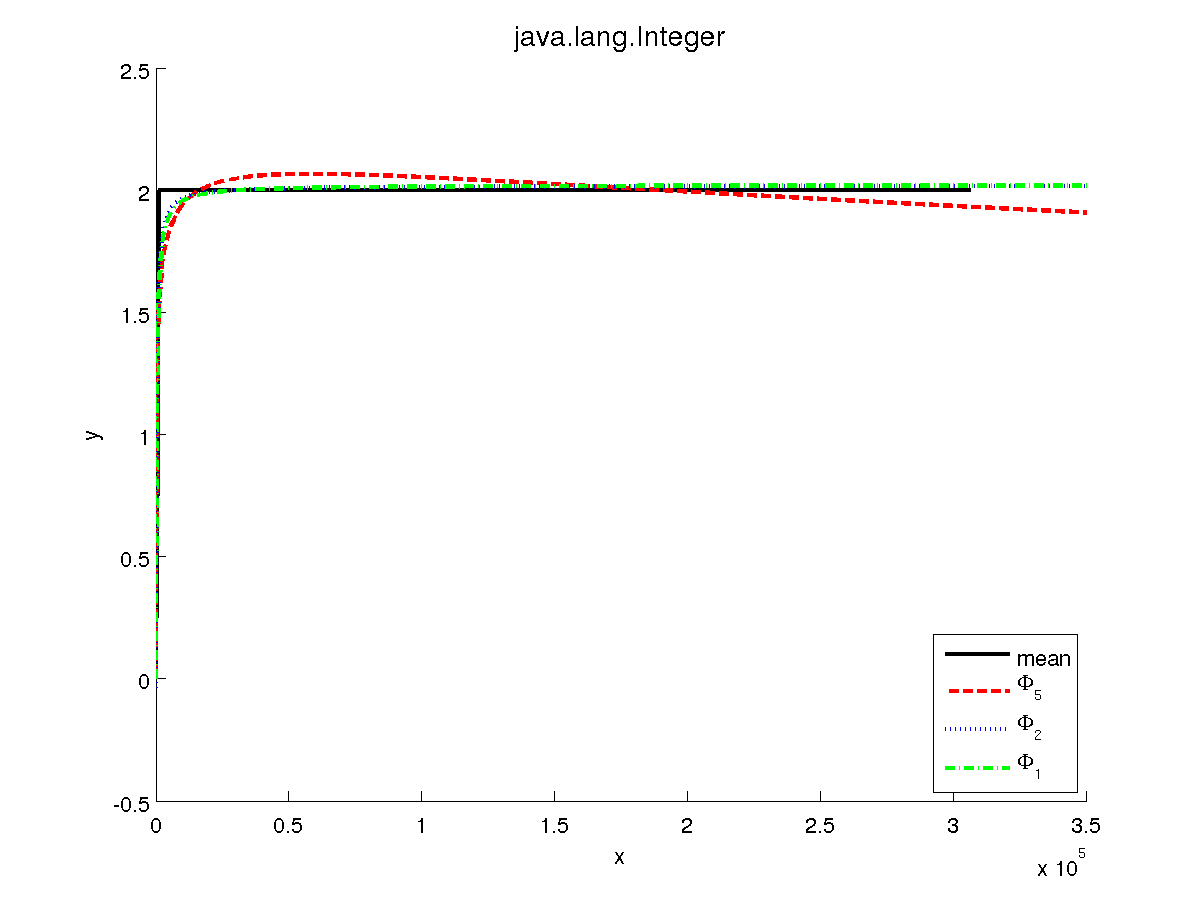}
    \end{array}$
\end{center}
\caption{Top 3 fits with mean for six Java classes (faults).}
\label{fig:Javafault-first}
\end{figure*}

\begin{figure*}[!p]
  \begin{center}$
    \begin{array}{cc}
\includegraphics[width={.48\textwidth}]{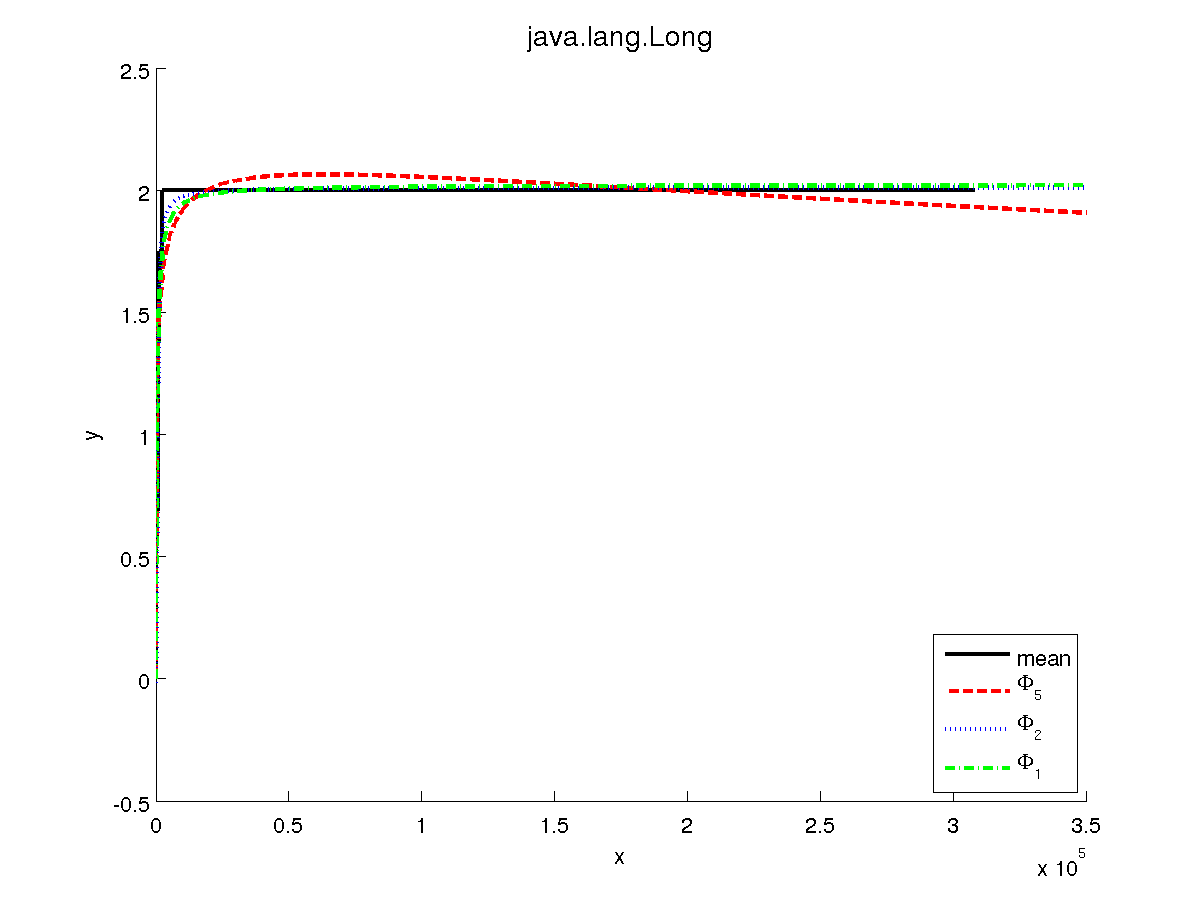} &
\includegraphics[width={.48\textwidth}]{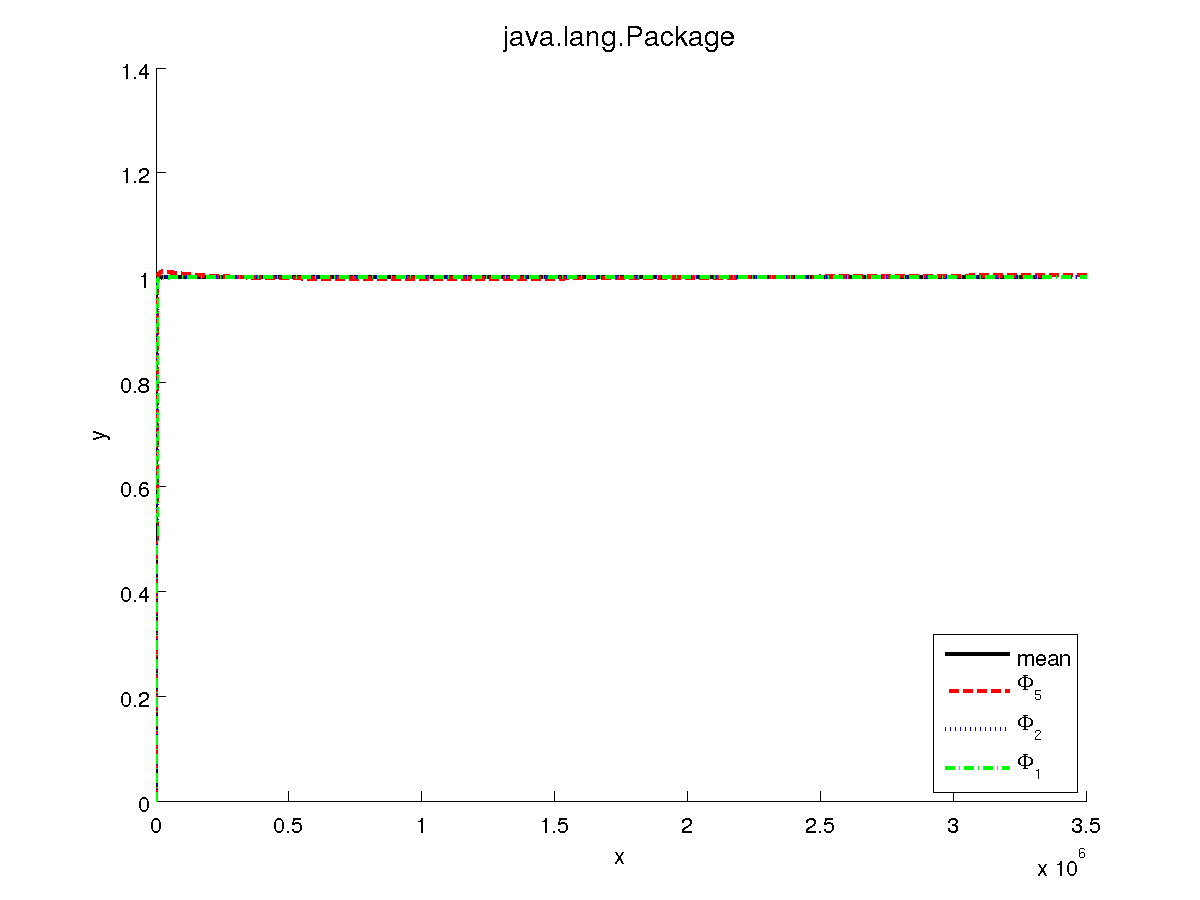} \\
\includegraphics[width={.48\textwidth}]{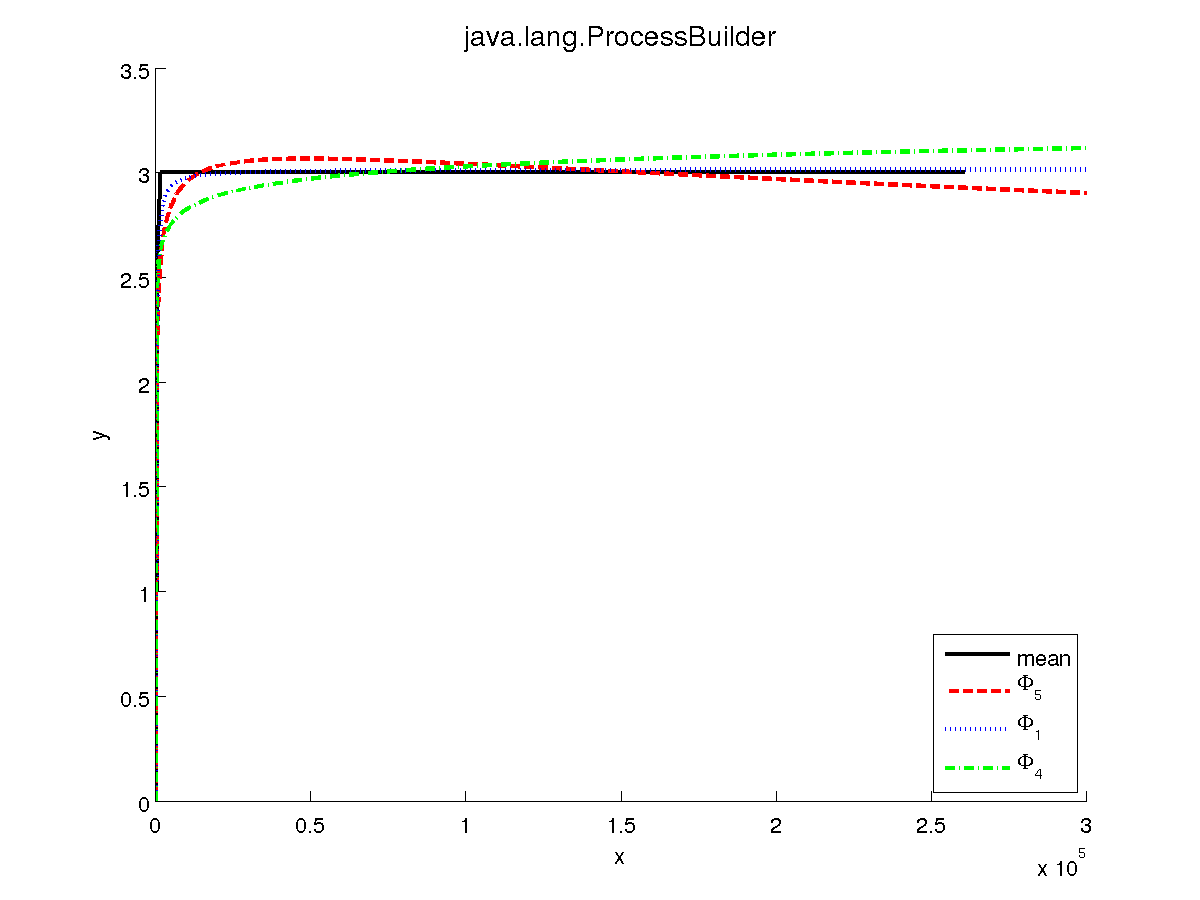} &
\includegraphics[width={.48\textwidth}]{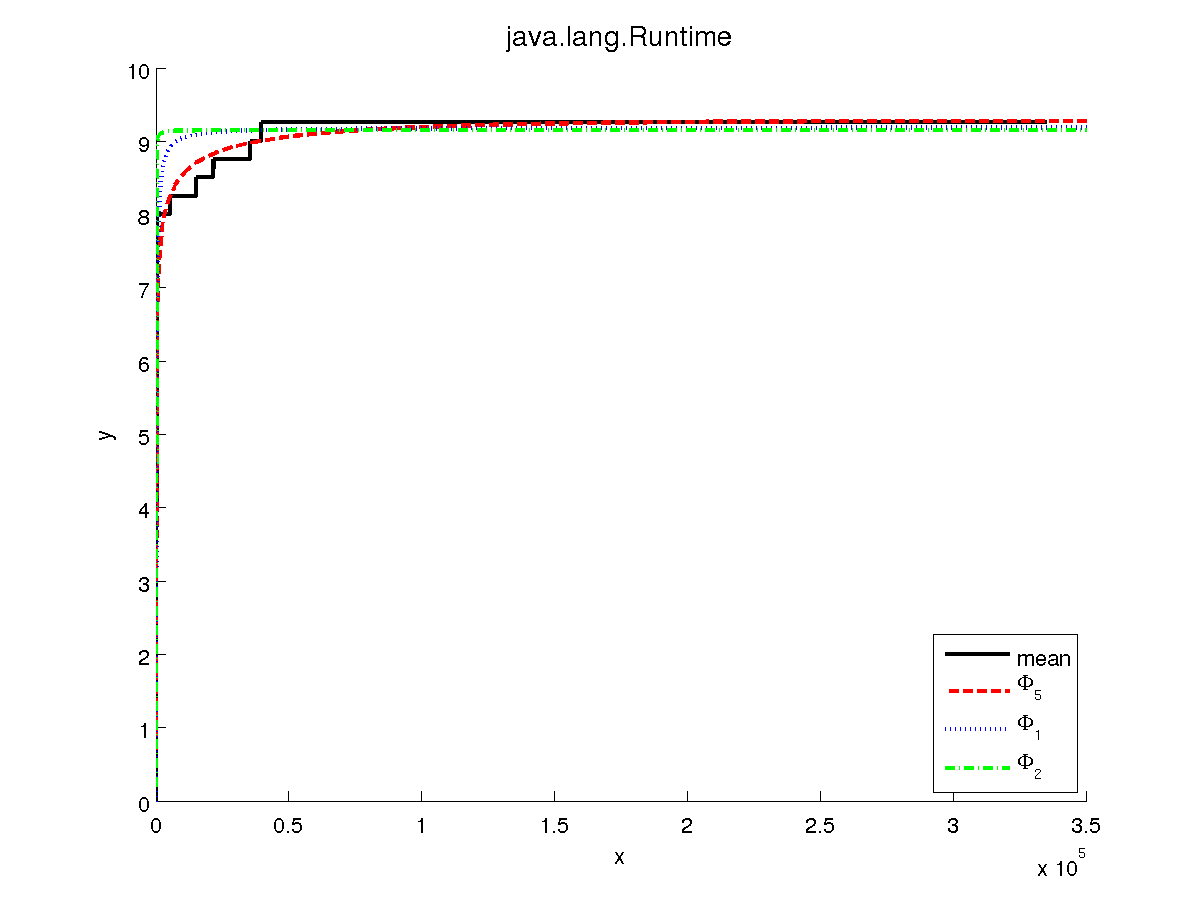} \\
\includegraphics[width={.48\textwidth}]{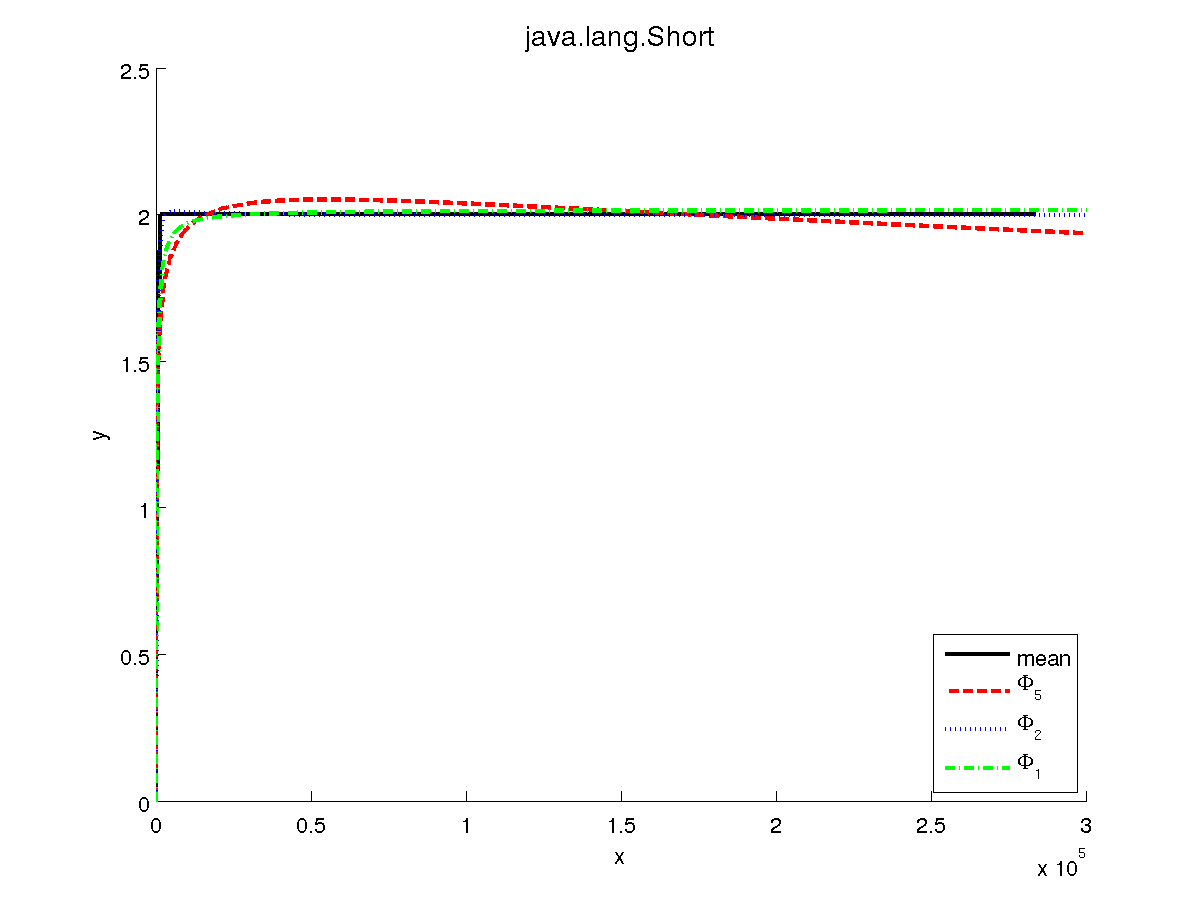} &
\includegraphics[width={.48\textwidth}]{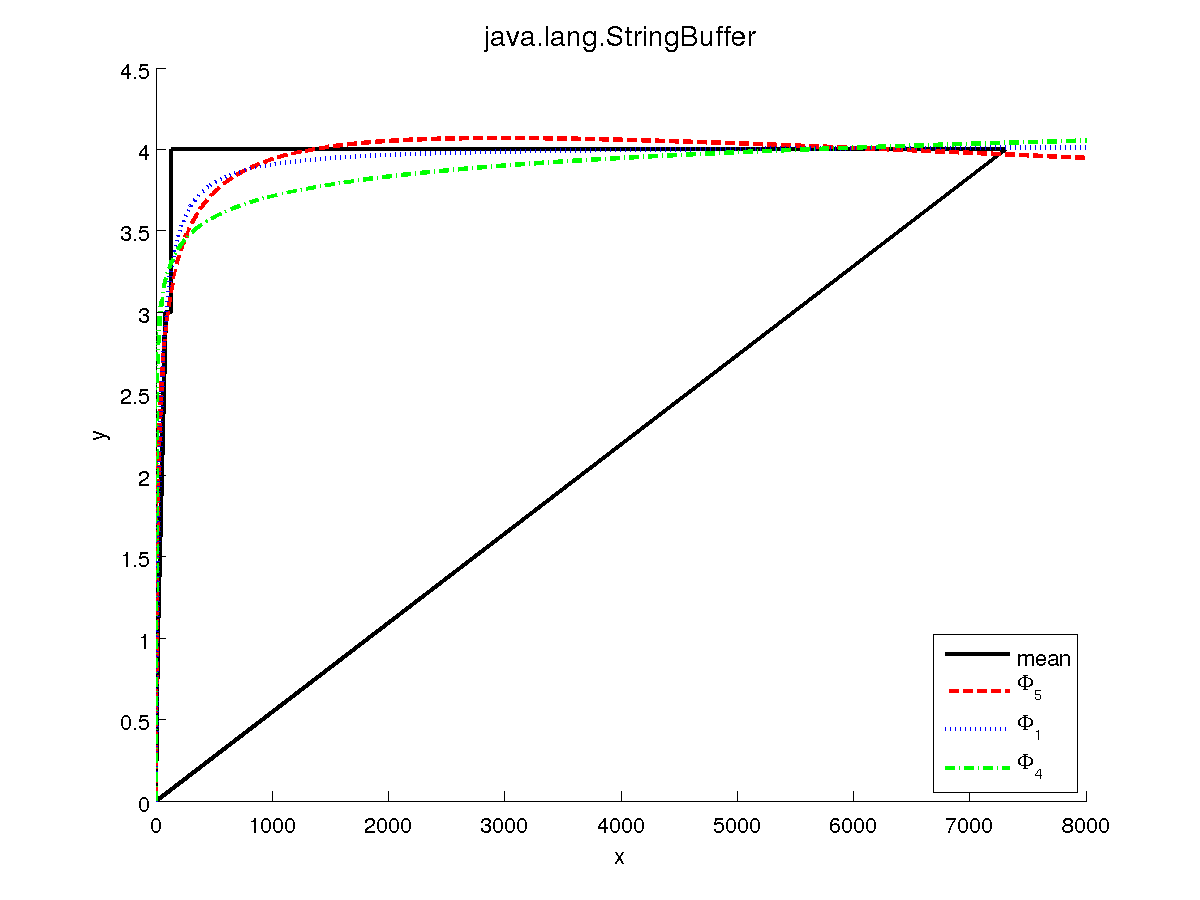} 
    \end{array}$
\end{center}
\caption{Top 3 fits with mean for six Java classes (faults).}
\end{figure*}

\begin{figure*}[!h]
  \begin{center}$
    \begin{array}{cc}
\includegraphics[width={.48\textwidth}]{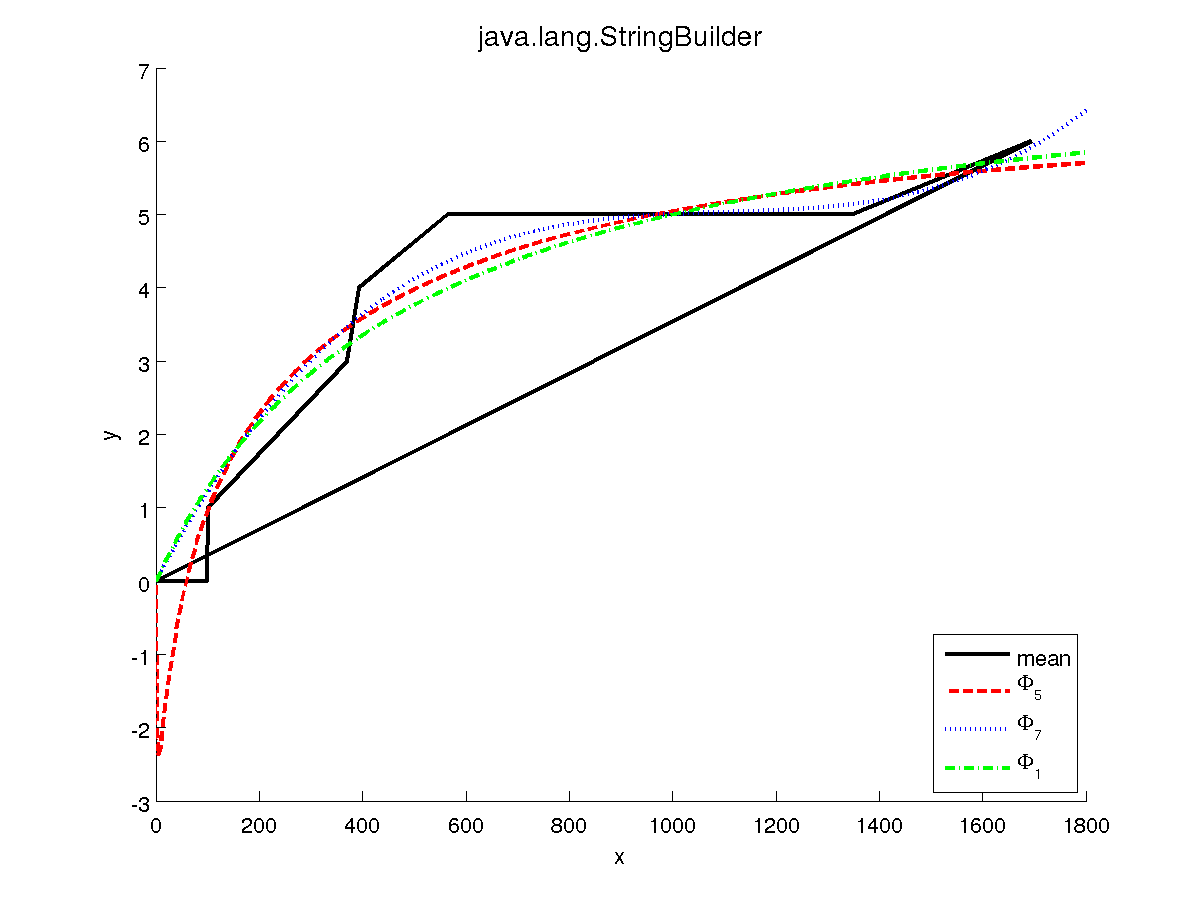} &
\includegraphics[width={.48\textwidth}]{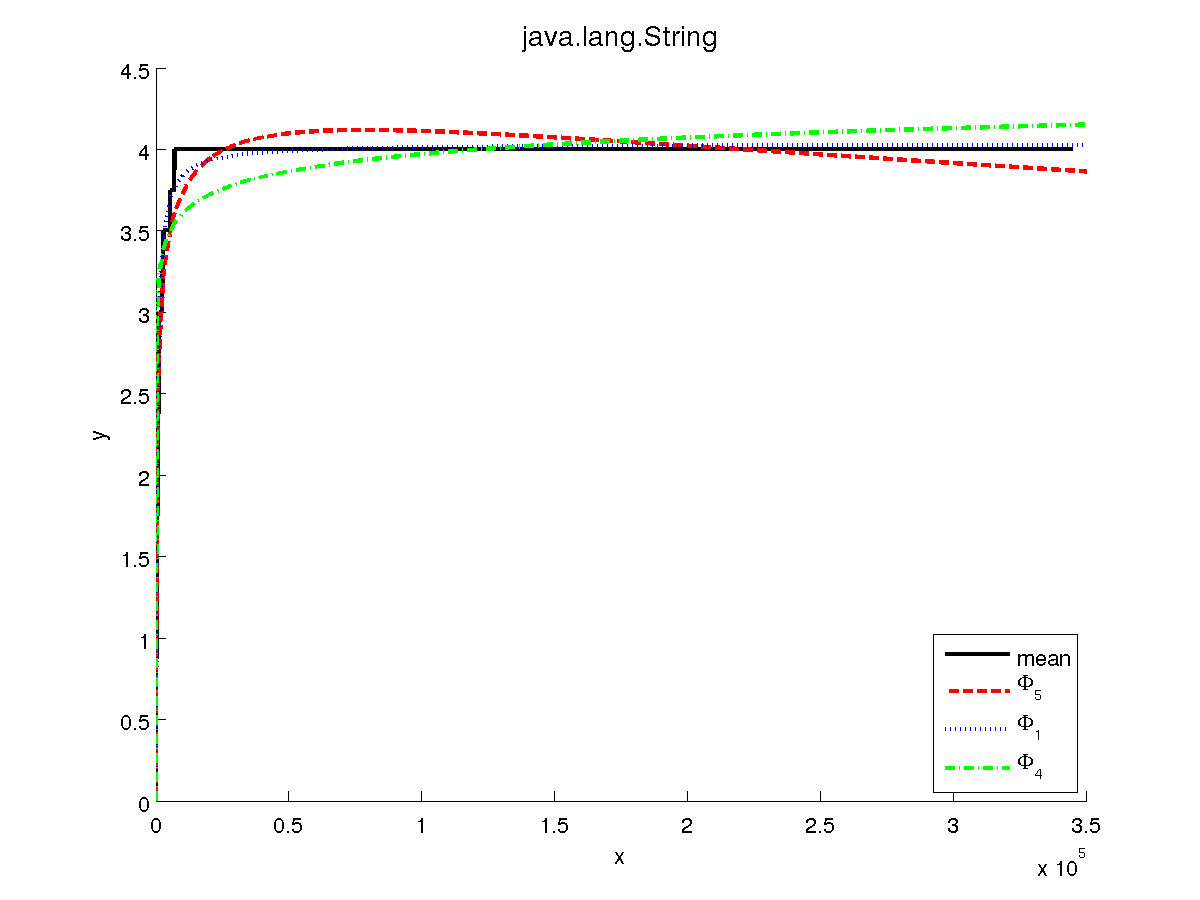}
    \end{array}$
\end{center}
\caption{Top 3 fits with mean for two Java classes (faults).}
\label{fig:Javafault-last}
\end{figure*}


\fi

\end{document}